\RequirePackage{ifpdf}
\documentclass[12pt,aps,amsmath,latexsym,amsfonts]{JHEP3}
\usepackage{graphicx}
\usepackage{epsfig}
\usepackage{amsfonts,amssymb,amsmath}
\usepackage[hang]{subfigure}
\usepackage{epstopdf}

\usepackage{color}
\let\normalcolor\relax

\newcommand{\be}{\begin{equation}}
\newcommand{\ee}{\end{equation}}
\newcommand{\bea}{\begin{eqnarray}}
\newcommand{\eea}{\end{eqnarray}}

\title{Multiple Reentrant Phase Transitions and Triple Points in Lovelock Thermodynamics}
\author{
Antonia M. Frassino$^{1,2}$\thanks{Email: frassino@fias.uni-frankfurt.de} , 
David Kubiz\v n\'ak$^{3,4}$\thanks{Email: dkubiznak@perimeterinstitute.ca} , 
Robert B. Mann$^{3,4}$\thanks{Email: rbmann@uwaterloo.ca} ,
Fil Simovic$^{4}$\thanks{Email: fil.simovic@gmail.com}\\
$^1${\it Frankfurt Institute for Advanced Studies, Ruth-Moufang-Stra\ss e 1, 
D-60438 Frankfurt am Main, Germany}\\
$^2${\it Institut  f\"{u}r Theoretische Physik, Johann Wolfgang
Goethe-Universit\"{a}t, Max-von-Laue-Stra\ss e 1, D-60438 Frankfurt am Main, Germany} \\
$^3${\it Perimeter Institute for Theoretical Physics, 31 Caroline St. N.,Waterloo, Ontario N2L 2Y5, Canada}\\
$^4${\it Department of Physics and Astronomy, University of Waterloo, Waterloo, Ontario N2L 3G1, Canada}
}

\abstract{We investigate the effects of higher curvature corrections from Lovelock gravity on the phase structure of asymptotically AdS black holes, treating the cosmological constant as a thermodynamic pressure. 
We examine how various thermodynamic phenomena, such as Van der Waals behaviour, reentrant phase transitions (RPT), and tricritical points are manifest for $U(1)$ charged 
black holes in Gauss--Bonnet and 3rd-order Lovelock gravities.  We furthermore observe a new phenomenon of `multiple RPT' behaviour, in which for fixed pressure the small/large/small/large black hole phase transition occurs as the temperature of the system increases. We also find that when the higher-order Lovelock couplings are related in a particular way, a peculiar isolated critical point emerges for hyperbolic black holes and is characterized by non-standard critical exponents. 
}


\begin{document}

\section{Introduction}\label{secIntroduction}

Black hole thermodynamics has remained a subject of interest for four decades. Its study continues to provide us with interesting
clues as to the underlying structure of quantum gravity.   The thermodynamics of (charged) black holes in asymptotically AdS spacetimes
is of particular interest, in part because it is straightforward to define thermodynamic equilibrium, and because of novel phenomena
that emerge,   such as thermal radiation/large AdS black hole phase transitions \cite{HawkingPage:1983}. Interest in asymptotically AdS black holes
increased further once it was seen that they admit a gauge duality description via a dual thermal field theory.

The proposal that the  mass of an AdS black hole should be interpreted as the enthalpy of  spacetime represents an  interesting new development in this subject.   The idea is that the cosmological constant $\Lambda$ be considered as a thermodynamic variable 
\cite{CreightonMann:1995} analogous to pressure in the first law \cite{CaldarelliEtal:2000, KastorEtal:2009, Dolan:2010, Dolan:2011a, Dolan:2011b, Dolan:2012, CveticEtal:2010, LarranagaCardenas:2012, LarranagaMojica:2012, 
Gibbons:2012, KubiznakMann:2012, GunasekaranEtal:2012, BelhajEtal:2012,  LuEtal:2012, SmailagicSpallucci:2012, HendiVahinidia:2012}. 
This proposal has been shown to 
provide a much richer panoply of thermodynamic behaviour for both AdS and dS black holes \cite{Dolan:2013ft}.  These include
the existence of reentrant phase transitions in rotating \cite{Altamirano:2013ane} and Born-Infeld \cite{GunasekaranEtal:2012} black holes
and the existence of a tricritical point in rotating black holes analogous to the triple point in water \cite{Altamirano:2013uqa}.  
Furthermore, it can be shown that there is a complete analogy between 4-dimensional 
Reissner-N\"ordstrom AdS black holes and the Van der Waals liquid--gas system, with the  critical exponents coinciding with those of the Van der Waals system and predicted by the mean field theory, significantly modifying previous considerations that emerged from the duality description \cite{ChamblinEtal:1999a,ChamblinEtal:1999b, Cvetic:1999ne, Cvetic:1999rb, Niu:2011tb, Tsai:2011gv}.  
 These phenomena are now under intensive study in a broad variety of contexts \cite{Dolan:2013dga,
Zou:2013owa,Zou:2014mha,Ma:2013aqa, Wei:2014hba,Mo:2014lza, Mo:2014wca, Mo:2014mba, Zhang:2014jfa, Kubiznak:2014zwa,Liu:2014gvf, Johnson:2014yja,
Dolan:2014mra, Dolan:2014cja}.

The extended thermodynamic phase space implied by this proposal is well motivated for a variety of reasons.  In the
extended phase space both the Smarr relation (which can be derived geometrically \cite{KastorEtal:2009}) and
the  first law of thermodynamics hold, whereas in the conventional phase space only the latter relation is satisfied  for nonzero $\Lambda$ .    Our understanding of the mathematical physics of
black holes is furthermore deepened insofar as a new conjecture emerges, namely that 
the  thermodynamic volume conjugate to the pressure satisfies a {\em reverse isoperimetric inequality} \cite{CveticEtal:2010};
thus far this conjecture is satisfied for all known cases. Furthermore, the use of an extended thermodynamic phase space is consistent with considering  more fundamental theories of physics that admit variation of physical constants  \cite{CreightonMann:1995,KastorEtal:2009, CveticEtal:2010}.
 Finally,  comparing the physics of black holes with
real world thermodynamic systems becomes a much more reasonable possibility \cite{Kubiznak:2014zwa}, insofar as  tricriticality, reentrant phase
transitions, and Van der Waals behaviour all have counterparts in laboratory physics \cite{KastorEtal:2009, Dolan:2010, Dolan:2011a, Dolan:2011b, KubiznakMann:2012,Altamirano:2013ane, Altamirano:2013uqa}; 
an extensive review of these issues in the context of rotating black holes was recently carried out in \cite{Altamirano:2014tva}.

Corrections to black hole thermodynamics from higher-curvature gravity theories has also been a subject of long-standing interest.  In these theories the entropy is no longer proportional to the area of the horizon but instead is given by a more complicated relationship depending on higher-curvature terms \cite{Iyer:1994ys}.
The most fruitful (and commonly explored) class of theories are the Lovelock gravity theories
 \cite{Lovelock:1971yv}, in large part because they yield differential equations for the metric functions that are second order.  While much work has been done in Lovelock theories for conventional phase space dynamics, the situation for extended phase space dynamics is not as well understood.  Previous work on this subject has been concerned primarily with 2nd-order Lovelock gravity (quadratic in the curvature) \cite{ Zou:2014mha,Wei:2014hba,Mo:2014mba, Zhang:2014jfa,Wei:2012ui,Cai:2013qga,Xu:2013zea}, better known as Gauss--Bonnet gravity, in which terms quadratic in the curvature are added to the 
action. Van der Waals behaviour has been shown to occur for this case \cite{Cai:2013qga,Xu:2013zea},  and tricriticality and reentrant behaviour have been observed \cite{Wei:2014hba} for charged spherically symmetric black holes.
 
3rd-order Lovelock gravity (cubic in the curvature)  has been considerably less explored.  So far the only
investigations have concentrated on a special subclass of theories in which the various coupling constants 
obey a particular relationship \cite{Dehghani:2005vh, Dehghani:2009zzb, Mo:2014qsa, Xu:2014tja, Belhaj:2014tga} (in our notation below $\alpha_2 = \sqrt{3\alpha_3}$). An investigation of this subclass  coupled to
 Born--Infeld electromagnetism in $d=7$ dimensions  was shown to reproduce VdW behavior \cite{Mo:2014qsa}.
For black holes with hyperbolic horizon geometry,\footnote{Such geometries generally yield horizons that are non-compact; however through an appropriate choice of identifications, they can be rendered compact 
\cite{Mann:1997iz,Smith:1997wx}.  We shall assume this throughout this paper.}  
an interesting `reverse VdW behavior' was noted. 
For the uncharged case \cite{Xu:2014tja}, a single critical point was observed for spherical 
horizon geometries in $d=7$, whereas two exist for dimensions $d=8,9,10,11$
(not necessarily with positive values of pressure/volume/temperature), and none exist for $d\geq 12$;  for hyperbolic
horizon geometries there is always only a single critical point in any $d\geq 7$.

In this paper we carry out a more extensive study of thermodynamics of Lovelock gravity.  We consider arbitrary
values of the coupling constants, and investigate $p-v$ criticality in $d=5,6$ and higher $d$ Gauss--Bonnet gravity and $d=7,8$
3rd-order Lovelock gravity.
In addition to recovering the aforementioned phenomena, we find a number of interesting new results.
We find that $d=6$ is unique to Gauss--Bonnet gravity insofar as it is the only dimension that
admits triple points for charged black holes, in accord with previous work \cite{Wei:2014hba}.
We find the existence of a maximal pressure, above which the black hole spacetime is no longer asymptotically AdS but
instead compactifies. 
We also show that a `thermodynamic singularity', characterized by $\frac{\partial p}{\partial t}|_{v=v_s}=0$, leading to `crossing isotherms' in the $p-v$ diagram and infinite `jump' in the Gibbs free energy as the specific volume $v\to v_s$, is generic to $\kappa=-1$ case in Lovelock gravities, whereas it is absent in Einstein--Maxwell gravity \cite{GunasekaranEtal:2012}. 
For $k$-th order Lovelock gravity there are $(k-1)$ such singular points.
This singularity---also known as a branch singularity---corresponds to an extremum of the entropy \cite{Maeda:2011ii} with respect to the volume.
Despite its presence, we are able to make sense of the thermodynamics of the system.
We also observe  for $d=7, 8$ hyperbolic Lovelock black holes
a new phenomenon of {\em multiple RPT} behaviour, in which for fixed pressure the small/large/small/large black hole phase transition occurs as the temperature of the system increases. 
Moreover, for special tuned Lovelock couplings in the hyperbolic case we find a new type of {\em isolated critical point}, characterized by new critical exponents. 
In other words, the thermodynamics of (charged) Lovelock black holes is much richer and more interesting than that of the Einstein(--Maxwell) theory.

The outline of our paper is as follows. We begin in Sec.~2 by reviewing the Lovelock gravity and its extended phase space thermodynamics and discuss  the corresponding $U(1)$ charged black hole solutions. 
Sec.~3 is devoted to the unified review of Gauss--Bonnet extended thermodynamics, with particular emphasis on the existence of various bounds in the hyperbolic case and the existence of thermodynamic singularities.  
In Sec.~4, specifying to black holes of 3-order Lovelock gravity in $d=7$ and $d=8$ dimensions, we study the corresponding equations of state and illustrate how tricriticality and (multiple) reentrant phase transitions can occur, a particular case of special tuned Lovelock couplings is discussed in detail. Our results are summarized in the concluding Sec.~5.

\section{Lovelock black hole thermodynamics}

\subsection{Lovelock Gravity}
Lovelock gravity \cite{Lovelock:1971yv} is a higher derivative theory that has received a lot of attention in recent years. In any attempt to perturbatively quantize gravity as a field theory, one should expect that the Einstein--Hilbert action is only an effective gravitational action valid for small curvature or low energies and that it will be modified by higher-curvature terms. Among higher derivative gravity theories, Lovelock gravities are the unique theories that give rise to field equations that 
are generally covariant and contain at most second order derivatives of the metric.

In $d$ spacetime dimensions, the Lagrangian of a Lovelock gravity theory is given by \cite{Lovelock:1971yv}
 \begin{equation}
\mathcal{L}=\frac{1}{16\pi G_N}\sum_{k=0}^{k_{max}}\hat{\alpha}_{\left(k\right)}\mathcal{L}^{\left(k\right)}\,. 
\label{eq:Lagrangian}
\end{equation}

The $\hat{\alpha}_{\left(k\right)}$
are the  Lovelock coupling constants and $\mathcal{L}^{\text{\ensuremath{\left(k\right)}}}$ are the $2k$-dimensional Euler densities, given by 
the contraction of $k$ powers of the Riemann tensor
\begin{equation}
\mathcal{L}^{\left(k\right)}=\frac{1}{2^{k}}\,\delta_{c_{1}d_{1}\ldots c_{k}d_{k}}^{a_{1}b_{1}\ldots a_{k}b_{k}}R_{a_{1}b_{1}}^{\quad c_{1}d_{1}}\ldots R_{a_{k}b_{k}}^{\quad c_{k}d_{k}}\,,
\end{equation}
where the `generalized Kronecker delta function' is totally antisymmetric in both sets of indices. The term $\mathcal{L}^{\left(0\right)}$ gives the cosmological constant term, $\mathcal{L}^{\left(1\right)}$ gives the Einstein--Hilbert action, and $\mathcal{L}^{\left(2\right)}$ corresponds to the quadratic Gauss--Bonnet term.  
The integer $k_{max}=\left[\frac{d-1}{2}\right]$ provides an upper bound on the sum, reflecting the fact that $\mathcal{L}^{\left(k\right)}$ 
contribute to the equations of motion for $d>2k$, whereas they are topological in $d=2k$, and vanish identically in lower dimensions. General relativity is recovered upon setting  $\hat{\alpha}_{(k)}=0$ for $k\geq 2$.\footnote{
Another interesting limit in odd dimensions arises for the following choice of Lovelock couplings
\be
\alpha_p =\frac{\ell^{2p-2n+1}}{2n-2p-1}\left( {n-1\atop p}  \right)  \qquad p=1,2,\ldots, n-1 = \frac{d-1}{2}
\ee
for which  the local Lorentz invariance of the Lovelock action is enhanced to a local (A)dS symmetry and the theory belongs to the class of Chern--Simons theories \cite{Zanelli:2005sa}.  We shall not consider this case here.}

The vacuum equations of motion for Lovelock gravity, following from the Lagrangian density \eqref{eq:Lagrangian}, are
\begin{equation}
\mathcal{G}_{\, b}^{a}=\sum_{k=0}^{k_{max}}\hat{\alpha}_{\left(k\right)}\mathcal{G}_{\,\,\quad b}^{\left(k\right)\, a}=0\,,
\end{equation}
where the Einstein-like tensors $\mathcal{G}_{\,\,\quad b}^{\left(k\right)\, a}$ are given by 
\begin{equation}
\mathcal{G}_{\,\,\quad b}^{\left(k\right)\, a}=-\frac{1}{2^{\left(k+1\right)}}\delta_{b\, e_{1}f_{1}\ldots e_{k}f_{k}}^{a\, c_{1}d_{1}\ldots c_{k}d_{k}}R_{c_{1}d_{1}}^{\quad e_{1}f_{1}}\ldots R_{c_{k}d_{k}}^{\quad e_{k}f_{k}}\label{eq:G}\,,
\end{equation}
and each of them independently satisfies a conservation law $\nabla_{a}\mathcal{G}_{\,\,\quad b}^{\left(k\right)\, a}=0\,$. 

Using the Hamiltonian formalism it is possible to derive the expression for  gravitational entropy in Lovelock gravity and the corresponding first law of black hole thermodynamics  \cite{Jacobson:1993xs}.  More recently, both the first law and the associated
Smarr formula in an extended phase space were obtained exploiting the Killing potential formalism \cite{Kastor:2010gq}.
In the extended thermodynamic phase space, all Lovelock coupling constants (including the cosmological constant $\hat{\alpha}_{\left(0\right)}$)
are considered as thermodynamic variables and allowed to vary in the first law of black hole thermodynamics. The physical meaning of these variables along with their conjugates, apart from the cosmological constant which has an interpretation of pressure and its conjugate variable is an associated volume, remains to be explored.\footnote{A similar situation was seen to occur in Born--Infeld electrodynamics, in which the thermodynamics conjugate to the Born--Infeld coupling constant was interpreted as  vacuum polarization \cite{GunasekaranEtal:2012}.
}

\subsection{Thermodynamic considerations}\label{secLCBH}
In what follows we shall consider Lovelock black holes, charged under a Maxwell field, $F=dA$, with the action given by 
[cf. Eq.\eqref{eq:Lagrangian}]
\begin{equation}
I =  \frac{1}{16\pi G_N} \int d^dx  \sqrt{-g}\Bigl({\sum_{k=0}^{k_{max}} }\hat{\alpha}_{\left(k\right)}\mathcal{L}^{\left(k\right)} - 4\pi G_N F_{ab}F^{ab}\Bigr)\,,
\label{eq:Loveaction}
\end{equation}
and the corresponding equations of motion
\begin{equation}
\sum_{k=0}^{k_{max}}\hat{\alpha}_{\left(k\right)}\mathcal{G}_{ab}^{\left(k\right)}=
8\pi G_N \Bigl(F_{a c} {F_b{}^{\,c}} -\frac{1}{4}g_{a b} F_{c d}F^{c d}\Bigr)\,.
 \label{eq:Graveq}
\end{equation}

For a black hole solution, characterized by mass $M$, charge $Q$, temperature $T$, and entropy $S$, the extended first law and the associated Smarr--Gibbs--Duhem relation read \cite{Jacobson:1993xs, Kastor:2010gq}  
\bea
\delta M&=&T\delta S-\frac{1}{16\pi G_N}\sum_{k}\hat \Psi^{\left(k\right)}\delta\hat{\alpha}_{\left(k\right)}
+\Phi\delta Q\,,\label{first}\\
\left(d-3\right)M&=&\left(d-2\right)TS+\sum_{k}2\left(k-1\right)\frac{\hat \Psi^{\left(k\right)}\hat{\alpha}_{\left(k\right)}}{16\pi G_N}+\left(d-3\right)\Phi Q\,.\label{Smarr}
\eea
In Lovelock gravity, the entropy is no longer given by  one quarter of the horizon area, but rather reads 
\begin{equation}\label{S}
S= \frac{1}{4G_N} \sum_k \hat{\alpha}_{k} {\cal A}^{(k)}\,,\quad   {\cal A}^{(k)}  = k\int_{\mathcal{H}} \sqrt{\sigma}{\mathcal{L}}^{(k-1)}\,.
\end{equation}
Here, $\sigma$ denotes the determinant of $\sigma_{ab}$,  the induced metric on the black hole horizon ${\mathcal{H}}$, and the Lovelock terms ${\mathcal{L}}^{(k-1)}$ are evaluated on that surface.  
Potentials $\hat \Psi^{\left(k\right)}$ are the thermodynamic conjugates to $\hat{\alpha}_{(k)}$'s and are given by 
\begin{equation}
\hat \Psi^{\left(k\right)}=4\pi T {\cal A}^{\left(k\right)}+ {\cal B}^{\left(k\right)}+\Theta^{\left(k\right)}\,,
\end{equation}
where  
\bea
{\cal B}^{\left(k\right)} &=& -\frac{16\pi k G_NM (d-1)!}{b(d-2k-1)! }\left( -\frac{1}{\ell^2} \right)^{k-1}\!\!\!\,, \quad 
b = \sum_k  \frac{\hat{\alpha}_k  k(d-1)!}{(d-2k-1)! }\left( -\frac{1}{\ell^2} \right)^{k-1}\!\!\!\,,\nonumber\\
\Theta^{\left(k\right)} &=& \int_{\Sigma} \sqrt{-g} {\mathcal{L}}^{(k)}[s] - \int_{\Sigma_{\textrm{AdS}}}\!\!\!\! \sqrt{-g_{\textrm{AdS}}} {\mathcal{L}}^{(k)}[s_{\textrm{AdS}}]\,,
\eea
and $\ell$ stands for the AdS radius; it is a non-trivial function of {the `bare' cosmological constant $\Lambda=-\hat \alpha_0/2$ and the higher-order} Lovelock couplings.  
$\Sigma$ is a spatial hypersurface spanning from the black hole horizon to spatial infinity, with timelike unit normal $n^a$ and induced metric $s_{ab} = g_{a b} + n_a n_b$; quantities with ``AdS subscript'' are pure AdS space counterparts of the corresponding black hole spacetime quantities, with no internal boundary and the same $\hat{\alpha_0}$.

In what follows we identify the (negative) cosmological constant $\Lambda=-\hat{\alpha}_{0}/2$ with the thermodynamic pressure and the conjugate quantity $\hat \Psi^{\left(0\right)}$ with the thermodynamic volume $V$, according to 
\begin{equation}\label{P}
P=-\frac{\Lambda}{8\pi G_N} = \frac{\hat{\alpha}_{0}}{16\pi G_N}\,,\quad V=-\hat \Psi^{\left(0\right)}\,.
\end{equation}
With this identification, the mass $M$ of the black hole is interpreted as an enthalpy rather than the internal energy \cite{KastorEtal:2009, Kubiznak:2014zwa}.

\subsection{Charged black holes}
Concentrating on the static charged spherically symmetric AdS Lovelock black holes, we employ the following ansatz  
\be\label{solution}
ds^{2} = -f\left(r\right)dt^{2}+f\left(r\right)^{-1}dr^{2}+r^{2}d\Omega_{(\kappa) d-2}^{2}\,,
\qquad  F= \frac{Q}{r^{d-2}} dt\wedge dr\,, 
\ee
where $d\Omega_{(\kappa) d-2}^{2}$ denotes the line element of a $\left( d-2 \right)$-dimensional compact space with constant curvature $(d-2)(d-3)\kappa$, with the horizon geometry corresponding to  $\kappa=0,+1,-1$ for flat, spherical, hyperbolic black hole horizon geometries respectively.
Denoting by $\Sigma_{d-2}^{(\kappa)}$  the finite volume of this compact space, a $(d-2)$-dimensional unit sphere $\Sigma_{d-2}^{(\kappa)}$ takes the following standard form: 
\begin{equation}
\Sigma_{d-2}^{(+1)}=\frac{2\pi^{\left(d-1\right)/2}}{\Gamma\left(\frac{d-1}{2}\right)}\,.
\end{equation}

In terms of the rescaled Lovelock coupling constants 
\begin{equation}
\alpha_{0}=\frac{\hat{\alpha}_{(0)}}{\left(d-1\right)\left(d-2\right)}\,,\quad{\alpha}_{1}={\hat \alpha}_{(1)}\,,\quad\alpha_{k}=\hat \alpha_{(k)}\prod_{n=3}^{2k}\left(d-n\right)  {\quad\mbox{for}\quad  k\geq2}\,,
\end{equation}
the field equations \eqref{eq:Graveq} reduce to the requirement that $f\left(r\right)$ solves the following polynomial equation of degree $k_{max}$
\cite{Boulware:1985wk, wheeler1986symmetric1, wheeler1986symmetric2, Cai:2003kt, 2013JHEP...07..164C,Camanho:2011rj,Takahashi:2011du}
\begin{equation} \label{eq:poly}
{\cal P}\left(f\right)=\sum_{k=0}^{k_{max}}\alpha_{k} \left(\frac{\kappa-f}{r^2}\right)^{k}=\frac{16\pi G_NM}{(d-2)\Sigma_{d-2}^{(\kappa)}r^{d-1}}-\frac{8\pi G_N Q^2}{ (d-2)(d-3)}\frac{1}{r^{2d-4}}\,.
\end{equation}
Here $M$ stands for the ADM mass of the black hole and $Q$ is the electric charge, given by  
\begin{equation}\label{MQ}
Q = \frac{1}{2 \Sigma_{d-2}^{(\kappa)}} \int * F\,. 
\end{equation}

Even without knowing $f=f(r)$ explicitly, it is possible, using the Hamiltonian formalism \cite{Cai:2003kt, Kastor:2011qp},  to find the thermodynamic quantities characterizing the black hole solution. Let $r_+$ denotes the radius of the event horizon, determined as the largest root 
of $f \left(r_{+}\right)=0$. 
The black hole mass $M$, the temperature $T$, the entropy $S$, and the gauge potential $\Phi$ are given by \cite{Cai:2003kt}
\bea
M&=&\frac{\Sigma_{d-2}^{(\kappa)}\left(d-2\right)}{16\pi G_N}\sum_{k=0}^{k_{max}}\alpha_{k}\kappa^kr_+^{d-1-2k}+\frac{\Sigma_{d-2}^{(\kappa)}}{2(d-3)}\frac{Q^2}{r_+^{d-3}}\,,\\
T &=&  \frac{\vert f^\prime(r_+)\vert}{4\pi} =\frac{1}{4\pi r_+ D(r_+)}\left[\sum_k\kappa\alpha_k(d\!-\!2k\!-\!1)\Bigl(\frac{\kappa}{r_+^2}\Bigr)^{k-1}\!\!\!-\frac{8\pi G_N Q^2}{(d-2)r_+^{2(d-3)}}\right]\,,\qquad  \label{T}\\
S&=&\frac{\Sigma_{d-2}^{(\kappa)}\left(d-2\right)}{4G_N}\sum_{k=0}^{k_{max}}\frac{k\kappa^{k-1}\alpha_{k}r_+^{d-2k}}{d-2k}\,, \quad 
 \Phi=\frac{\Sigma_{d-2}^{(\kappa)}Q}{(d-3)r_+^{d-3}}\,, \label{eq:Entropy}
\eea
where 
\be
{D(r_+)=\sum_{k=1}^{k_{max}}k\alpha_{k}\left(\kappa r_{+}^{-2}\right)^{k-1}\,.} 
\ee
The leading term in the expression for $S$ is one-quarter the horizon area; the  other terms come from higher-curvature contributions. Note that $S$ does not explicitly depend on the cosmological constant or the charge $Q$.

Using the expression for $M$, we find the following formulae for the 
potentials $\Psi^{\left(k\right)}$ conjugate to $\alpha_k$,  
$\delta M=T\delta S+\Phi \delta Q+\sum_k \Psi^{(k)} \delta \alpha_k$:
\be
\Psi^{(k)}=\frac{\Sigma_{d-2}^{(\kappa)}(d-2)}{16\pi G_N}\kappa^{k-1}{r^{d-2k}_+}\left[\frac{\kappa}{r}-\frac{4\pi kT}{d-2k}\right]\,, \quad k\geq 0\,.
\ee
Returning back to the $\hat \alpha_{(k)}$ couplings we get the thermodynamic volume
\be\label{V}
V=-\hat \Psi^{(0)}=\frac{16\pi G_N \Psi^{(0)}}{(d-1)(d-2)}=\frac{\Sigma_{d-2}^{(\kappa)}r_+^{d-1}}{d-1}\,,
\ee
while the other potentials read 
\be
\hat \Psi^{(1)}=\Psi^{(1)}\,,\quad \hat \Psi^{(k)}=-16\pi G_N \prod_{n=3}^{2k}(d-n)\Psi^{(k)}\,, \quad k\geq 2\,.
\ee
One can then verify that the Smarr relation \eqref{Smarr} and the first law \eqref{first} are satisfied. 
Note that $\hat \alpha_{(1)}$ effectively captures a possible change in the gravitational constant $G_N$; the corresponding potential 
$\hat \Psi^{(1)}$ does not contribute to the Smarr relation but modifies the first law.

In what follows we shall treat the couplings $\alpha_k$ for $k\geq 1$ as fixed external parameters and treat only the cosmological constant and hence $\alpha_0$ as a thermodynamic variable. Using \eqref{P} and \eqref{V}, this allows as to reinterpret equation \eqref{T} as the ``fluid equation of state''
\bea\label{state}
P&=&P(V,T,Q,\alpha_1,\dots, \alpha_{k_{max}})\nonumber\\
&=&\frac{d-2}{16\pi G_N}\sum_{k=1}^{k_{max}}\frac{\alpha_k}{r_+^2}\Bigl(\frac{\kappa}{r_+^2}\Bigr)^{k-1}
\Bigl[4\pi k r_+T-\kappa(d-2k-1)\Bigr]+\frac{Q^2}{2\alpha_1 r_+^{2(d-2)}}\,,
\eea
and study the possible phase transitions based on the behavior of the Gibbs free energy in the `canonical ensemble',
given by \cite{Kastor:2010gq} (see also  \cite{Kofinas:2007ns} for the Euclidean action calculation) 
\be\label{G}
G=M-TS=G(P,T, Q,\alpha_1,\dots, \alpha_{k_{max}})\,.
\ee
The thermodynamic state corresponds to the global minimum of this quantity for   fixed parameters $P, T, Q$ and $\alpha$'s.

Before we start our thermodynamic analysis, let us pause to make a few remarks regarding the parameter space for the Lovelock couplings. In principle, one may consider $\alpha_{k}$ to have arbitrary positive or negative values.
However, for cubic (or higher order) Lovelock gravities, there are entire regions of the parameter space where there are no black holes at all, even if there is a well defined AdS vacuum. This is a feature that does not take place in Gauss--Bonnet gravity. Up to now, most of the information needed to clarify the existence of black hole solutions for different values of the Lovelock couplings has relied on the behaviour of cubic polynomials, see, e.g., Eq. (\ref{eq:poly}) in  \cite{PhysRevD.79.064015}, or on the AdS/CFT calculations \cite{Camanho:2009hu,Camanho:2010ru}. 
In particular, for  3rd-order Lovelock gravity, negative values of $\alpha_k$ yield solutions with naked singularities over a broad parameter range \cite{Myers:1988ze}.  Moreover, in the low energy effective action of heterotic string theory $\alpha_{2}$ is proportional to the inverse string tension and hence is positive. For all these reasons we limit ourselves in this paper to considering only positive Lovelock couplings $\alpha_2>0$ and $\alpha_3>0$; from now on we also set  $\alpha_1=1$ to recover general relativity in the small curvature limit and put $G_N=1$.   

We close this section by noting two phenomena present for hyperbolic black holes, $\kappa=-1$.  First, for such black holes  a `{\em thermodynamic singularity}' (also known as a {\em branch singularity} \cite{Maeda:2011ii}), characterized by 
\be
{\frac{\partial P}{\partial T}\Bigr|_{V=V_s}=0\,,}
\ee
will be present.  {In the $P-V$ diagram, which is a 2d projection of the 3d $P=P(V,T)$ relation, the isotherms will cross at $V=V_s$. The above relations lead to the condition
\be\label{TDsing}
D(r_+)|_{\mbox{\tiny TD singularity}}=\sum_{k=1}^{k_{max}}k\,\alpha_{k}\left(\kappa^{d-1}  
\bigg[\frac{\Sigma_{d-2}^{(\kappa)}}{(d-1)V_{s}}\bigg]^{2}\right)^{\frac{{k-1}}{d-1}} =0\,,
\ee
upon using (\ref{V}).
For $\kappa=-1$, this equation will have at least one real solution for $k\geq 2$, and will have up to $(k-1)$ distinct solutions
in $k$-th order Lovelock gravity. This generic situation stands in contrast to  $k=0$  Einstein--Maxwell gravity, where it is absent 
\cite{GunasekaranEtal:2012}.   
}

{It is clear upon inspection of equations \eqref{T} and \eqref{G} that both the temperature and the Gibbs free energy 
diverge at $V=V_s$ ($M$ and $S$ are finite for all values of $P$ and $V$), apart from a very special choice of the pressure $P=P_s$ for which both $T$ and $G$ are finite and $T$ takes the special value $T=T_s$.
By taking the derivative of the entropy \eqref{eq:Entropy} 
with respect to $r_+$ it straightforward to show that the entropy is maximized here as 
well \cite{Maeda:2011ii}. It is easy to see that, apart from $P=P_s$, the thermodynamic singularity reflects the presence of a curvature singularity of the Riemann tensor, since the Kretschmann scalar at the horizon 
is
\be\label{Kretsch}
K(r_+)=R_{abcd}R^{abcd}|_{r=r_+}=\left[\Bigl(\frac{d^2f}{dr^2}\Bigr)^2 + \frac{2(d-2)}{r^2} \Bigl(\frac{df}{dr}\Bigr)^2 +\frac{2(d-2)(d-3)}{r^4}\right]_{r=r_+}\,.
\ee
However by analytically continuing around $(P_s, V_s, T_s)$ one can not only make sense of thermodynamics but also show that the Kretschmann scalar is finite. We shall discuss this for the concrete example of $d=5$ Gauss--Bonnet black hole in the next section.  
We examine another very interesting special case in Sec.~\ref{Sec:alpha3}. 
}

Second, we note that when $\kappa=-1$ the entropy $S$, given in \eqref{eq:Entropy}, is not always positive. Demanding its positivity imposes the following condition:
\be\label{Spos}
\sum_{k=1}^{k_{max}}\frac{k\kappa^{k-1}\alpha_{k}r_+^{d-2k}}{d-2k}
=\sum_{k=1}^{k_{max}}\frac{k\kappa^{k-1}\alpha_{k}}{{d-2k}}\left[\frac{(d-1)V}{\Sigma_{d-2}^{(\kappa)}}\right]^{\frac{d-2k}{d-1}}\geq 0\,.
\ee
 In what follows we consider the black holes with negative entropy as unphysical and exclude them from thermodynamic considerations.

\section{$P-v$ criticality in Gauss--Bonnet gravity}

Before we proceed to 3rd-order Lovelock gravity, in this section we recapitulate and slightly elaborate on the behaviour of 
$U(1)$ charged Gauss--Bonnet black holes  \cite{Zou:2014mha,Wei:2014hba,Mo:2014mba, Zhang:2014jfa,Cai:2013qga,Xu:2013zea}. We first concentrate on $d=5$ and $d=6$, the lowest possible two dimensions where the Gauss--Bonnet theory brings new qualitative features, and then briefly discuss what happens in higher dimensions.
We will show that only for $d=6$ does the phase diagram admit a triple point. In all higher dimensions `only' the standard VdW behaviour is observed.

\subsection{Maximal pressure and other conditions}
\begin{figure*}
\centering
\begin{tabular}{cc}
\rotatebox{-90}{
\includegraphics[width=0.34\textwidth,height=0.28\textheight]{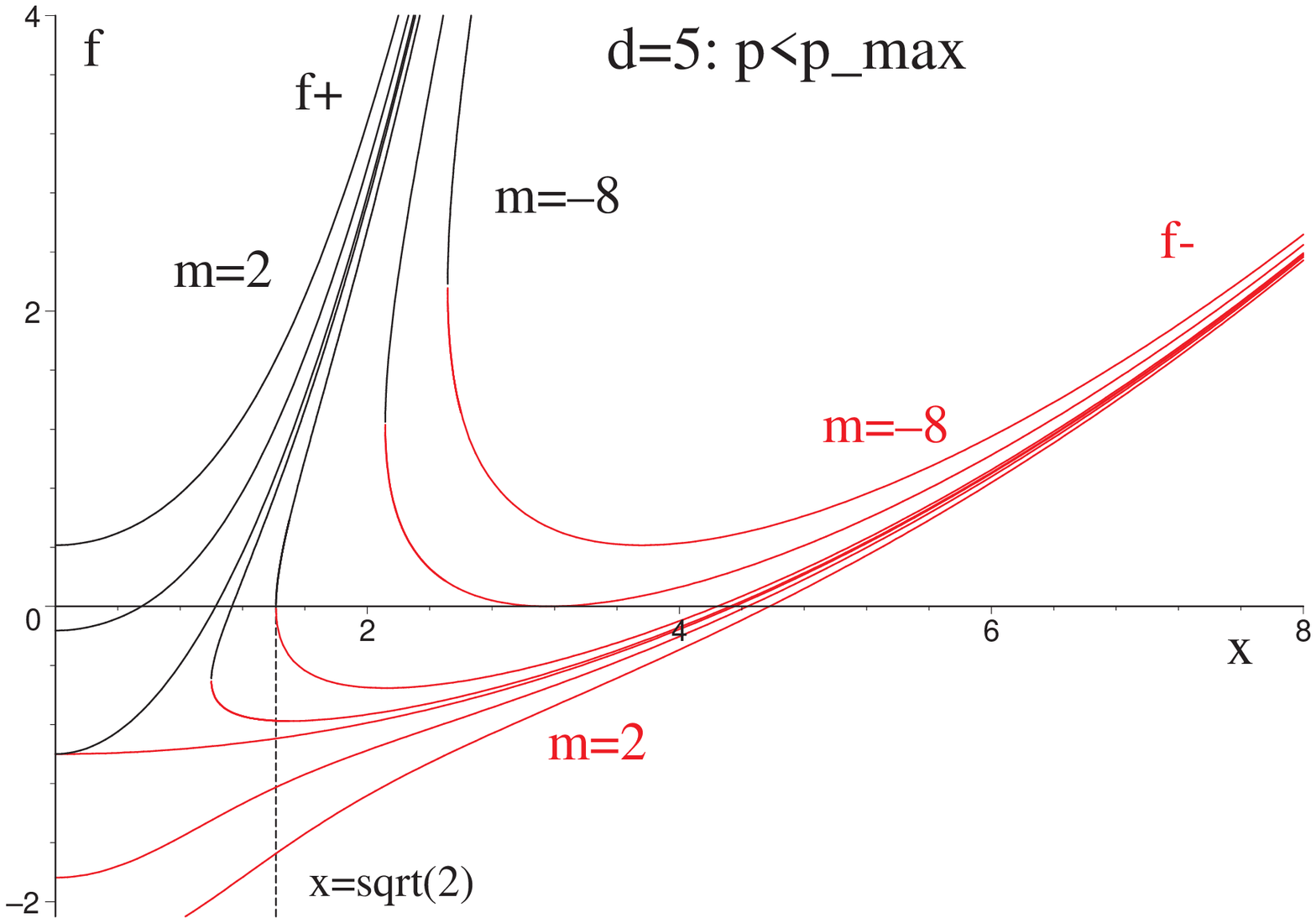}} &
\rotatebox{-90}{
\includegraphics[width=0.34\textwidth,height=0.28\textheight]{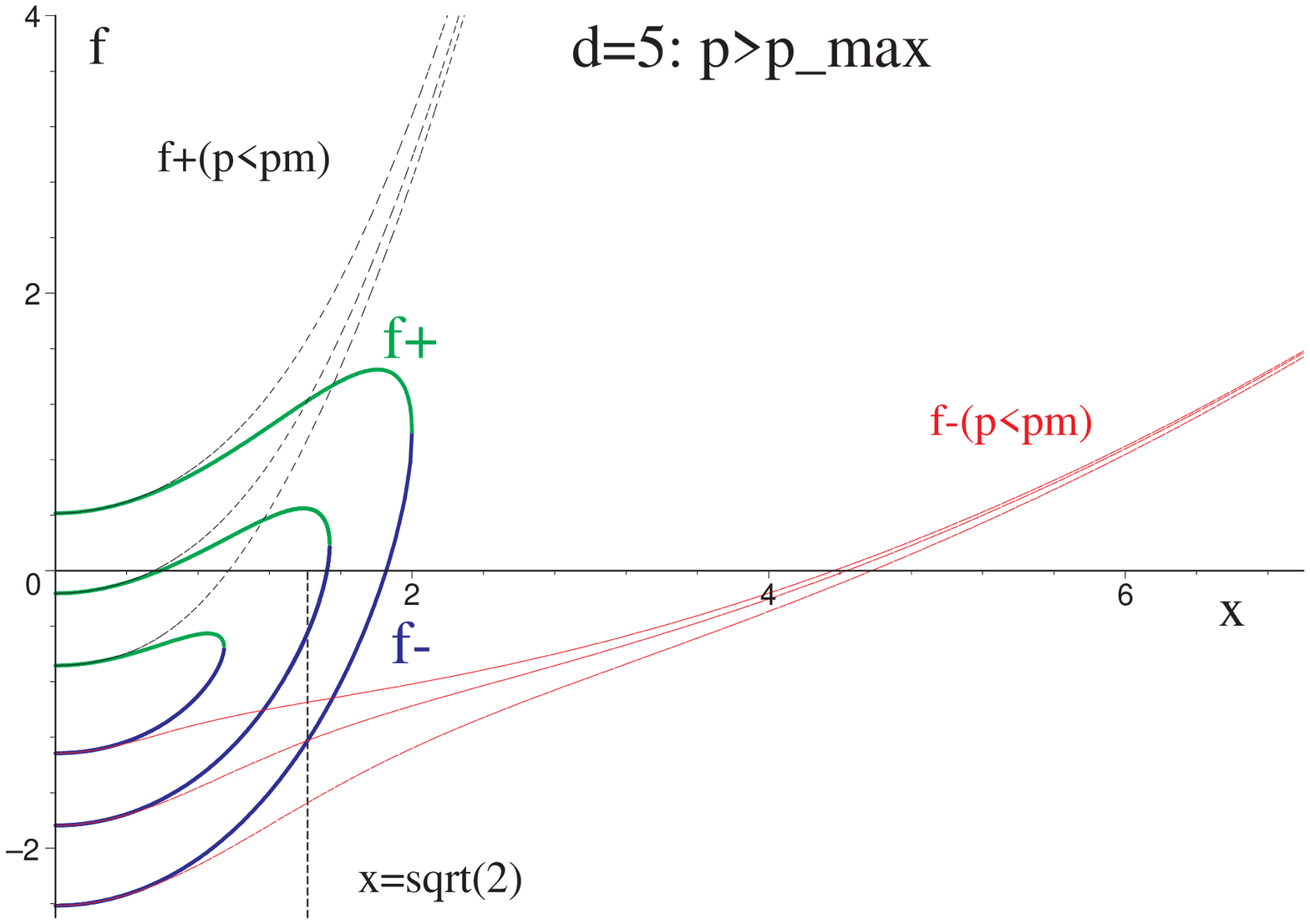}}\\
\end{tabular}
\caption{{\bf Metric functions $f_\pm$ for $d=5, q=0, \kappa=-1$.}
Metric functions $f_\pm$ are displayed as function of $x=r/\sqrt{\alpha_{2}}$ for fixed pressure $p=\frac{1}{5}p_{max}$ on the {\em left} by red and black curves 
for $m=-8,-4,-4/5,-0.2, 0, 0.7, 2$ (following red curves from up down) and {\em right} for $p=\frac{3}{2}p_{max}$ by green and blue curves for $m=0.1, 0.7, 2$ from `inside out' (these 
only exist for $m>0$).
For $p<p_{max}$ we have an asymptotic AdS region, with the 
 red curve for $f_-$ and black curve for $f_+$ each growing like $r^2$ for large $r$.
 However, for $p>p_{max}$ 
the nonlinear curvature is too strong, and the space becomes compact---there
is no asymptotic `AdS region', and the  green and blue curves now portray this situation.
 Note that for $p<p_{\max}$, $f_-$ has a zero for $x\geq \sqrt{2}$; $x=\sqrt{2}$ is  displayed by the vertical black dashed line. For $\kappa=+1$ all the curves are shifted upwards by 2:  consequently there are no zeros for $f_+$ and there is no minimal bound on horizon radius of the Einstein branch.
}  
\label{Fig:1}
\end{figure*} 

\begin{figure*}
\centering
\begin{tabular}{cc}
\rotatebox{-90}{
\includegraphics[width=0.34\textwidth,height=0.28\textheight]{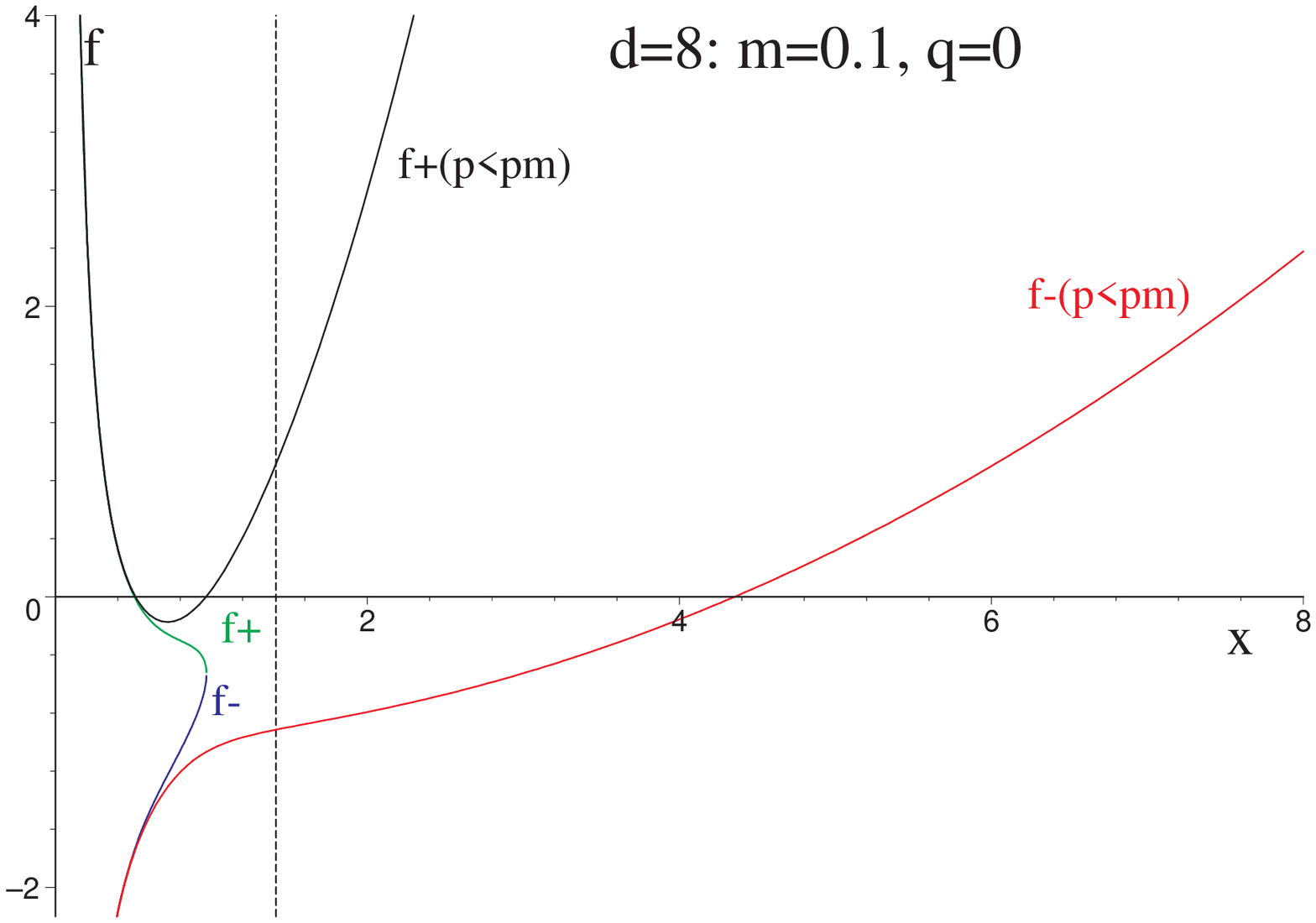}} &
\rotatebox{-90}{
\includegraphics[width=0.34\textwidth,height=0.28\textheight]{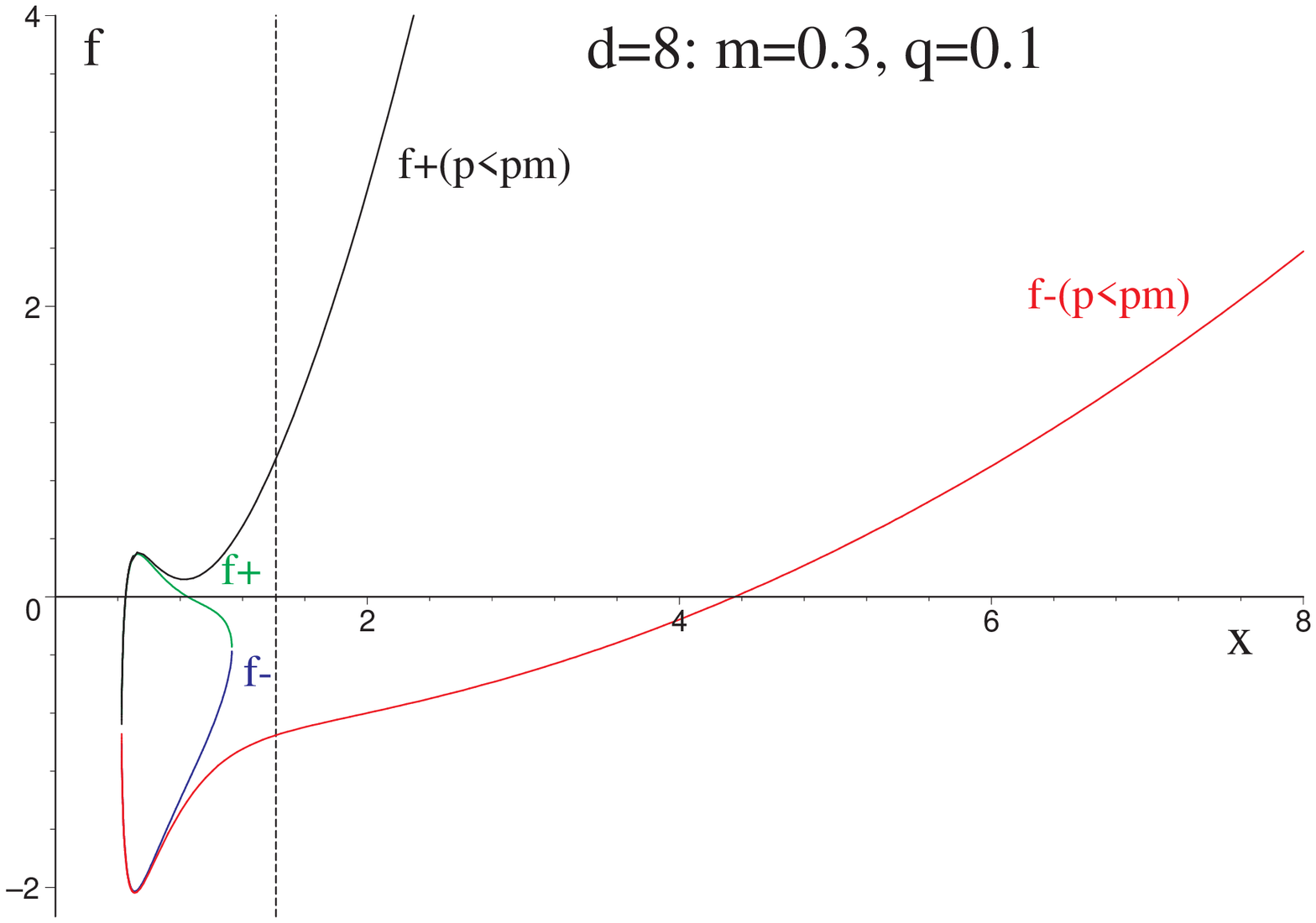}}\\
\end{tabular}
\caption{{\bf Metric functions $f_\pm$ for $d=8$ and $\kappa=-1$.}
{\em Left:} The behavior of $f_\pm$ for $d>5$ remains qualitatively similar to $d=5$ except $f_+$ approaches infinity as $x\to 0$. {\em Right:} When the charge is added,  {functions $f_\pm$ do not extend all the way to $x\to 0$ but rather terminate}  at a finite $x_{\min}$. Similar to the previous figure we have set $p=\frac{1}{5}p_{max}$ {for red and black curves and $p=\frac{3}{2}p_{max}$ for blue and green curves}.
}  
\label{Fig:2}
\end{figure*} 

Only for sufficiently small pressures does the solution \eqref{solution} possesses an asymptotic `AdS region'. To see this, consider equation \eqref{eq:poly}, which for   Gauss--Bonnet gravity reduces to a quadratic equation for $f$
\bea
\alpha_2 \frac{(\kappa-f)^2}{r^4}+\frac{(\kappa-f)}{r^2}+\alpha_0  -\frac{16\pi M}{(d-2)\Sigma_{d-2}^{(\kappa)}{r^{d-1}}}+\frac{8\pi Q^2}{(d-2)(d-3)r^{2d-4}}=0\,,
\eea
resulting in two solutions, $f_\pm$; it is $f_-$ that approaches the Einstein case when the Gauss--Bonnet coupling $\alpha_2\to 0$.  We call the corresponding branch the `{\em Einstein branch}' whereas we refer to the black holes corresponding to $f_+$ as the `{\em Gauss--Bonnet branch}'. 

The requirement for the existence of $f_\pm$ for large $r$ yields the following necessary condition:  
\be
1-4\alpha_0\alpha_2\geq 0\,. 
\ee
When this inequality is violated, the nonlinear curvature is too strong, and the space becomes compact:  there
is no asymptotic `AdS region' and hence no 
 `proper' black hole with `standard' asymptotics, as illustrated in Figs.~\ref{Fig:1} and \ref{Fig:2}. {In terms of the
following dimensionless variables}
\begin{equation}
r_+=v\,\alpha_{2}^{\frac{1}{2}}\,,\quad T=\frac{t\alpha_{2}^{-\frac{1}{2}}}{d-2}\,,
\quad m=\frac{16\pi M}{(d-2)\Sigma_{d-2}^{(\kappa)}\alpha_{2}^{\frac{d-3}{2}}}\, ,
\quad  Q= \frac{q}{\sqrt{2}} \alpha_{2}^{\frac{d-3}{2}}\,,\quad  P=\frac{p}{4\alpha_{2}}\,, 
\label{subst}
\end{equation}
this imposes the important bound (see also \cite{Cai:2013qga})
\be
p\leq p_{\max}=\frac{(d-1)(d-2)}{16\pi} 
\ee
on the validity of any $p-v$ analysis. This bound applies to both $\kappa=\pm1$ cases, and is the same bound
that for ensuring that transverse-traceless excitations of the metric about an AdS vacuum are not ghostlike \cite{Myers:2010ru}, and forms a critical threshold beyond which the qualitative behaviour of the masses and
angular momenta of boson stars changes from its spiral-like AdS pattern \cite{Henderson:2014dwa}. 
Since in our considerations the pressure is allowed to vary, one might wonder if there a phase transition from the AdS case to the compact case as we cross $p=p_{max}$. We shall return to this speculation in the discussion section.

 For $\kappa=-1$ Gauss--Bonnet black holes, the thermodynamic singularity discussed above (see Eq.~\eqref{TDsing}) occurs for 
\be
v=v_s=\sqrt{2}\,,
\ee 
for which the black hole temperature and  the Gibbs free energy both diverge. This corresponds to a branch-cut singularity between the Einstein and Gauss--Bonnet branches, which meet where the metric function has an infinite slope, {as shown in Fig.~\ref{Fig:1} for $q=0$ for which   $m=m_s=-1+4\pi p/3$.  Black holes in the Einstein branch always have $v>\sqrt{2}$, whereas $v\leq \sqrt{2}$ corresponds to the Gauss--Bonnet branch; 
`crossing' from $v>\sqrt{2}$ to $v<\sqrt{2}$ corresponds to `jumping' from the Einstein to the Gauss--Bonnet branch. 
We depict the behaviour of metric functions $f_\pm$ for $q=0$ in Fig.~\ref{Fig:1} and for more general case refer the reader to the thorough discussion of spacetime structure in \cite{Torii:2005xu, Torii:2005nh}; see also recent papers \cite{Izumi:2014loa,Reall:2014pwa} for 
a discussion of causal structures.

Finally, the non-negativity of $S$, Eq.~\eqref{Spos}, requires that for $\kappa=-1$ we must have 
\be
v\geq v_{S+}= \sqrt{2+\frac{4}{d-4}}\,.
\ee 
Since this implies that in any dimension $v> \sqrt{2}$, we see that the Gauss--Bonnet branch black holes never satisfy this bound and 
all have negative entropy; these black holes are known to be unstable \cite{Boulware:1985wk} {and we eventually exclude them from thermodynamic considerations}.

\subsection{Equation of state}

For the Gauss--Bonnet gravity, the equation of state \eqref{state} reads  
\be
P=\frac{(d-2)T}{4 r_+}-\frac{(d-2)(d-3)\kappa}{16\pi r_+^2}+
\frac{(d-2)\alpha_2\kappa T}{2r_+^3}-\frac{(d-2)(d-5)\alpha_2\kappa^2}{16\pi  r_+^4}+\frac{Q^2}{2r_+^{2(d-2)}}\,,
\ee
or in terms of the variables \eqref{subst}, the dimensionless equation of state is   
\be\label{PvGB}
p=\frac{t}{v}-\frac{(d-2)(d-3)\kappa}{4\pi v^2}+
\frac{2\kappa t}{v^3}-\frac{(d-2)(d-5)\kappa^2}{4\pi v^4}+\frac{q^2}{v^{2(d-2)}}\,.
\ee
The crucial information about the equation of state is encoded in the number of critical points it admits. A critical point occurs when $p=p(v)$ has an inflection point, i.e., when
\be\label{cp}
\frac{\partial p}{\partial v}=0\,,\quad \frac{\partial^2 p}{\partial v^2}=0\,.
\ee
Together with the equation of state \eqref{PvGB} this determines the critical values 
$(p_c, v_c, t_c)$ as functions of $q$ and $\kappa$. Obviously, no critical points occur 
when $\kappa=0$: the planar Gauss--Bonnet black holes, and in fact any planar black holes of higher-order Lovelock gravity in arbitrary number of spacetime dimensions do not admit  critical behavior.  In what follows, we therefore concentrate on $\kappa=\pm 1$, 
each of them bringing us new qualitative features.

To find a critical point we have to solve (higher-order polynomial) Eqs. \eqref{cp} for $t_c, v_c$ and insert the result  into the equation of state \eqref{PvGB} to find $p_c$, subject to the restriction that  $p_c, v_c, t_c$ are all positive in order the critical point be physical. Solving the first equation in \eqref{cp} for $t_c$ yields
\be\label{crit1}
t_c=\frac{(d-2)v_c}{2\pi(v_c^2+6\kappa)}\Bigl[(d-3)\kappa+\frac{2(d-5)}{v_c^2}-\frac{4\pi q^2}{v_c^{2(d-3)}}\Bigr]\,.
\ee
The second equation in \eqref{cp} then becomes
\be\label{crit2}
(d-3)v_c^{2d-4}-12\kappa v_c^{2d-6}+12(d-5)v_c^{2d-8}
-4\kappa\pi q^2(2d-5)v_c^2-24\pi q^2(2d-7)=0\,.
\ee
To get an upper estimate on the number of critical points, without actually having to solve this equation, we apply Descartes' rule of signs.\footnote{This rules states that  the maximum number of positive roots of a polynomial, when written in descending order of exponents, is  no more than the number of sign  variations in consecutive coefficients, and differs from this upper bound by an even integer \cite{marden1966geometry, Descartes1998}.} For example, we see that for $\kappa=+1$ we have at most three sign variations and hence can have at most three critical points with positive $v_c$, whereas we have at most one critical point for $\kappa=-1$.  
To determine the exact number and nature of physical critical points we proceed numerically (or analytically if possible), checking the positivity of $(p_c,v_c,t_c)$.

The dimensionless  counterpart $g$, of the  Gibbs free energy, \eqref{G} is
\be
g=\frac{1}{\Sigma_{d-2}^{(\kappa)}} \alpha_2^{\frac{3-d}{2}} G=g(t,p,q)\,
\ee
and reads
\bea\label{gGB}
g&=&-\frac{1}{16\pi(2\kappa+v^2)}\Bigl[\frac{4\pi p v^{d+1}}{(d-1)(d-2)}
-\kappa v^{d-1}+\frac{24\pi\kappa p v^{d-1}}{(d-1)(d-4)}-\frac{(d-8)v^{d-3}}{d-4}\nonumber\\
&&-\frac{2\kappa(d-2)v^{d-5}}{d-4}\Bigr]
+\frac{q^2\bigl[(2d-5)(d-4)v^2+2\kappa(d-2)(2d-7)\bigr]}{4(d-2)(d-3)(d-4)(2\kappa+v^2)v^{d-3}}\,.\quad
\eea
The thermodynamic state corresponds to the global minimum of this quantity for fixed $t, p$ and $q$.

\subsection{Five dimensions}
In five dimensions we have the following equation of state:  
\bea\label{p5}
p&=&\frac{t}{v}-\frac{3\kappa}{2\pi v^2}+\frac{2\kappa t}{v^3}+\frac{q^2}{v^6}\,,
\eea
and Eqs.~\eqref{crit1} and ~\eqref{crit2} reduce to 
\be\label{vc5d}
t_c=\frac{3(\kappa v_c^4-2\pi q^2)}{\pi v_c^3(v_c^2+6\kappa)}\,,\quad v_c^6-6\kappa v_c^4-10\pi\kappa q^2 v_c^2-36 \pi q^2=0\,.
\ee
The maximal pressure reads 
\be
p_{\tiny \mbox{max}} = \frac{3}{4\pi} \approx 0.2387\,.
\ee
and for $\kappa=-1$ black holes we demand $v\geq v_{S+}=\sqrt{6}$ to have positive entropy.

\subsubsection{Spherical case: VdW behavior}
In the uncharged $(q=0)$ case, an analytic solution for the critical point is possible and reads  
\be\label{CPoint5}
p_c= \frac{1}{12\pi}\,,\quad t_c =\frac{\sqrt{6}}{4\pi}\,,\quad v_c=\sqrt{6}\,.
\ee
Note that $p_c = p_{\max}/9$.
We get standard VdW-like behavior---similar to the Einstein--Maxwell case \cite{KubiznakMann:2012, GunasekaranEtal:2012}--- as shown in Fig. \ref{fig1}. Namely, we observe the characteristic $p-v$ diagram and the classical swallowtail  behavior of the Gibbs free energy. For $p<p_c$ there is a first order phase transition between small and large black holes that eventually terminates at a critical point \eqref{CPoint5} where the 
phase transition becomes second order and is characterized by the standard swallowtail mean field theory critical exponents 
\be\label{exponents}
\tilde \alpha=0\,,\quad \tilde \beta=\frac{1}{2}\,,\quad \tilde \gamma=1\,,\quad \tilde \delta=3\,.
\ee 
The whole situation is reminiscent of the liquid/gas phase transition of the Van der Waals fluid \cite{KubiznakMann:2012, GunasekaranEtal:2012}.

\begin{figure*}
\centering
\begin{tabular}{cc}
\rotatebox{-90}{
\includegraphics[width=0.34\textwidth,height=0.28\textheight]{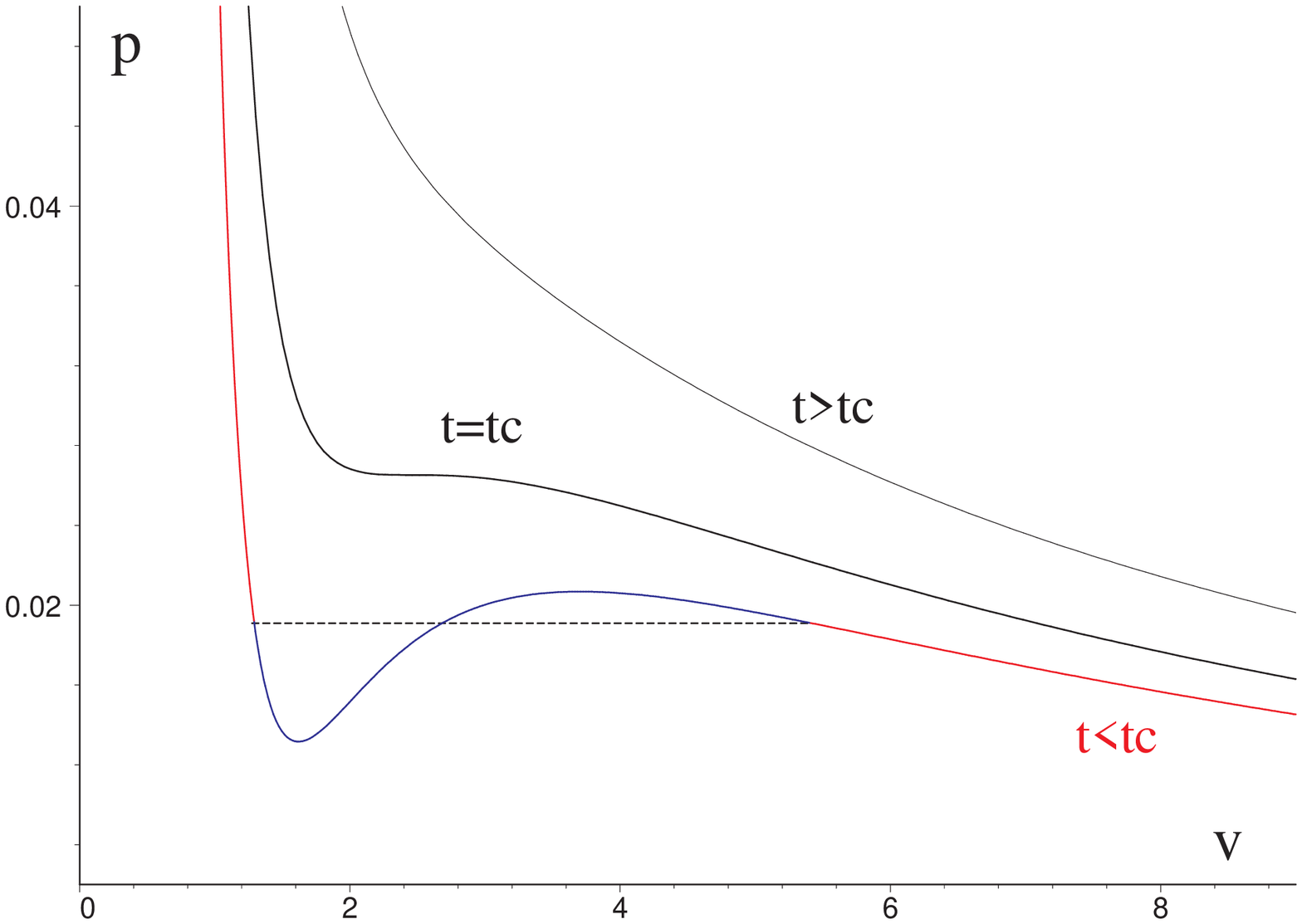}} &
\rotatebox{-90}{
\includegraphics[width=0.34\textwidth,height=0.28\textheight]{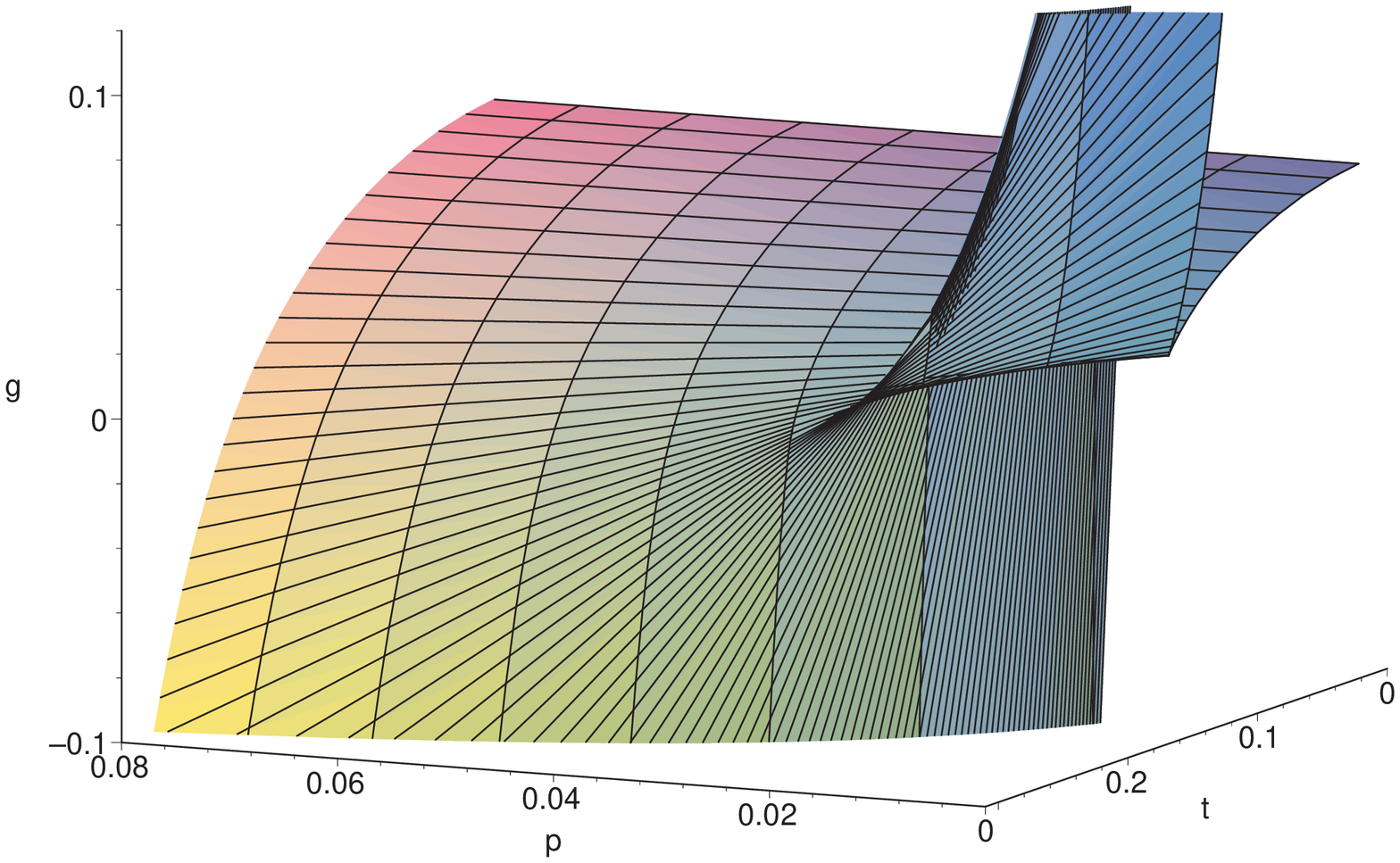}}\\
\rotatebox{-90}{
\includegraphics[width=0.34\textwidth,height=0.28\textheight]{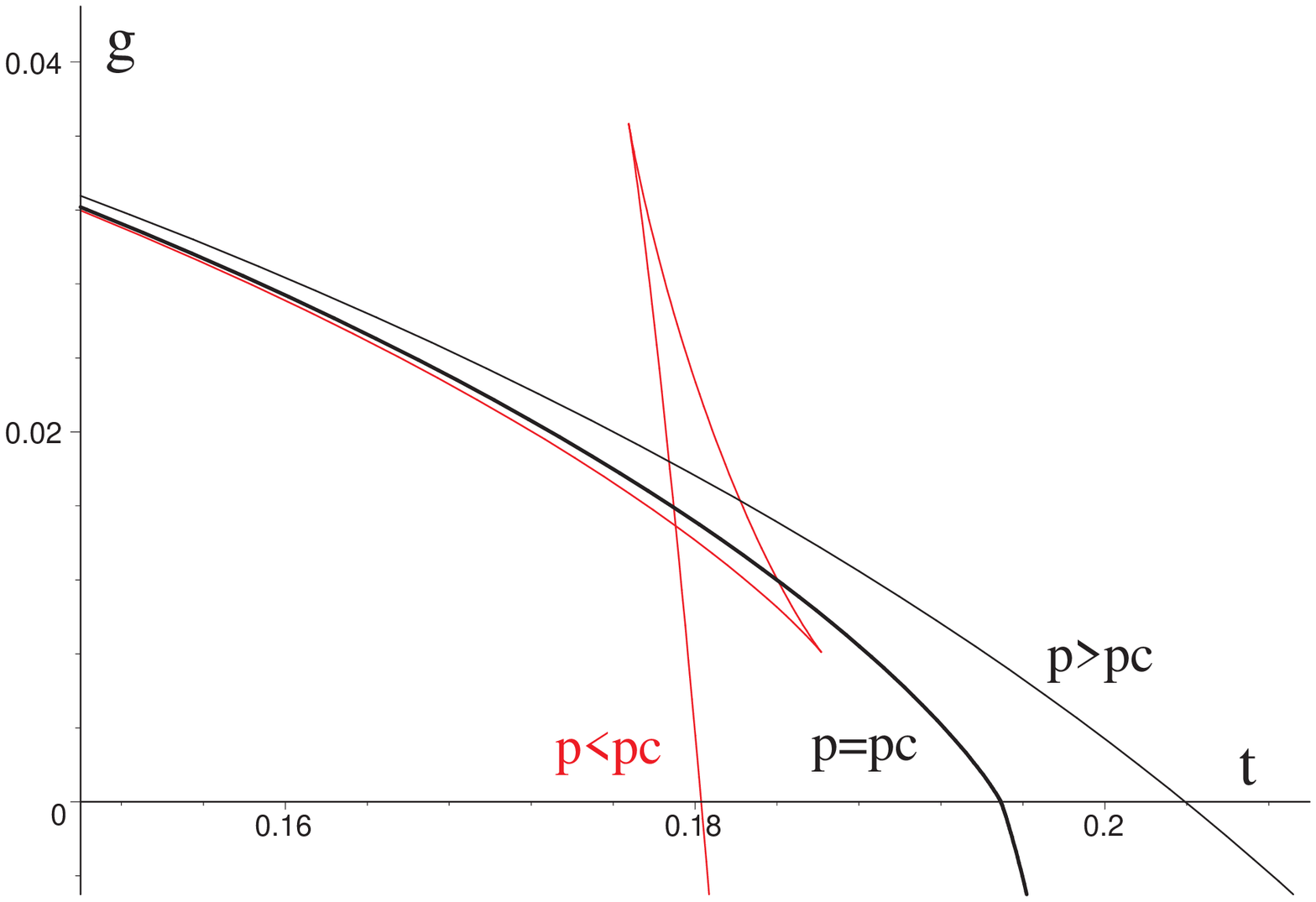}}&
\rotatebox{-90}{
\includegraphics[width=0.34\textwidth,height=0.28\textheight]{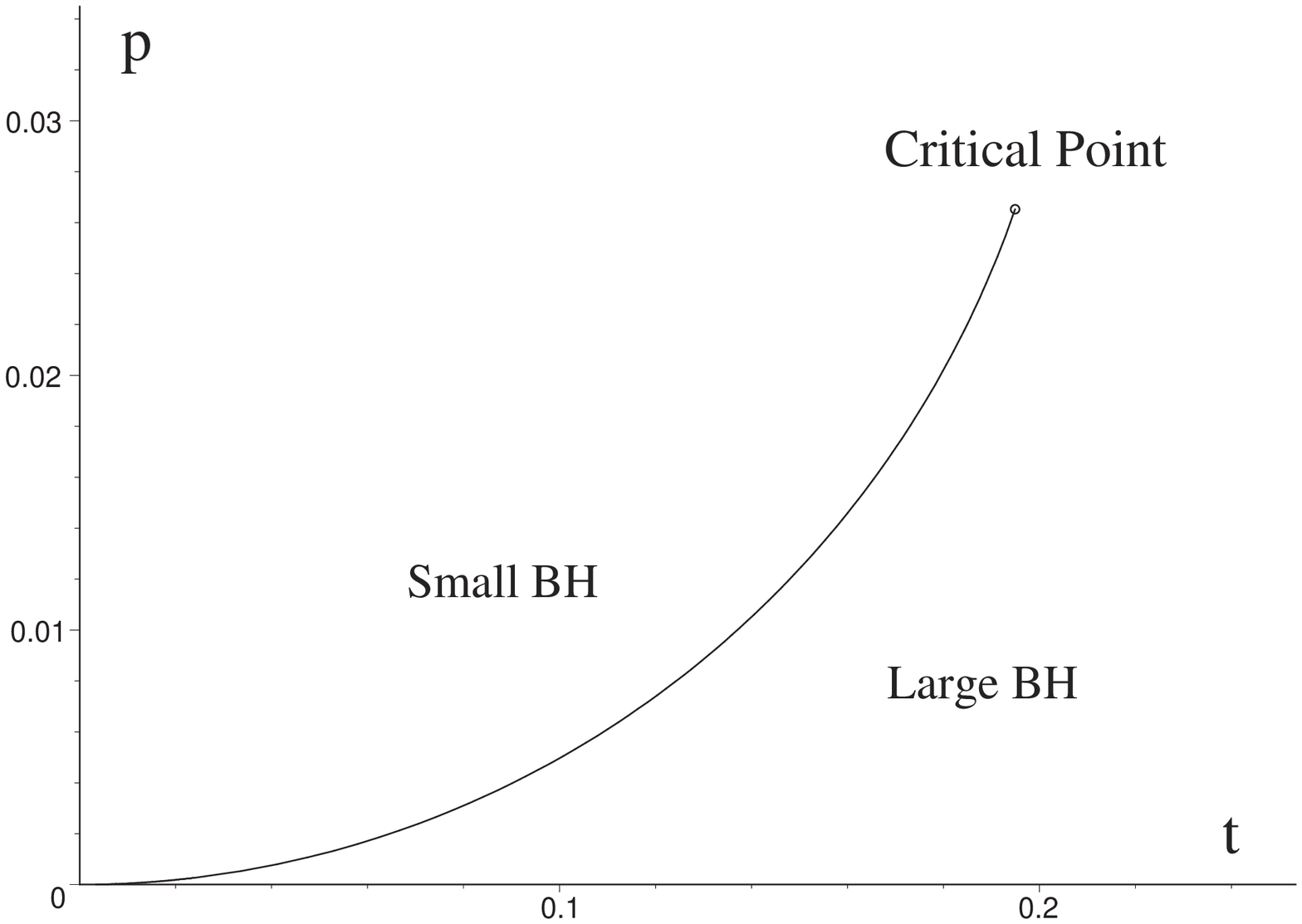}}\\
\end{tabular}
\caption{{\bf Van der Waals behaviour: $d=5, q=0, \kappa=1$.}
{\em Top left}: {The $p-v$ diagram. The oscillatory part of isotherm $t<t_c$ (blue curve) is replaced 
by the curves joined via the intermediate straight line
according to Maxwell's equal area law. {\em Top right:} The Gibbs free energy $g=g(p,t)$ demonstrates the classical swallowtail behavior for $p<p_c$. 
{\em Bottom left:} The $g-t$ diagram displays the $p=\mbox{const.}$ slices of the Gibbs free energy. Note that $p_{max}$ is much larger than $p_c$ and hence is not displayed in this figure. {\em Bottom right:} The $p-t$ diagram. The coexistence line of the first-order phase transition terminates at a critical point. The overall situation in all these graphs is reminiscent of the liquid/gas phase transition of the Van der Waals fluid.}
}  
\label{fig1}
\end{figure*} 

Even in the charged case an analytic solution for a critical point is possible--Eq.~\eqref{vc5d} effectively represents a cubic equation for $v^2_c$. However it is easy to see using the rule of signs that  there is exactly one sign change between the successive coefficients of \eqref{vc5d}, and hence we may have up to one critical point. Similar to the uncharged case, we observe (but do not display) standard VdW behavior.

\subsubsection{Hyperbolic case: thermodynamic singularity}
\begin{figure*}
\centering
\begin{tabular}{cc}
\rotatebox{-90}{
\includegraphics[width=0.34\textwidth,height=0.28\textheight]{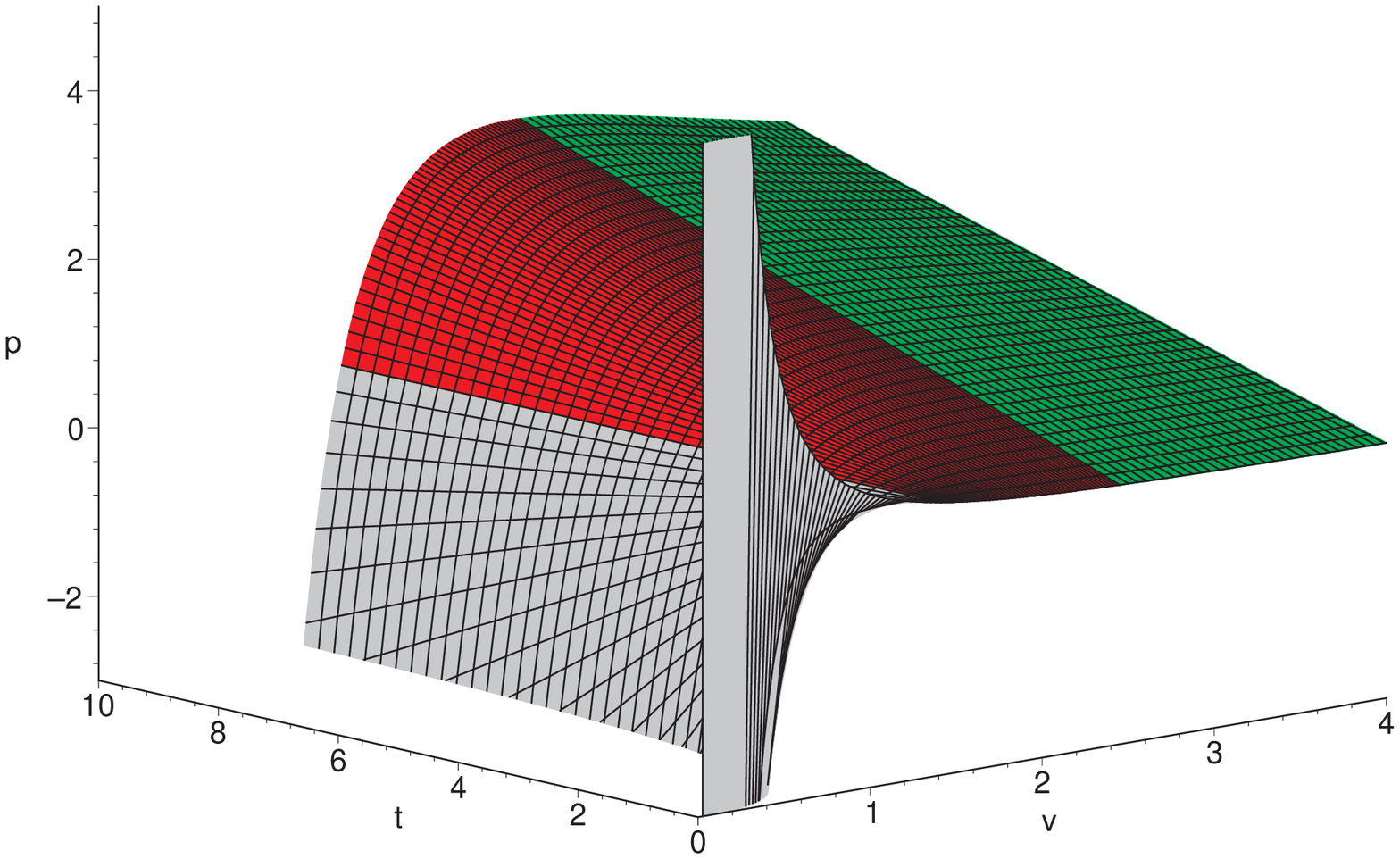}} &
\rotatebox{-90}{
\includegraphics[width=0.34\textwidth,height=0.28\textheight]{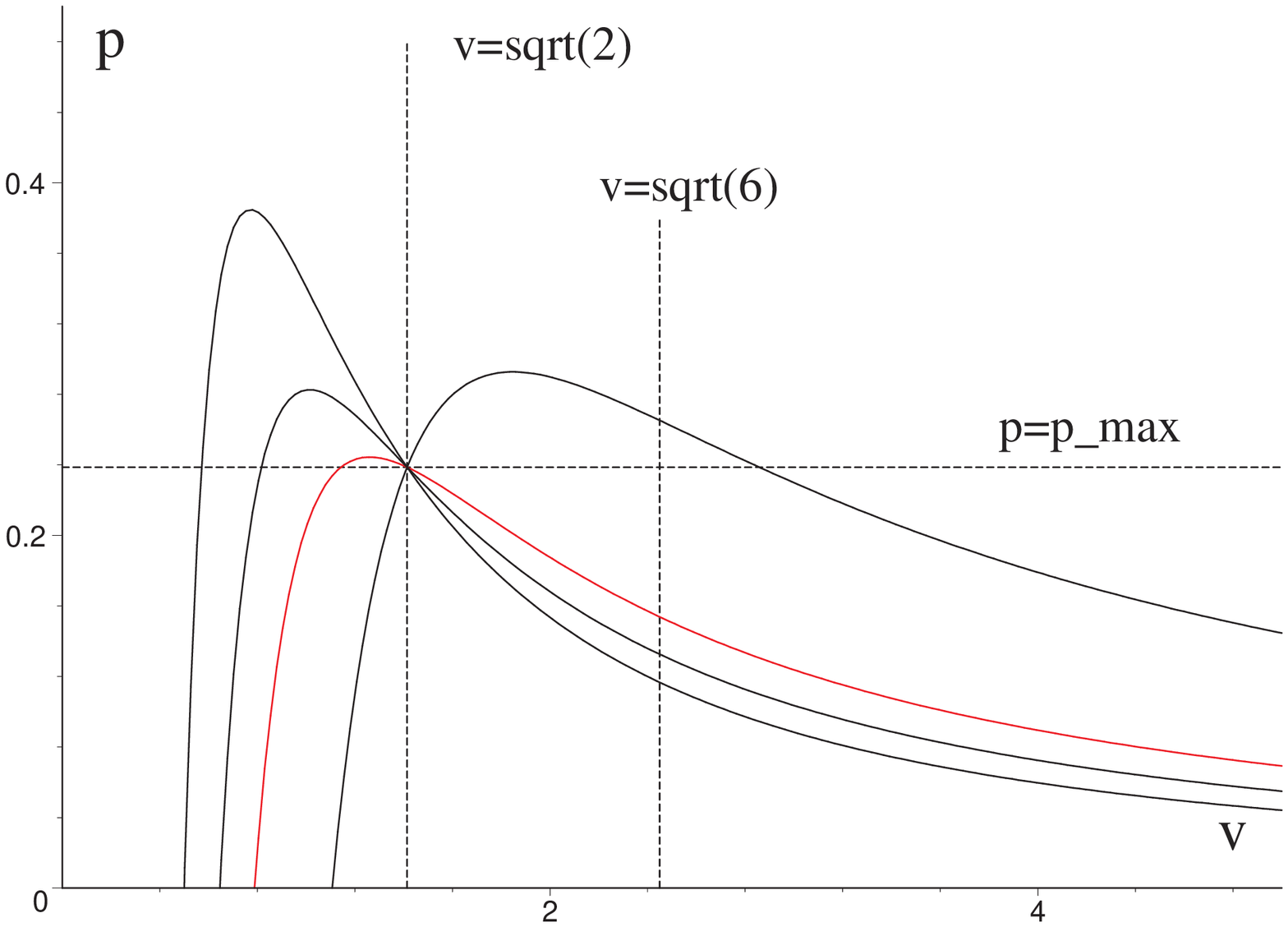}} \\
\rotatebox{-90}{
\includegraphics[width=0.34\textwidth,height=0.28\textheight]{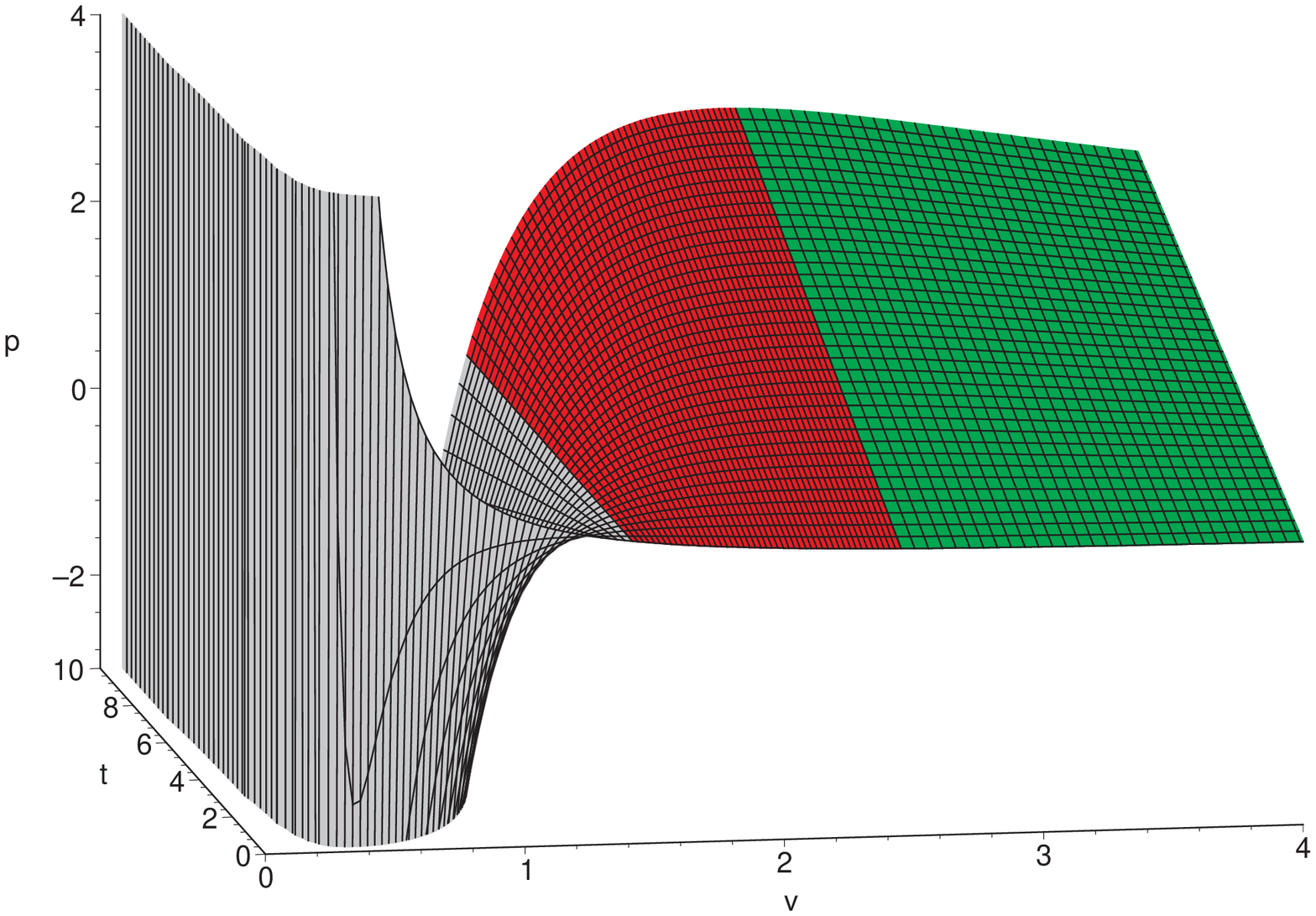}} &
\rotatebox{-90}{
\includegraphics[width=0.34\textwidth,height=0.28\textheight]{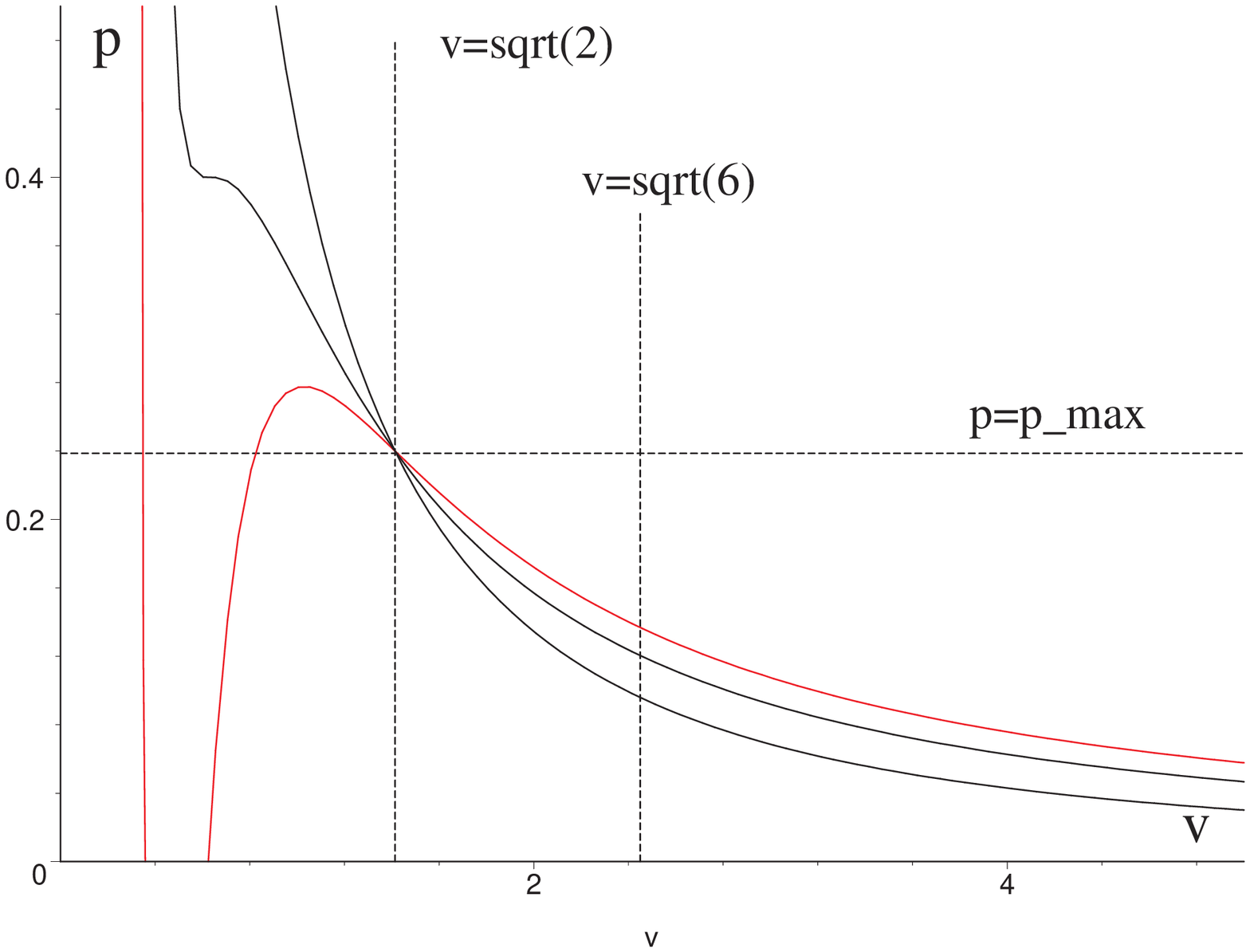}}\\
\end{tabular}
\caption{ {\bf $p(v,t)$ and $p-v$ diagrams: Gauss--Bonnet $d=5, \kappa=-1$.} 
{\em Left:} $p(v,t)$ diagrams displayed for $q=0$ (top) and $q=0.1$ (bottom). The grey colour corresponds to $v<v_s=\sqrt{2}$, red to $v\in(v_s, v_{S+})$, and green to physical black holes with $v>v_{S_+}=\sqrt{6}$. Note that at $v=v_s$ we have $\partial p/\partial t=0$ showing the presence of thermodynamic singularity. {\em Right:} corresponding $p-v$ diagrams. In this 2d ($t=\mbox{const.}$) projection the thermodynamic singularity 
at $v=v_s$ is manifest as a `singular point' where all the isotherms cross and reverse. Note also 
the maximal pressure $p_{max}$  and the reverse VdW behavior for $q=0.1$. 
}  
\label{Fig:5dPV23-5}
\end{figure*}

\begin{figure*}
\centering
\begin{tabular}{cc}
\rotatebox{-90}{
\includegraphics[width=0.34\textwidth,height=0.28\textheight]{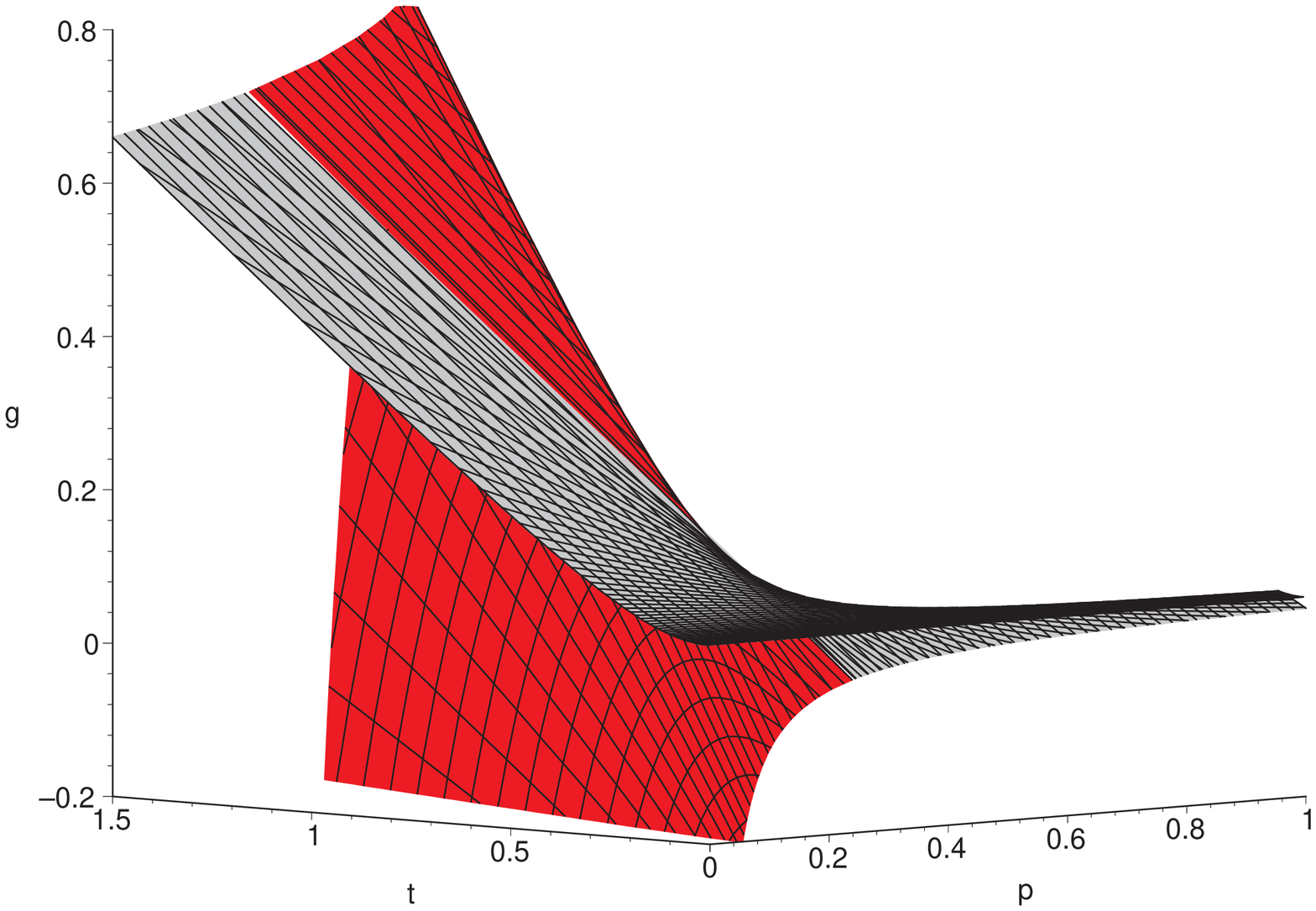}} &
\rotatebox{-90}{
\includegraphics[width=0.34\textwidth,height=0.28\textheight]{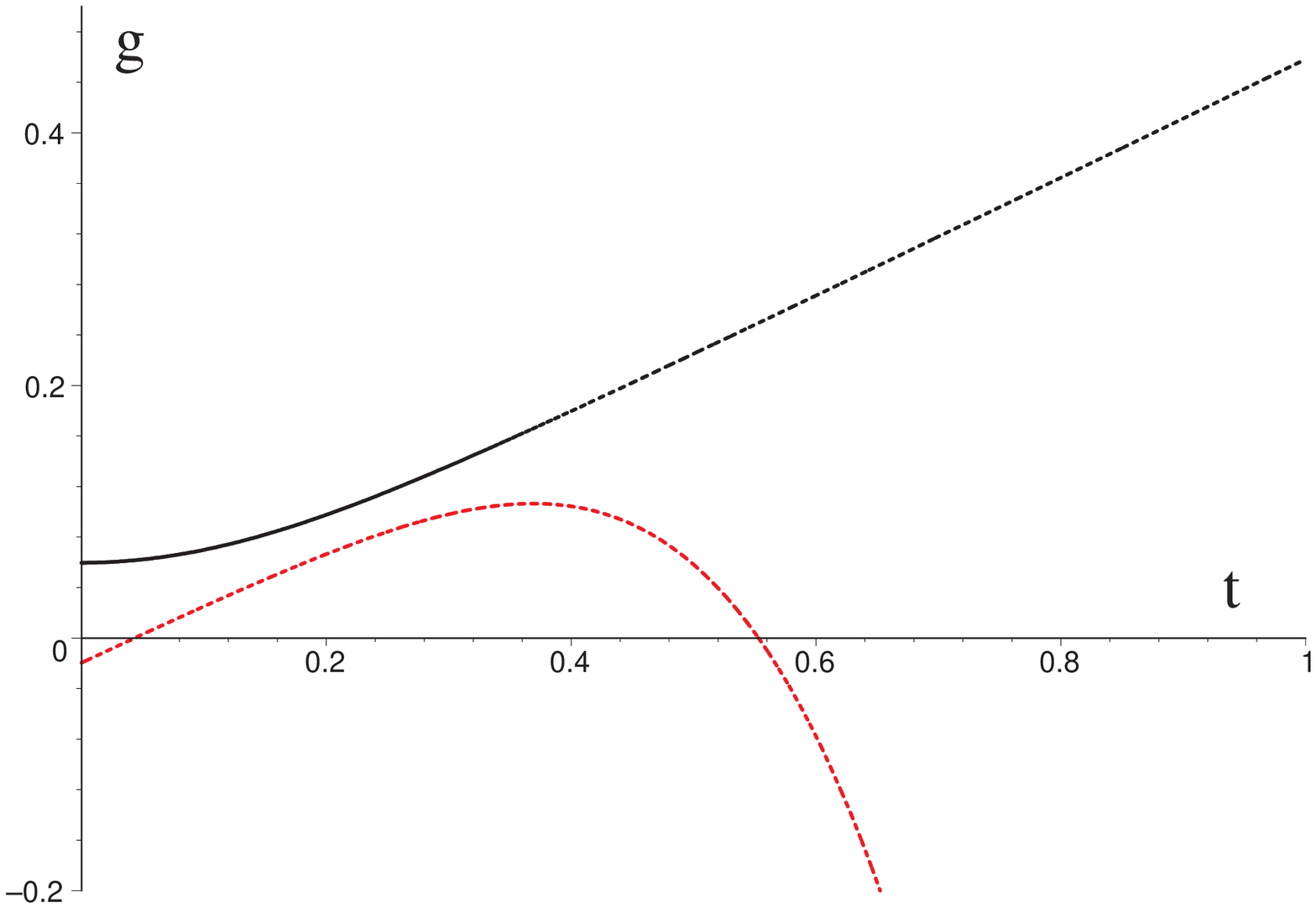}} \\
\rotatebox{-90}{
\includegraphics[width=0.34\textwidth,height=0.28\textheight]{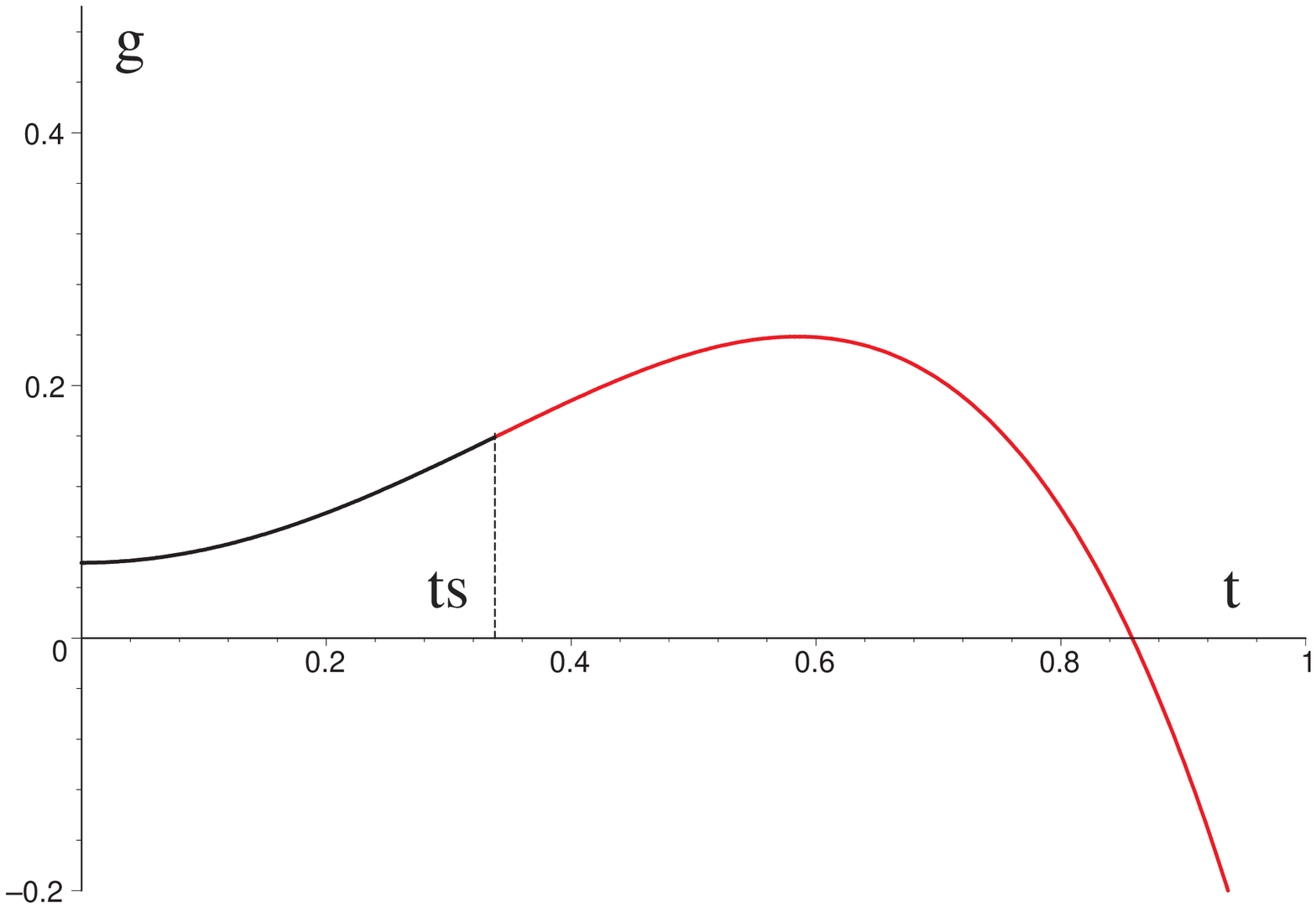}} &
\rotatebox{-90}{
\includegraphics[width=0.34\textwidth,height=0.28\textheight]{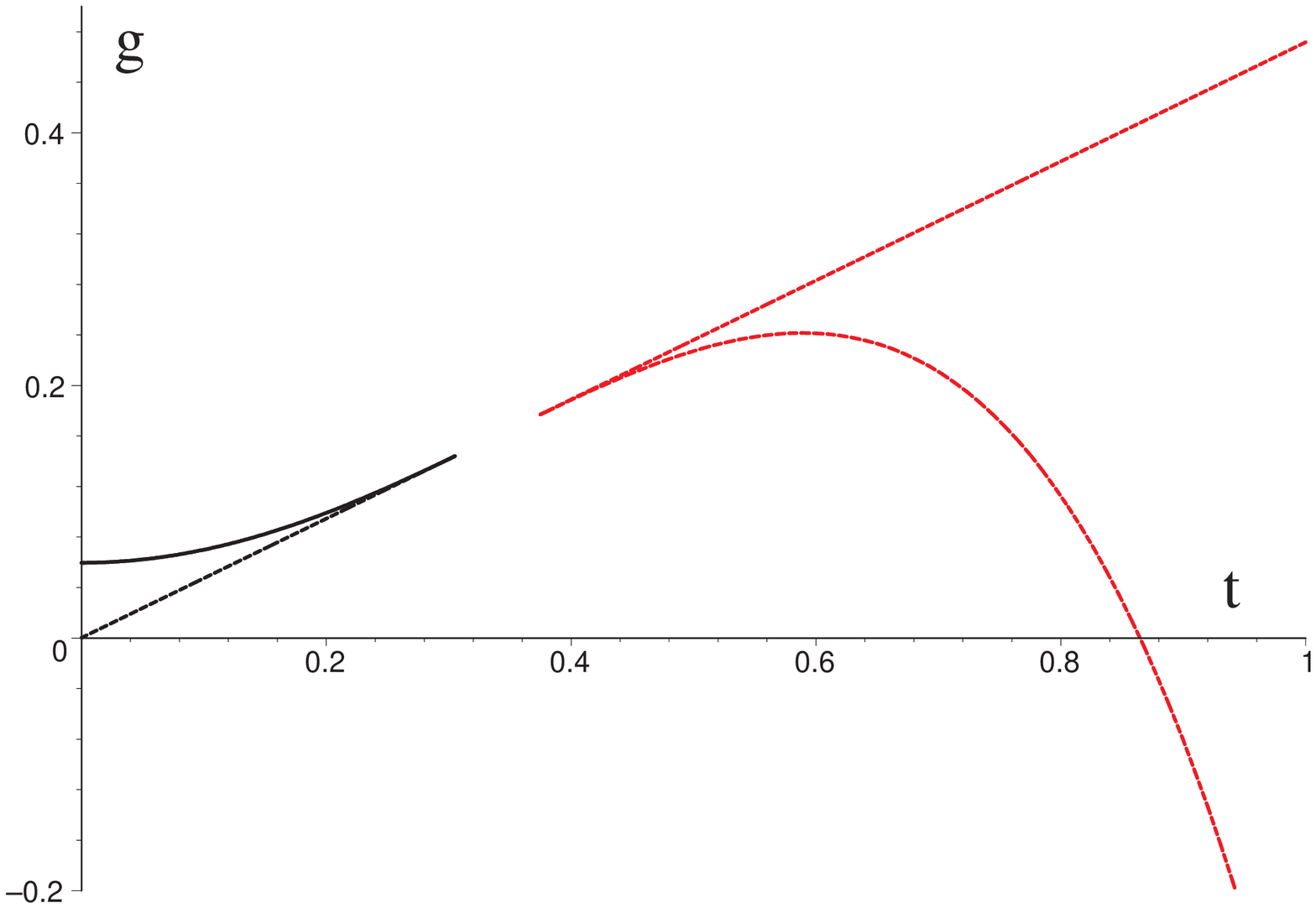}} 
\\
\end{tabular}
\caption{ {\bf Gibbs free energy and its characteristic `reconnection'.} The first diagram displays the Gibbs free energy $g(p,t)$  
for $d=5, \kappa=-1$ and $q=0$. The grey colour corresponds to black holes with $v<v_s$, the red to black holes with $v>v_s$. 
The remaining diagrams show the corresponding $g-t$ diagrams for $p=0.18, p_s, 0.245$, from top right to bottom right; 
the middle pressure is the reconnection pressure for which the Gibbs free energy no longer diverges at $v=v_s$. As the pressure is increased, the two branches, black ($v<v_s$) and red ($v>v_s$), touch and then reconnect forming a new asymptotic structure. Qualitatively similar behavior happens at any thermodynamic singularity.}  
\label{Fig:5dPV23-6}
\end{figure*} 

In the hyperbolic case there is no critical point when $q=0$, whereas we have one critical point in the charged case. Imposing the maximal pressure  $p<p_{max}$ and positive entropy ( $v>v_{S+}=\sqrt{6}$)
constraints, we observe in either case  `ideal gas' behaviour---without any criticality \cite{Cai:2013qga}.
This is because  i) the thermodynamic singularity always occurs for the branch of black holes with negative entropy and ii) the critical point always has $p_c>p_{max}$. These two statements remain true for Gauss--Bonnet black holes in all higher dimensions; in particular, the thermodynamic singularity does not play any significant role in their thermodynamics\footnote{ For the $d>5$ charged case, we may, however, observe the VdW-like with $p_c>p_{max}$ behavior.}.  
However it is instructive to look at what happens as we approach the thermodynamic singularity, as it will be present for physical (positive entropy) 3rd-order Lovelock black holes studied in the next section.

 To get a grasp of what is happening at the thermodynamic singularity we display the $p(v,t)$ diagram for $q=0$ and $q=0.1$ and its 2-dimensional $t=\mbox{constant}$ slices ($p-v$ diagrams) in 
 Fig.~\ref{Fig:5dPV23-5}. Ostensibly there is nothing unusual happening in the $p(v,t)$ graph. Upon  closer inspection, however,  we find that at $v=v_s=\sqrt{2}$, the pressure remains constant for any temperature, i.e., we have 
\be
\frac{\partial p}{\partial t}\bigg|_{v=v_s}=0\,.
\ee
Consequently the corresponding $p-v$ diagrams display `singular points' where all the isotherms {\em cross and `reverse'}. Namely, 
whereas for $v>v_s$ the hotter isotherm corresponds to a higher pressure, the converse holds after `crossing' the thermodynamic singularity, i.e., for  $v<v_s$ the cooler isotherms correspond to higher pressure. 
In fact, for a generic thermodynamic singularity one can define a {\em thermodynamic singular point} at $(p_s,v_s,t_s)$   from the following
equations:
\be
p_s=p_s(v_s,t_s)\,,\quad \frac{\partial p}{\partial t}\bigg|_{v=vs}=0\,,\quad 
\frac{\partial p}{\partial v}\bigg|_{t=ts}=0\,,
\ee
with the first equation being the equation of state. In particular, for  Gauss--Bonnet gravity in $d$ dimensions we get
\be
p_s=p_{max}+\frac{4q^2}{2^d}\,,\quad v_s=\sqrt{2}\,,\quad t_s=(d-2)\sqrt{2}\bigg[\frac{1}{4\pi}+\frac{q^2}{2^{d-2}}\bigg]
\ee
which in turn implies that the mass of the black hole vanishes 
 for $q=0$. This singularity point plays a very important role for the behavior of the Gibbs free energy and its characteristic reconnection as discussed in the next paragraph. Note also that $t_s$ is the correspondingly rescaled $T_s$, defined previously by the limiting process, cf. discussion before \eqref{Kretsch}. (The same is also true for 3rd-order Lovelock black holes discussed in the next section.)

{In the presence of thermodynamic singularity the Gibbs free energy $g=g(p,t)$ displays   complicated behavior, depicted for $q=0$ and $d=5$ in Fig.~\ref{Fig:5dPV23-6}, such that  the 2d $p=\mbox{constant}$ slices   {\em `reconnect'}. As the size of the black hole $v$ approaches $v_s$ from left or right, both the Gibbs and the temperature diverge to plus minus infinity. However both are finite at $p=p_s$;
 one can show that at $p=p_s$ the Gibbs free energy is `smooth' and finite
at $v=v_s$, whereas it diverges here for pressures slightly different from $p_s$. At the same time, as we increase the pressure, the two asymptotic branches ($v<v_s$ and $v>v_s$) approach each other, merge together at ($p=p_s, t=t_s$), reconnect, and change their asymptotic behaviour, as displayed in 
Fig.~\ref{Fig:5dPV23-6}. 

Let us consider two examples. First we consider the case $q=0, d=5$. In this case
\be
p_s=\frac{3}{4\pi}\,,\quad v_s=\sqrt{2}\,,\quad t_s=\frac{3\sqrt{2}}{4\pi}\,,
\ee
and the expansion of the Gibbs free energy and the temperature, expressed as functions of $p$ and $v$, yields,
\bea
g(v_s+dv,p_s+dp)&=&\frac{\sqrt{2}}{3}\frac{dp}{dv}+\frac{1}{2\pi}+O(dp,dv)\,,\nonumber\\
t(v_s+dv,p_s+dp)&=&\frac{dp}{dv}+t_s+O(dp,dv) 
\eea
as claimed. However, the Gibbs free energy is naturally expressed as function of $p$ and $t$ rather than $p$ and $v$, giving 
\be
g(t_s+dt,p_s+dp)=\frac{\sqrt{2}}{3}dt+\frac{1}{2\pi}+O(dp,dt)\,,
\ee
which is  smooth. At the same time it can be shown that $f''(r_+)$ (and thus the Kretschmann scalar) on the black hole horizon is finite at $p_s$ and so the tidal forces on the horizon are finite.
}

 As a second example, consider a very special case when the critical point 
coincides with the thermodynamic singular point. This happens for 
\be
q_c=\sqrt{\frac{2}{\pi}}\,,\quad p_c=\frac{1}{\pi}\,,\quad v_c=\sqrt{2}\,,\quad t_c=\frac{3\sqrt{2}}{2\pi}\,.
\ee
Around this point we have the following expansion of the Gibbs and temperature, expressed as functions of $p$ and $v$:
\bea
g(v_c+dv,p_c+dp)&=&\frac{\sqrt{2}}{3}\frac{dp}{dv}+\frac{19}{16 \pi}+O(dp,dv)\,,\nonumber\\
t(v_c+dv,p_c+dp)&=&\frac{dp}{dv}+t_c+O(dp,dv)\,,
\eea
Expressing again $g$ as a function of $p$ and $t$, we get
\be
g(t_c+dt,p_c+dp)=\frac{\sqrt{2}}{3}dt+\frac{19}{16 \pi}+O(dp,dt)\,,
\ee
which is  smooth. Despite  the thermodynamics appearing to be well defined near this critical point, 
contrary to the previous case (and in fact whenever the charge $q$ is non-zero) one can show that the black hole {\em horizon becomes singular}, as the Kretschmann scalar is not well behaved here and diverges \eqref{Kretsch}.

Qualitatively similar interesting features, such as reconnection and finiteness of the Gibbs free energy at $p=p_s$, are observed for thermodynamic singularities in higher dimensions and higher-order Lovelock gravity. The very special case discussed in Sec.~\ref{Sec:alpha3} where two thermodynamic singularities coincide together allows for even more interesting behaviour as we shall later demonstrate.
We stress again that in the case of Gauss--Bonnet black holes, one of the black hole branches is always unphysical as it possess a negative entropy. Hence the  behaviour described above is only a toy example of what we shall see in the case of 3rd-order Lovelock gravity.

Finally we discuss  the {\em `reverse VdW behavior'} that happens in the charged case, for $q<q_c=\sqrt{\frac{2}{\pi}}$, as observed in Fig.~\ref{Fig:5dPV23-5}. For such charges the critical point occurs for $v<v_s$, i.e., in the region of reversed isotherms. Consequently, the characteristic VdW-like oscillations in the $p-v$ diagram now occur for $t>t_c$ rather than $t<t_c$ as is the case of normal VdW. All this is of course unphysical as $p_c>p_{max}$ and the corresponding black holes have negative entropy. However, we shall see in the next section that a physical reverse VdW behaviour is observed for 3rd-order hyperbolic Lovelock black holes---noted previously for a restricted class of
3rd-order Lovelock gravity theories  \cite{Mo:2014qsa,Xu:2014kwa}.

\subsection{Six dimensions}
\begin{table}
\begin{centering}
\begin{tabular}{|c|c|c|c|}
\hline
q & \# apparent critical points & \# physical critical points & behavior\tabularnewline
\hline
0&1 & 0& cusp \tabularnewline
\hline
$(0,q_{\min})$&3 & 0& VdW with $p_c>p_{max}$ \tabularnewline
\hline
$(q_{min}, q_1)$& 3 & 1& VdW \tabularnewline
\hline
$(q_1,q_2)$& 3 & 2 & triple point\tabularnewline
\hline
$(q_2,q_3)$&3& 1& VdW \tabularnewline
\hline
$q>q_3$&1& 1& VdW \tabularnewline
\hline
\end{tabular}
\protect\caption{Types of physical behavior in $d=6$ Gauss--Bonnet $\kappa=+1$ case.\label{tab:casesGB}}\label{TableOne}
\end{centering}
\end{table} 

In $d=6$ we obtain  the following equation of state: 
\bea
p&=&\frac{t}{v}-\frac{3\kappa}{\pi v^2}+\frac{2\kappa t}{v^3}-\frac{1}{\pi v^4}+\frac{q^2}{v^8}\,,
\eea
while the critical point satisfies
\be\label{vc6d}
t_c=\frac{2(3\kappa v_c^6+2v_c^4-4\pi q^2)}{\pi v_c^5(v_c^2+6\kappa)}\,,\quad 
3v_c^8-12\kappa v_c^6+12v_c^4-4\pi q^2(7\kappa v_c^2+30)=0\,.
\ee
The maximum pressure is $p_{max}=\frac{5}{4\pi}\approx 0.3979$ and for $\kappa=-1$ only black holes with $v>v_{S+}=2$ have positive entropy.

\subsubsection{Spherical case: triple point}

\begin{figure*}
\centering
\begin{tabular}{cc}
\rotatebox{-90}{
\includegraphics[width=0.34\textwidth,height=0.28\textheight]{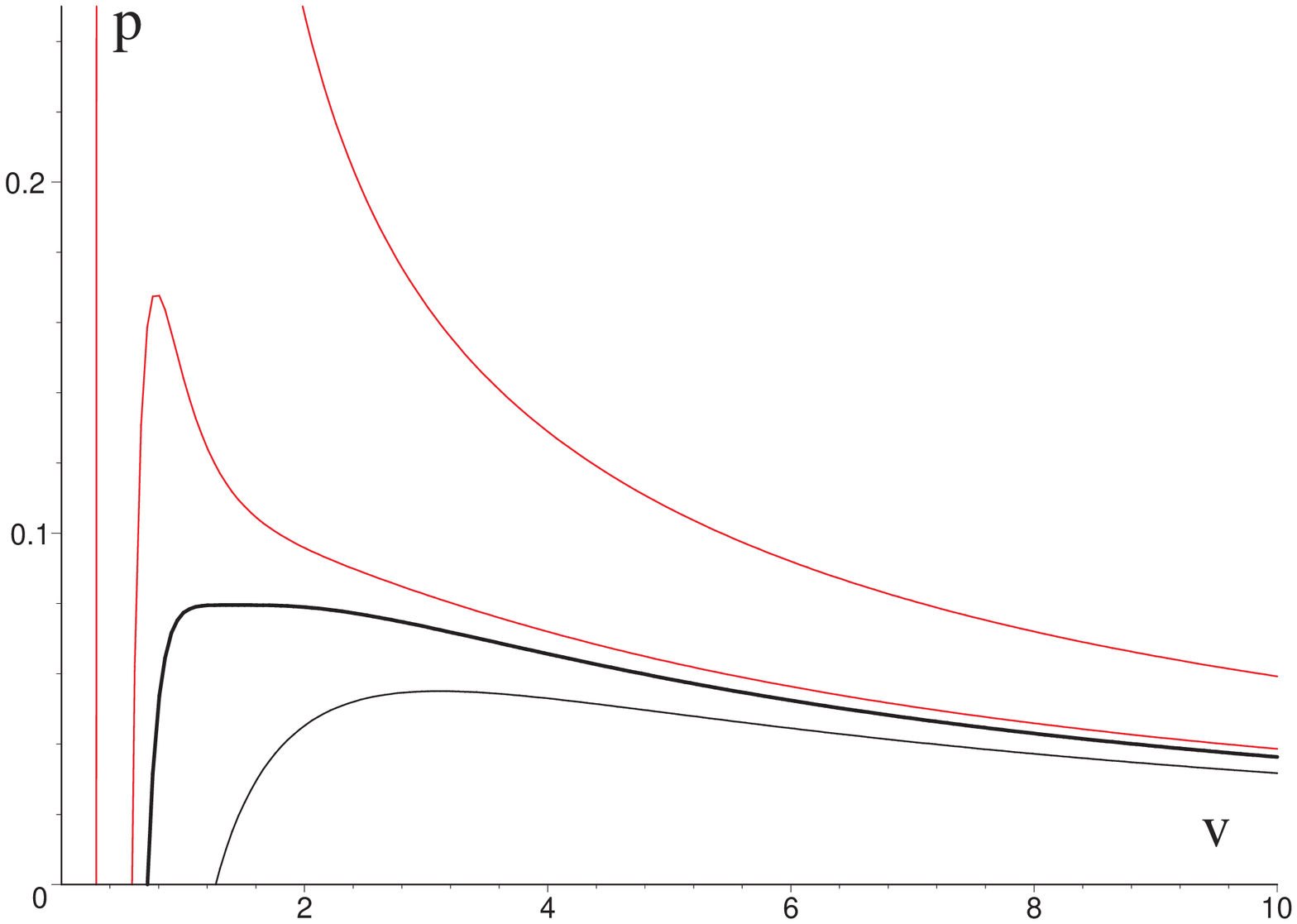}} &
\rotatebox{-90}{
\includegraphics[width=0.34\textwidth,height=0.28\textheight]{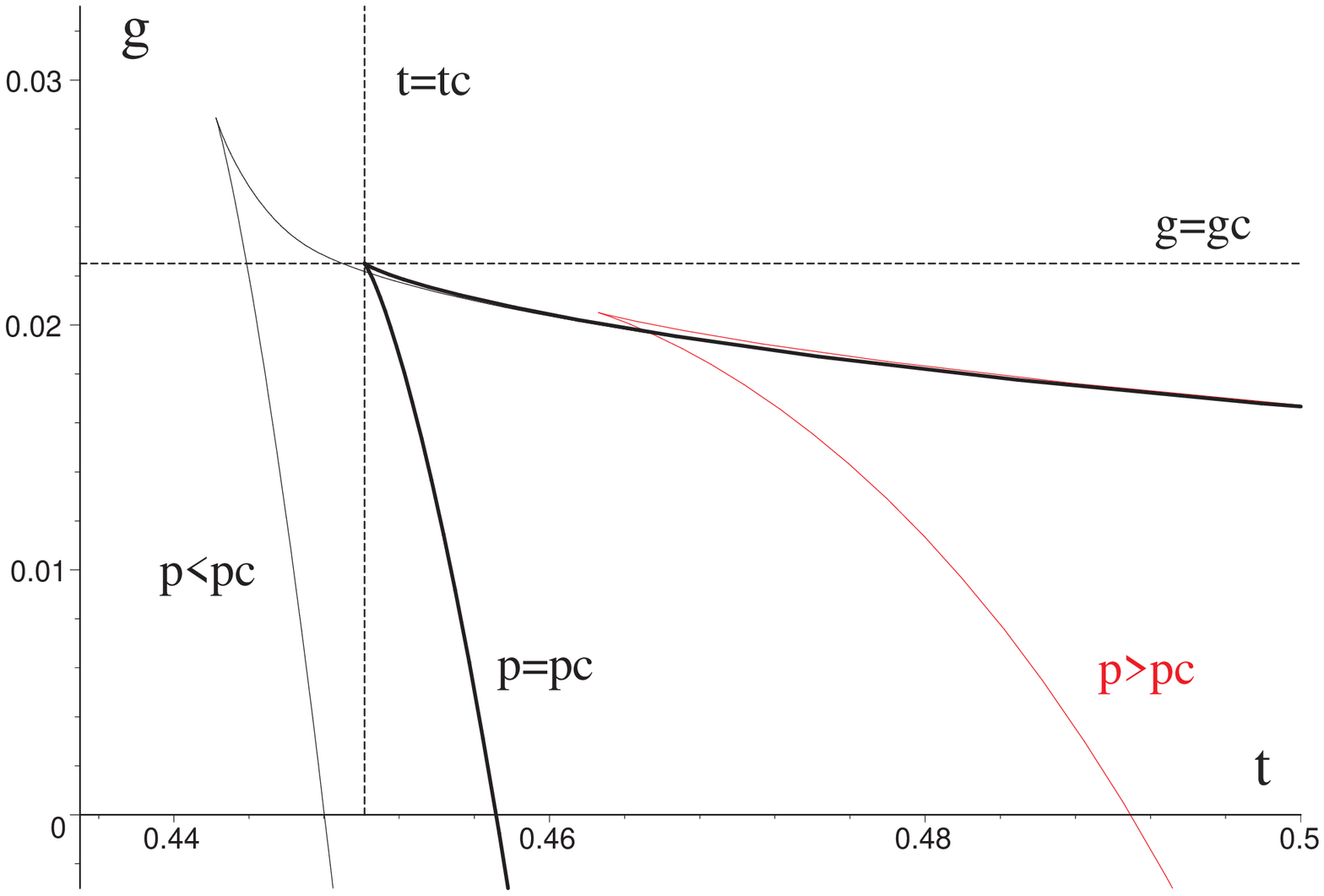}}\\
\end{tabular}
\caption{{\bf Thermodynamics in $d=6$: $q=0, \kappa=1$.}
The $p-v$ diagram (left) admits an inflection point (with the corresponding isotherm highlighted by the thick solid black curve) which however does not correspond to a critical point but rather to a cusp in $g$ as displayed in $g-t$ diagram on the right.  
}  
\label{fig6d}
\end{figure*}

The types of physical behavior for $d=6$ Gauss--Bonnet black holes with spherical horizon topology are summarized in Table~\ref{TableOne}.
In the uncharged case  an analytic solution of \eqref{vc6d} is possible and gives
\be\label{crit6d} 
p_c= \frac{1}{4\pi}\,,\quad 
t_c = \frac{\sqrt{2}}{\pi}\,,\quad v_c=\sqrt{2}\,.
\ee
Here we find a curious situation insofar as the solution \eqref{crit6d} is  an inflection point in the $P-v$ diagram that 
corresponds to a cusp in $g$ at $g_c=\frac{\sqrt{2}}{20\pi}$.  Hence it does not correspond 
to a real critical point and no phase transition takes place.  We illustrate this in  Fig.~\ref{fig6d}.

\begin{figure*}
\centering
\begin{tabular}{cc}
\rotatebox{-90}{
\includegraphics[width=0.34\textwidth,height=0.28\textheight]{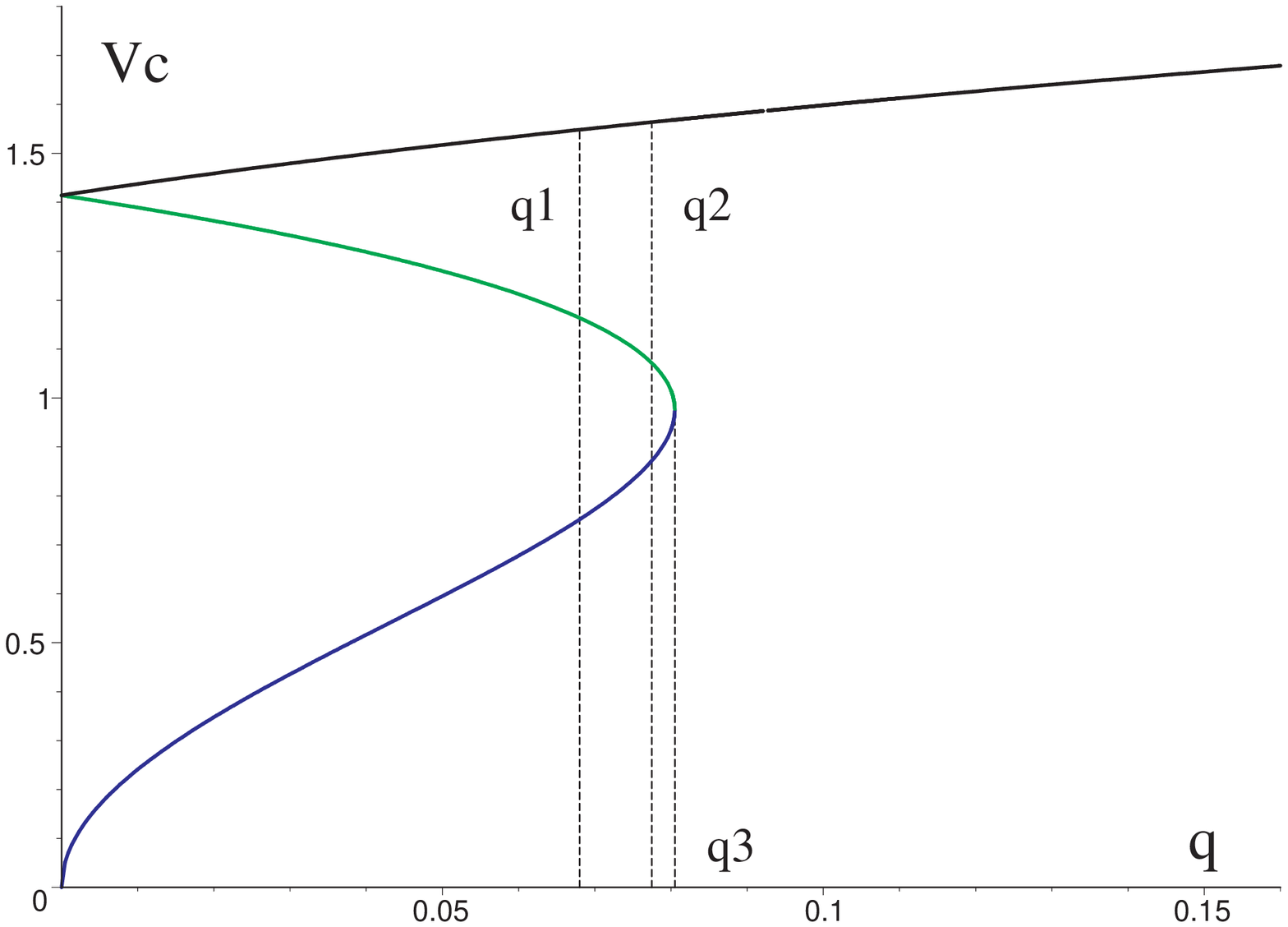}} &
\rotatebox{-90}{
\includegraphics[width=0.34\textwidth,height=0.28\textheight]{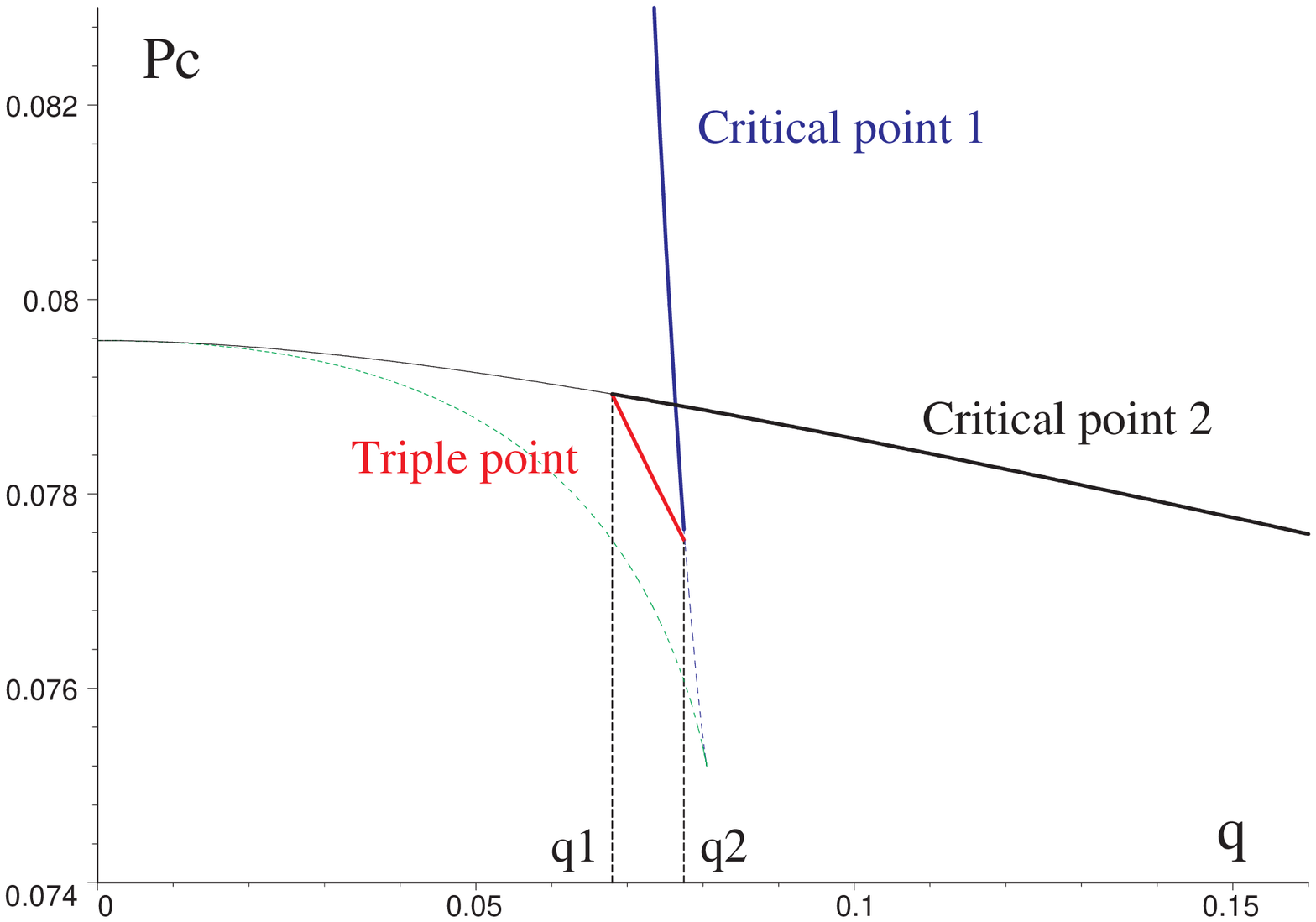}}\\
\end{tabular}
\caption{{\bf Critical points and a triple point: $d=6, \kappa=1$.}
{\em Left:} Critical volume $v_c$ is displayed for up to three critical points. Whereas for $q<q_3$ there exists 3 (not necessarily physical) critical points, for larger $q$ there exists only one. {\em Right:} Critical pressure is displayed by thick solid lines for various physical critical points; the dashed curves correspond to unphysical critical points which do not globally minimize the Gibbs free energy $g$. For $q<q_1$ the system exhibits one critical point, displayed by solid thick blue curve. Note that as $q\to 0$ the corresponding $p_{c_1}$ rapidly rises and quickly exceeds the maximal pressure $p_{max}$ resulting in a minimal $q=q_{min}$ below which the critical point 1 is no longer physical. For $q\in (q_1,q_2)$ we observe two critical points; the corresponding first-order phase transitions eventually merge in a triple point displayed by a thick solid red curve. Finally, for $q>q_2$ the system again admits only one physical critical point, displayed by solid thick black curve.   
}  
\label{Fig:8}
\end{figure*}

{\em Triple point.}
When $q>0$, Eq.~\eqref{vc6d} admits three sign changes and so we may have up to 3 critical points.
In fact, one can show that for 
for $q>q_3 =\frac{2}{147}\sqrt{\frac{110}{\pi}}\approx 0.08051$ there is only one critical point   corresponding to the standard VdW behavior,
whereas for $q<q_3$, there exist 3 critical points with positive $p_c, v_c$ and $t_c$. However, not all of these critical points occur for a branch globally minimizing the Gibbs free energy or have $p_c<p_{max}$ (see Fig.~\ref{Fig:8}); hence these
points are not `physical'.
Consequently only for the much smaller range, $q\in (q_1, q_2)\approx (0.0680,0.0775)$, may we observe two first-order small/intermediate and intermediate/large black hole phase transitions that, as the pressure is decreased, eventually merge in a tricritical (triple) point.
The triple point for example occurs for 
\be
q=0.075\,,\quad p_{3c}=0.07791\,,\quad t_{3c}=0.4486\,,
\ee
where small ($v_1=0.72165$), intermediate ($ v_2=1.07614$) and large ($v_3=1.9491$) black holes  coalesce.  
In the $p-v$ diagram this corresponds to a special isotherm for which we observe two Van der Waals oscillations such that the two equal area laws saturate for the same tricritical pressure $p_{3c}$, as shown in Fig.~\ref{Fig:9} on the left. 
(A similar interpretation takes place in the $T-v$ plane, not-shown.) 
Alternatively, for a fixed pressure the triple point corresponds to a `double swallow tail' of the Gibbs free energy, as displayed in Fig.~\ref{Fig:9b}. 
The situation is similar to what was observed for rotating black holes \cite{Altamirano:2013uqa} and in some sense is reminiscent of a solid/liquid/gas phase transition  \cite{Altamirano:2014tva}. In the context of Gauss--Bonnet gravity this phenomenon was first observed in \cite{Wei:2014hba}.  

\begin{figure*}
\centering
\begin{tabular}{cc}
\rotatebox{-90}{
\includegraphics[width=0.34\textwidth,height=0.28\textheight]{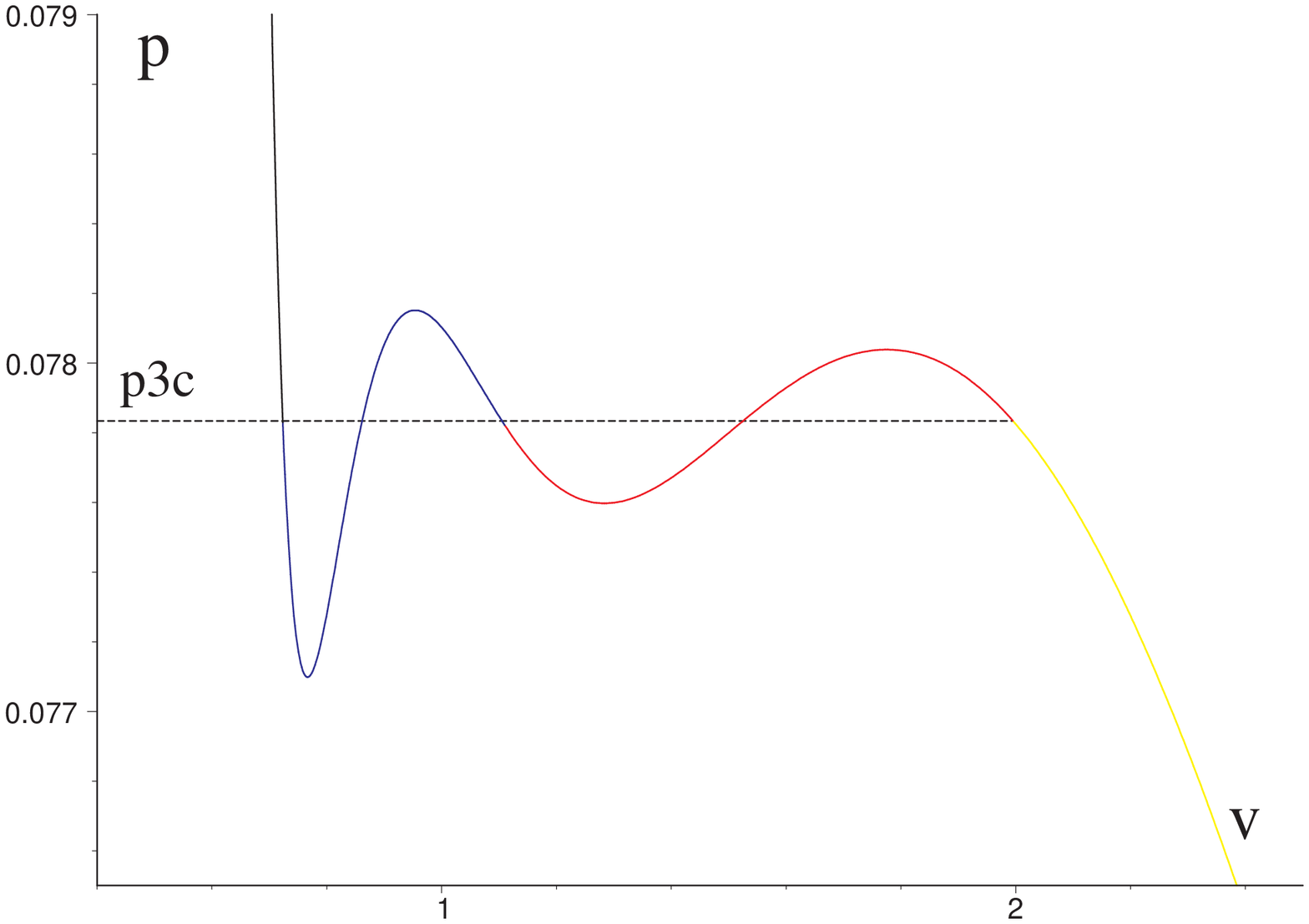}} &
\rotatebox{-90}{
\includegraphics[width=0.34\textwidth,height=0.28\textheight]{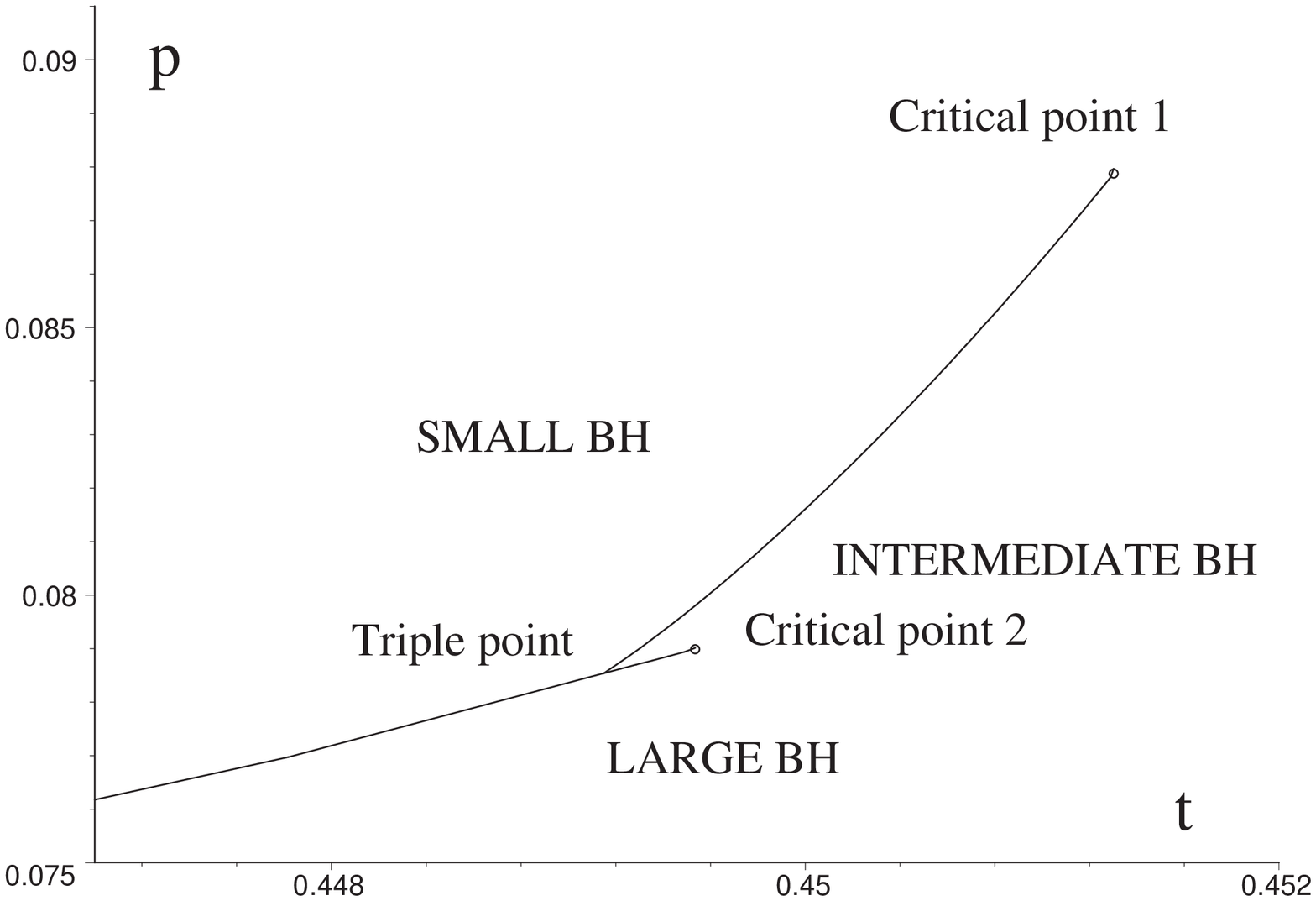}}\\
\end{tabular}
\caption{{\bf A triple point: $d=6, \kappa=1$.}
{\em Left}: The triple point in the $p-v$ diagram corresponds to an isotherm for which a `double' Maxwell equal area holds for the same pressure; 
$q=0.075$. 
{\em Right:} $p-t$ phase diagram. We observe two critical points and a triple point; $q=0.071$. The first-order phase transition in the left low corner eventually terminates at $[0,0]$.
}  
\label{Fig:9}
\end{figure*}

\begin{figure*}
\centering
\begin{tabular}{cc}
\rotatebox{-90}{
\includegraphics[width=0.34\textwidth,height=0.28\textheight]{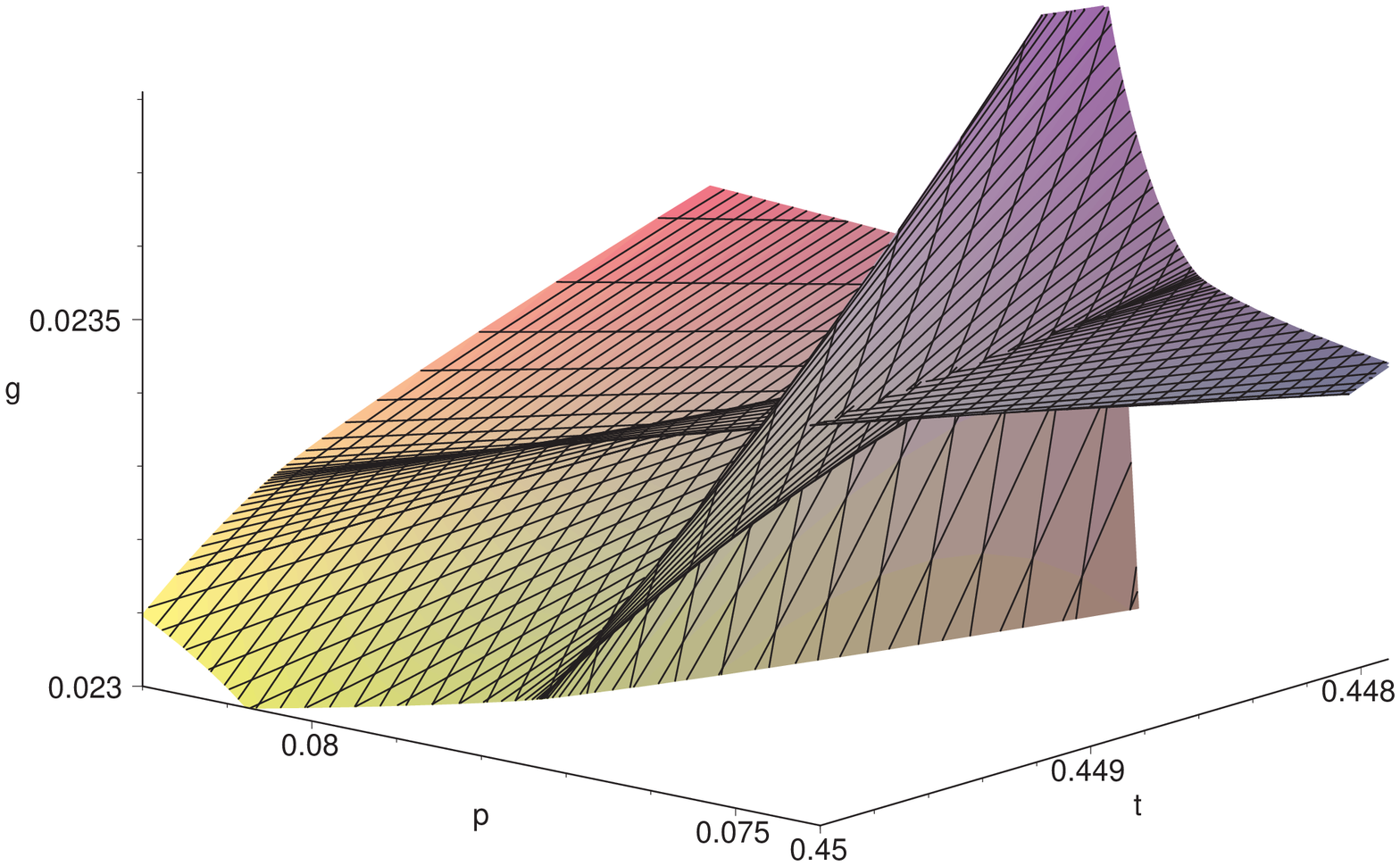}} &
\rotatebox{-90}{
\includegraphics[width=0.34\textwidth,height=0.28\textheight]{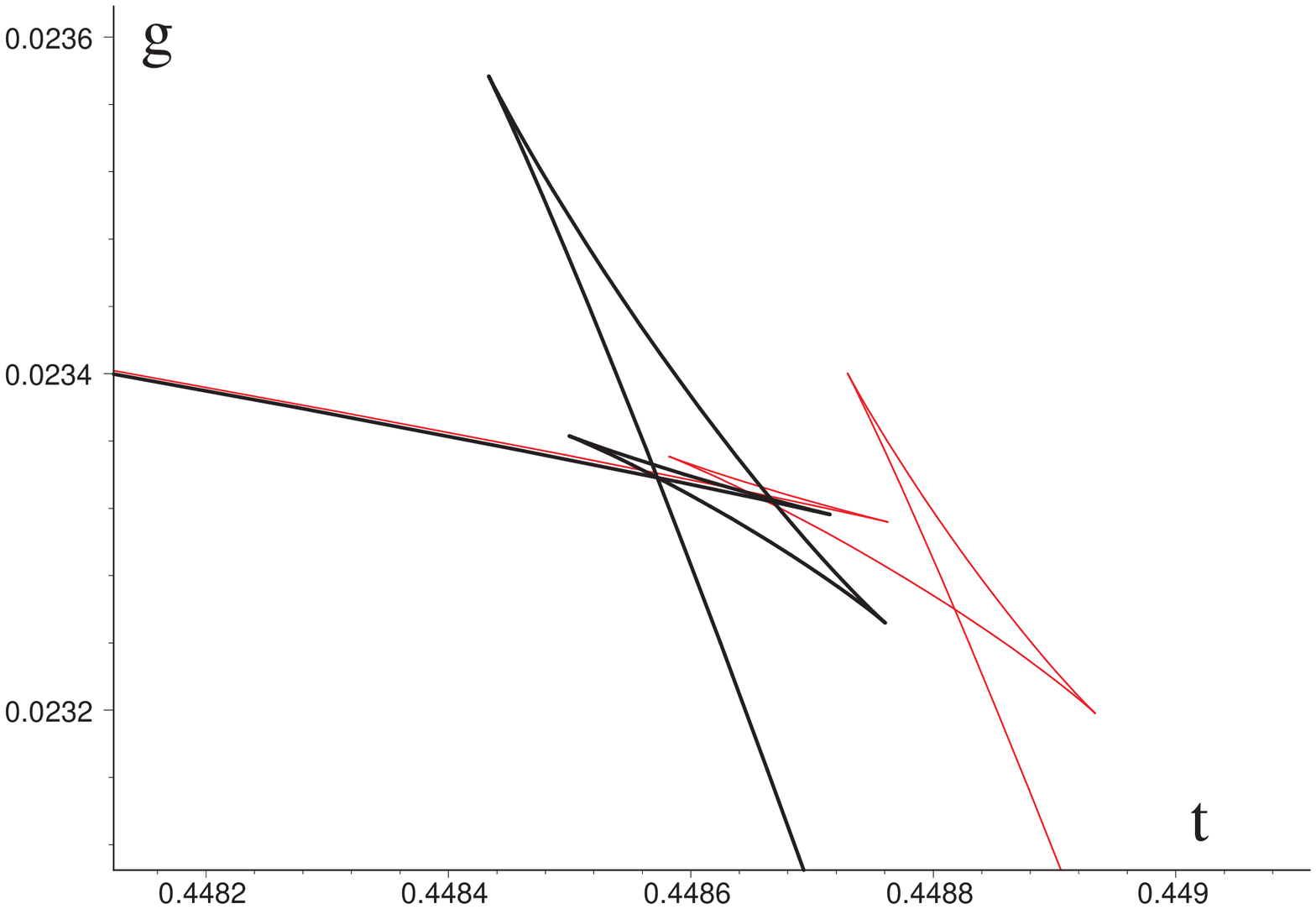}}\\
\end{tabular}
\caption{{\bf The Gibbs free energy of a triple point: $d=6, q=0.075, \kappa=1$.}
{\em Left:} 3d graph of the Gibbs free energy exhibits two swallow tails merging together to form a single swallow tail.
{\em Right:} a 2d $p=\mbox{const}$ slice of the left figure. The red curve shows two swallow tails indicating the presence of the two first-order phase transition that as the pressure decreases eventually merge (black curve) to form a triple point. 
}  
\label{Fig:9b}
\end{figure*}

\subsubsection{Hyperbolic case}
The thermodynamic behavior of (un)charged hyperbolic black holes in $d=6$ dimensions is very similar to what happens 
in the $d=5, \kappa=-1$ (un)charged case. The only difference is that in $d>5$ dimensions and for sufficiently large charge, $v_c$ may occur for positive entropy black holes, $v_c>v_{S+}$. Consequently, we may observe a VdW-like behavior, however with $p_c>p_{max}$.

\subsection{Higher dimensions}
\begin{figure*}[h]
\centering
\begin{tabular}{cc}
\includegraphics[width=0.49\textwidth]{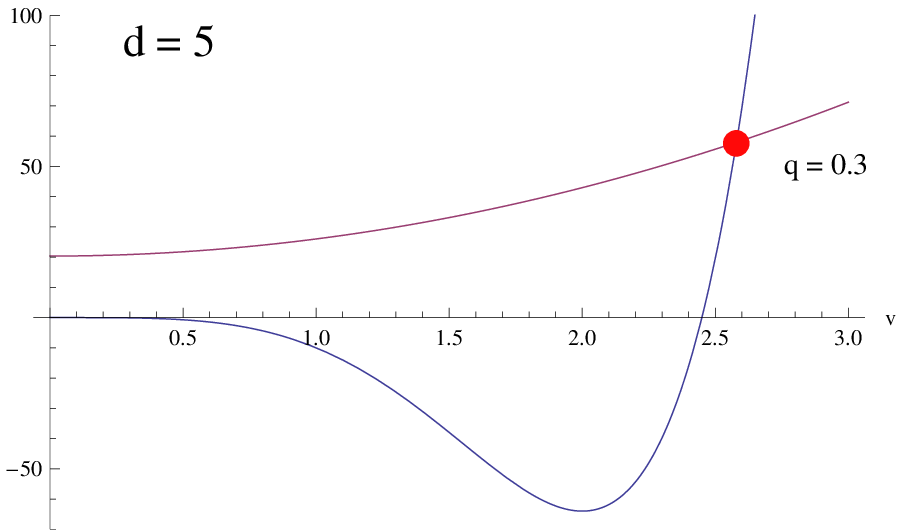} &
\includegraphics[width=0.48\textwidth]{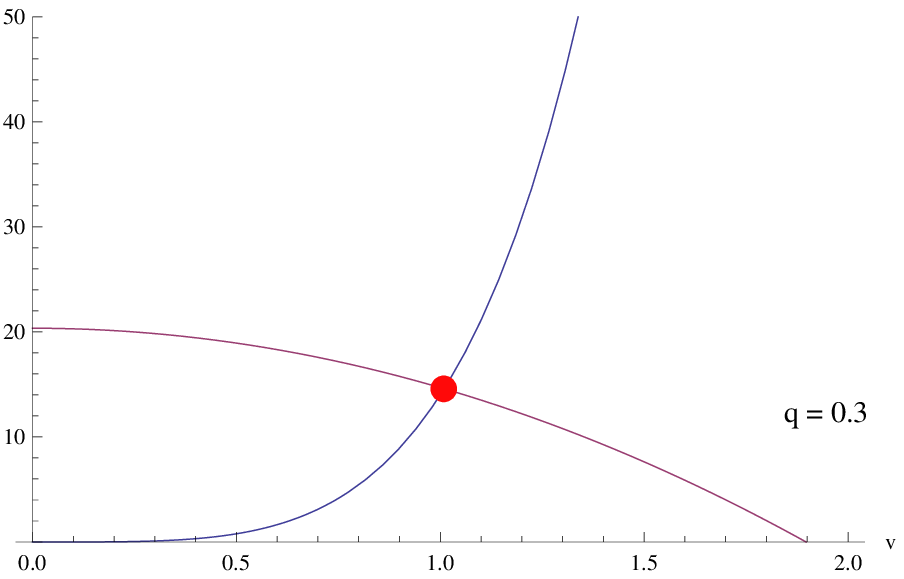}\\
\includegraphics[width=0.46\textwidth]{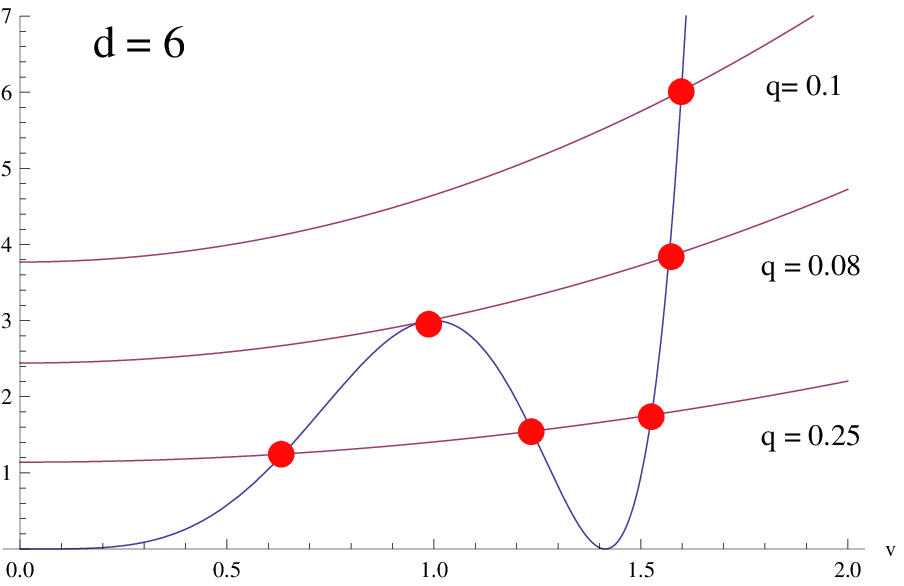} &
\includegraphics[width=0.48\textwidth]{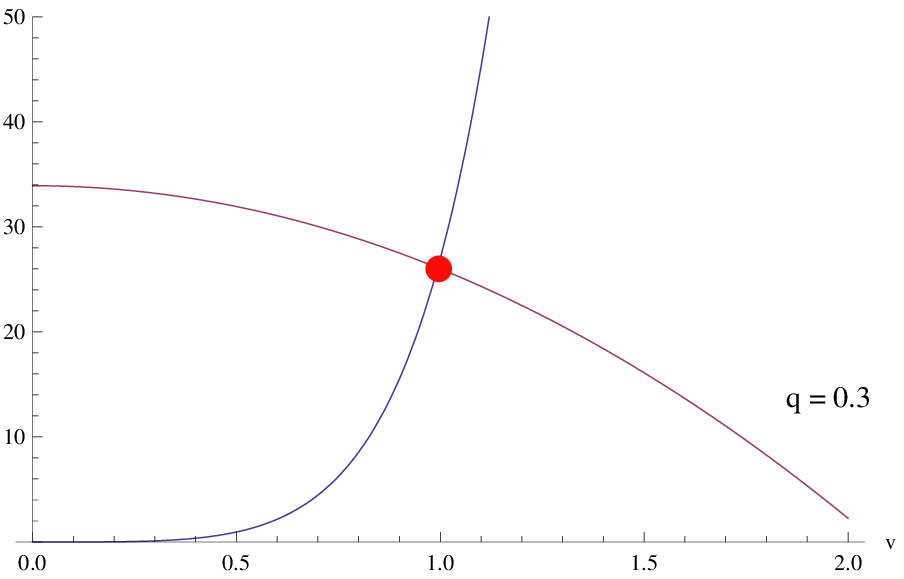}\\
\includegraphics[width=0.48\textwidth]{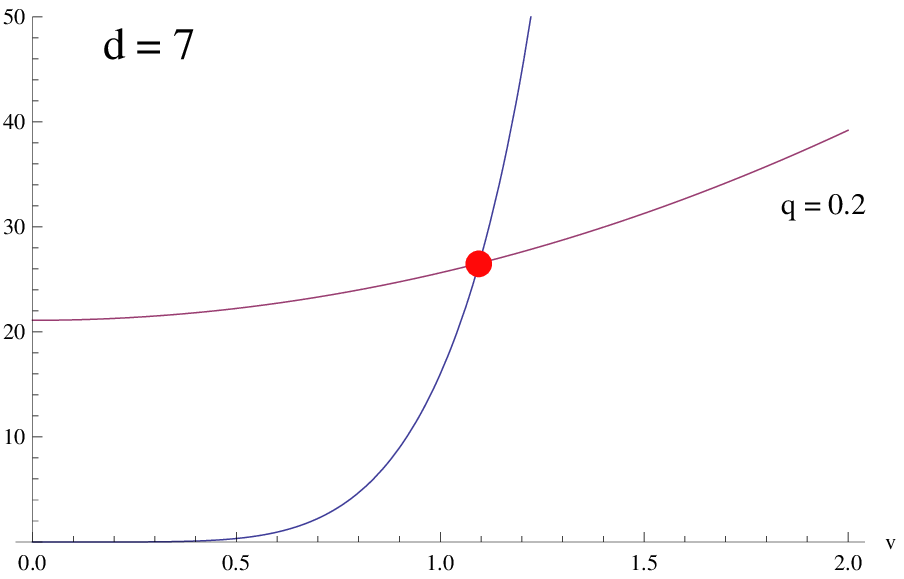} &
\includegraphics[width=0.48\textwidth]{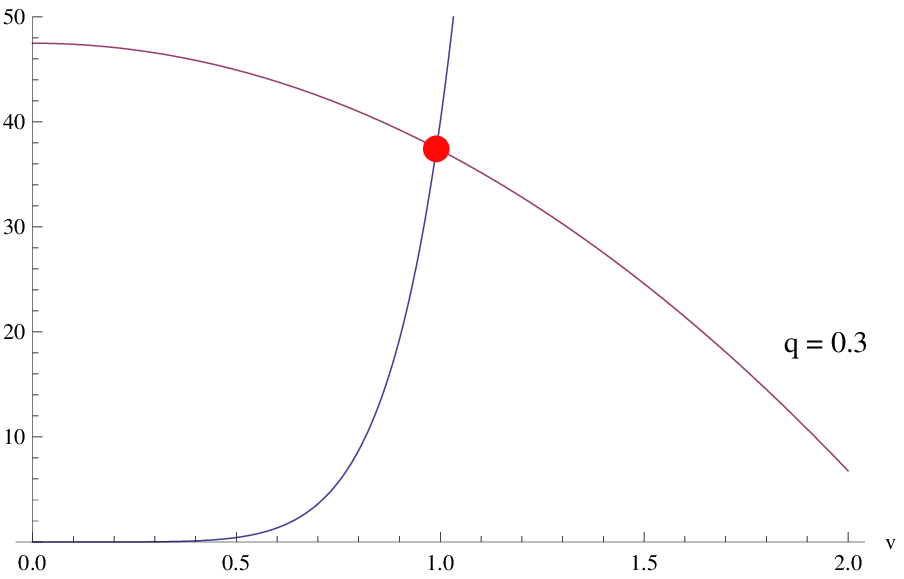}
\end{tabular}
\caption{$w_1$ and $w_2$ for $\kappa=1$ (left column) and $\kappa=-1$ (right column). Only in the case of $d=6$, $\kappa=1$ is there more than one intersection of $w_1$ and $w_2$, allowing for more than one critical point. Note that all the displayed critical points in the $\kappa=-1$ case have negative entropy. This is not necessarily the case in $d>5$ dimensions for sufficiently large $q$.
}
\label{Fig:GBwd567}
\end{figure*}

Remarkably, one can show that the existence of a triple point in $d=6$ is an exceptional case that has  no counterpart in higher dimensions.  
In fact, in all $d>6$ dimensions one observes at most one critical point, accompanied in the $\kappa=+1$ case by  standard VdW behavior and in the $\kappa=-1$ case by the ideal gas or potentially VdW-like with $p_c>p_{max}$ behavior (the reverse VdW may also be present but it is unphysical as it involves black holes with negative entropy.) 

To prove that no triple point can exist, let us return back to the equation \eqref{crit2}, which we equivalently
write as $w_1=w_2$, where 
\bea
w_1&\equiv&(d-3)v_c^{2d-4}-12\kappa v_c^{2d-6}+12(d-5)v_c^{2d-8}\,,\nonumber\\
w_2&\equiv&4\pi q^2\left[\kappa(2d-5)v_c^2+6(2d-7)\right]\,,
\eea
with $w_2$ describing the charge-dependent part. This is illustrated in Fig.~\ref{Fig:GBwd567} 
for $d=5,6,7$ and $\kappa=\pm 1$; we clearly see that only in the second ($d=6$, $\kappa=+1$) case more than one critical point is possible. 

More generally in any $d$, since   $g$  is monotonic, 
 the existence of multiple intersections of $w_1$ and $w_2$ requires that $w_1$ oscillates over a certain 
 range of  sufficiently small $v_c$.   The existence of  maxima/minima of $w_1$ requires that 
\be
\dfrac{\partial w_1}{\partial v_c}=0\quad \Rightarrow\quad 
0=(d-2)(d-3)v_c^4-12\kappa(d-3)v_c^2+12(d-4)(d-5)\,.
\label{GBfmaxmin}
\ee
This is a quadratic equation in $v_c^2$ whose discriminant is independent of $\kappa=\pm1$ and reads
\be
\Delta_{GB}=-48(d-3)(d^3-11d^2+35d-31)\,.
\label{GBdisc}
\ee
It is positive for $d=5, 6$ and negative for $d>6$. Therefore we have shown that  
in all dimensions $d>6$ there can only be at most one critical point. In fact, this critical point always exists and is characterized 
by positive $(p_c, v_c, t_c)$.

However, as noted in \cite{Cai:2013qga}, for $\kappa=+1$ and any fixed $d\geq 6$, as $q$ is lowered the corresponding critical pressure increases and for $q<q_m$ it exceeds the maximal pressure.  We therefore have a limit on how small the charge $q$ of the black hole in a given 
dimension $d$ can be in order that criticality (and hence  VdW behavior) exists. The situation is illustrated in Fig.~\ref{Fig:qmin}.  For $\kappa=-1$ we always have $p_c>p_{max}$.

\begin{figure}
\begin{center}
\rotatebox{-90}{
\includegraphics[width=0.34\textwidth,height=0.31\textheight]{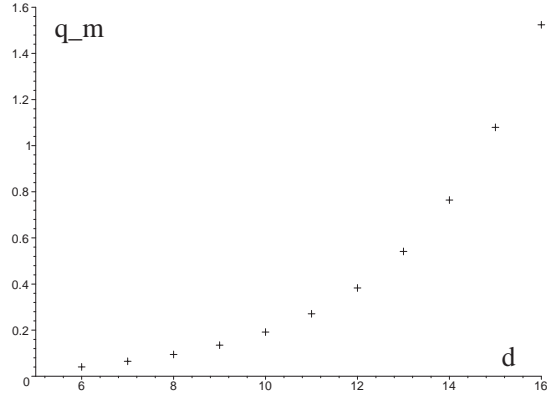}
}
\caption{{\bf Minimum $q$, $\kappa=+1$.} For $d\geq 6$ there exists a minimal $q_m$ such that for $q<q_m$ the critical pressure $p_c$ of the critical point exceeds the maximal pressure $p_{max}$, resulting in an unphysical critical point. This is displayed for $d=6,\dots, 16$.
}  
\label{Fig:qmin}
\end{center}
\end{figure}

\section{$P-v$ criticality in 3rd-order Lovelock gravity}
We now consider thermodynamic phenomena of $U(1)$ charged black holes in 3rd-order Lovelock gravity. 
After some general considerations we concentrate on $d=7$ and $d=8$, the lowest two dimensions for which 3rd-order Lovelock theory brings new qualitative features.  

\subsection{Maximal pressure and other conditions}
We have seen in the Gauss--Bonnet gravity that the requirement for the existence of an asymptotic `AdS region' imposes an important bound on how high the cosmological pressure can be in order the solution not to collapse into a compact region. Here we discuss an analogous condition for the 3rd-order Lovelock.  

In the Lovelock case we have the following cubic equation for the metric function $f$:
\bea\label{cubic}
&&\alpha_3 \frac{(\kappa-f)^3}{r^6}+\alpha_2 \frac{(\kappa-f)^2}{r^4}+\frac{(\kappa-f)}{r^2}+A_0(r)=0\,,\nonumber\\
&&A_0(r) = \alpha_0 
-\frac{16\pi M}{(d-2)\Sigma_{d-2}^{(\kappa)}{r^{d-1}}}+\frac{8\pi Q^2}{(d-2)(d-3)r^{2d-4}}\,,
\eea
whose solution is  
\be
f = \kappa + \frac{r^2}{\sqrt{\alpha_3}} X\,,
\ee
where  $X$ is a solution to the  equation $X^3 - \frac{\alpha_2}{\sqrt{\alpha_3}} X^2 + X - \sqrt{\alpha_3}A_0(r) = 0$.
We label the resulting three solutions $f_\textsc{e}, f_\textsc{gb}$ and $f_\textsc{l}$. Upon successively taking
the limit  $\alpha_3\to 0$ and then  $\alpha_2\to 0$,  the solution $f_\textsc{e}$ approaches the Einstein branch $f_-$ studied in the previous section, $f_\textsc{gb}$ approaches the Gauss--Bonnet branch $f_+$, and $f_\textsc{l}$ represents a new branch which does not have a smooth limit when $\alpha_3\to 0$.  
 We call the corresponding branches of black holes the {\em Einstein branch} ($f_\textsc{e}$), the {\em Gauss--Bonnet branch} ($f_\textsc{gb}$), and the {\em Lovelock branch} ($f_\textsc{l}$). 

When the following condition is satisfied,
\be\label{cond}
4 \alpha_3-\alpha_2^2+27 \alpha_0^2\alpha_3^2-18\alpha_0\alpha_2\alpha_3+4\alpha_0 \alpha_2^3\leq 0\,.
\ee
all three branches admit an 
asymptotic `AdS region'. 
Equality represents a quadratic equation for $\alpha_0$, whose solution is
\be
p_\pm=\frac{(d-1)(d-2)}{108\pi}\Bigl[9\alpha-2\alpha^3\pm 2(\alpha^2-3)^{3/2}\Bigr]\,,
\ee
in terms of the following dimensionless parameters: 
\be\label{alphaDef}
{\alpha}=\frac{\alpha_2}{\sqrt{\alpha_3}}\,,\quad p=4\sqrt{\alpha_3}P=\frac{\alpha_0(d-1)(d-2)\sqrt{\alpha_3}}{4\pi}\,.
\ee

We may now define two parameter regions according to the asymptotics of solutions of
the various branches.\footnote{Note that whereas the asymptotic structure is independent of
the values of the charge and mass, the existence of horizons is affected by it.}    
{\em Region I } occurs for values of ${\alpha}$ and $p$ outside the envelope of the region bounded by the
$p_\pm$ curves. In this region only the Lovelock branch has proper AdS asymptotics whereas the Gauss--Bonnet and Einstein branches either do not exist as real solutions or represent a compact space. 
{\em Region II} is inside the region bounded by the $p_\pm$ curves. In this region all three branches admit proper AdS asymptotics.
The situation is displayed in Fig.~\ref{LovelockAsympt}.

Regions I and II were defined according to the asymptotic structure. Further restriction ensue when horizons are taken into consideration. Consider for simplicity the $q=0$ case. When $\kappa=-1$ it can be shown that region 
I always admits Lovelock black holes at least in a certain range of parameters, whereas region II admits all three kinds of black holes.  The situation is more restrictive when $\kappa=+1$.  Region II then admits only Einstein type black holes whereas region I splits into a {\em region Ia} and {\em region Ib}.  In region Ia, 
defined by (${\alpha}<\sqrt{3}$) or ($\sqrt{3}<{\alpha}<2$ and $p<p_-$), Lovelock black holes   exist in a certain parameter range.
In  {\em region Ib}, defined by (${\alpha}>\sqrt{3}$ and $p>p_+$),  no black holes can exist. This is illustrated in Fig.~\ref{LovelockBHAsympt}.

\begin{figure}
\begin{center}
\rotatebox{-90}{
\includegraphics[width=0.34\textwidth,height=0.31\textheight]{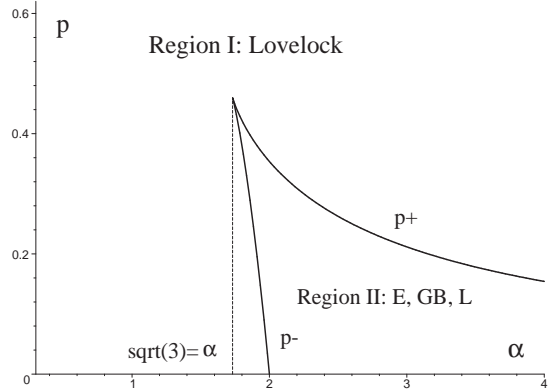}
}
\caption{{\bf Branches with `AdS' asymptotics.} According to the asymptotics of solutions
of the various branches the ${\alpha}-p$ space splits into 2 parameter regions. In region I only the Lovelock branch 
admits AdS asymptotics. In region II all three branches admit the correct asymptotics. 
We have set $d=7$, for other dimensions the behavior is qualitatively similar ($p_\pm$ are appropriately scaled by $d$-dependent factor). 
}  
\label{LovelockAsympt}
\end{center}
\end{figure} 

\begin{figure*}
\centering
\begin{tabular}{cc}
\rotatebox{-90}{
\includegraphics[width=0.34\textwidth,height=0.28\textheight]{Figures/AAdSregions.eps}} &
\rotatebox{-90}{
\includegraphics[width=0.34\textwidth,height=0.28\textheight]{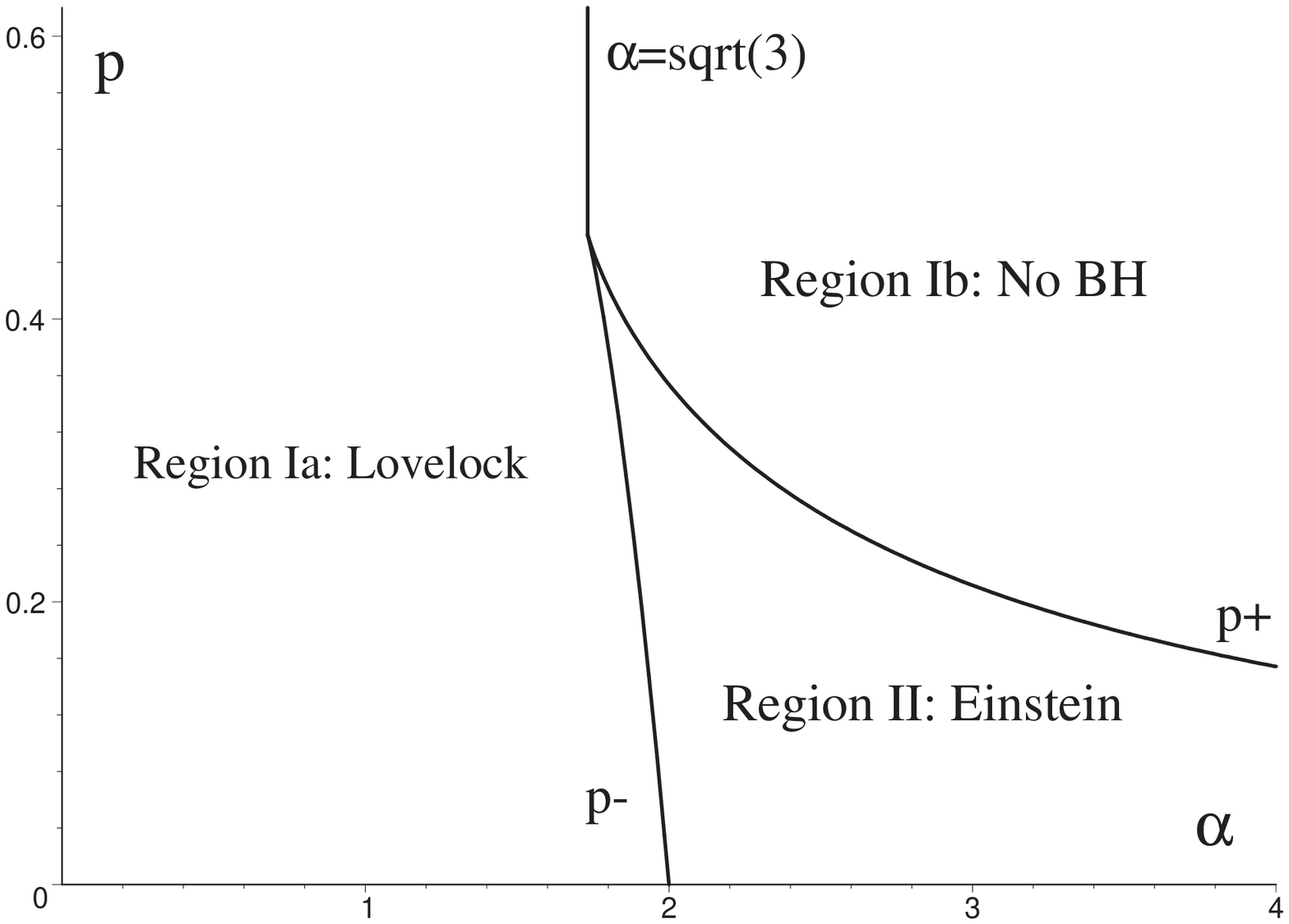}}\\
\end{tabular}
\caption{{\bf Possible uncharged black holes.}
{\em Left: $\kappa=-1$ case}. The Lovelock black holes may exist in region I for a certain range of parameters; all 3 kinds may exist in region II. {\em Right: $\kappa=+1$ case.} Region I now splits into 2 regions: the Lovelock region Ia and the no black hole region Ib. In region II only Einstein branch possesses horizons and can represent a black hole.  We have displayed $d=7$ case, for $d=8$ the situation is qualitatively similar. 
}
\label{LovelockBHAsympt}
\end{figure*}

\subsection{Equation of state}
In terms of \eqref{alphaDef} and the following dimensionless quantities:
\be\label{dimLov}
r_+=v\,\alpha_{3}^{\frac{1}{4}}\,,\quad  T=\frac{t\alpha_{3}^{-\frac{1}{4}}}{d-2}\,,\quad
m=\frac{16\pi M}{(d-2)\Sigma_{d-2}^{(\kappa)}\alpha_3^{\frac{d-3}{4}}}\,,\quad 
Q= \frac{q}{\sqrt{2}} \alpha_{3}^{\frac{d-3}{4}}\,,
\ee
the equation of state \eqref{state} for 3rd-order Lovelock $U(1)$ charged black holes now reduces to ($\kappa=\pm 1$) 
\bea\label{EQSTATELOV}
p&=&\frac{t}{v}-\frac{(d-2)(d-3)\kappa}{4\pi v^2}+
\frac{2{\alpha}\kappa t}{v^3}-\frac{(d-2)(d-5){\alpha}}{4\pi v^4}+\frac{3t}{v^5}\nonumber\\
&&-\frac{(d-2)(d-7)\kappa}{4\pi v^6}+\frac{q^2}{v^{2(d-2)}}\,. 
\eea
Similar to the Gauss--Bonnet case we shall also investigate a 
dimensionless  counterpart of the Gibbs free energy,  
\be
g=\frac{1}{\Sigma_{d-2}^{(\kappa)}} \alpha_3^{\frac{3-d}{4}} G=g(t,p,q, \alpha)\,,
\ee
which now reads
\bea\label{gLovelock}
g&=&-\frac{1}{16\pi(3+2\alpha\kappa v^2+v^4)}\Bigl[\frac{4\pi p v^{d+3}}{(d-1)(d-2)}
-\kappa v^{d+1}+\frac{24\pi\kappa p \alpha v^{d+1}}{(d-1)(d-4)}\nonumber\\
&&-\frac{\alpha v^{d-1}(d-8)}{d-4}+\frac{60\pi p v^{d-1}}{(d-1)(d-6)}-\frac{2\kappa \alpha^2 v^{d-3}(d-2)}{d-4}
+\frac{4\kappa v^{d-3}(d+3)}{d-6}\nonumber\\
&&-\frac{3\alpha v^{d-5}(d-2)(d-8)}{(d-4)(d-6)}-\frac{3\kappa v^{d-7}(d-2)}{d-6}\Bigr]\\
&&+\frac{q^2}{4(3+2\alpha\kappa v^2+v^4)(d-3)v^{d-3}}
\Bigl[\frac{v^4(2d-5)}{d-2}+\frac{2\alpha\kappa(2d-7)v^2}{d-4}+\frac{3(2d-9)}{d-6}\Bigr]\,.\quad\nonumber
\eea
The thermodynamic state corresponds to the global minimum of this quantity for its fixed parameters $t, p, q$ and $\alpha$.

Analyzing its critical points, the analogues of Eqs.~\eqref{crit1} and \eqref{crit2} now read
\be\label{crit3}
t_c=\frac{(d-2)}{2\pi v_c(v_c^4+6\alpha\kappa v_c^2+15)}\Bigl[3\kappa(d-7)+2\alpha(d-5)v_c^2+(d-3)\kappa v_c^4-\frac{4\pi q^2}{v_c^{2(d-5)}}\Bigr]\,,
\ee
and 
\bea\label{crit4}
0&=&\left( d-3 \right) v_c^{2d-2}-12\,\alpha\,\kappa\,v_c^{2d-4}+6v_c^{2d-6} \bigl( 2{\alpha}^{2} \left( d-5 \bigr) +5-5d
 \right) \nonumber\\
&&+12\,\kappa\,\alpha\, \left( 2d-19 \right) v_c^{2\,d-8}+45\left( d-7 \right) v_c^{2\,d-10}\nonumber\\
&&-4\pi\kappa{q}^{2} \bigl[ 
 \left( 2\,d-5 \right) v_c^{4}+6\kappa\alpha\left( 2d-7
 \right) v_c^{2}+30d-135 \bigr] \,.
\eea

The thermodynamic singularity for the $\kappa=-1$ 3rd-order Lovelock black holes occurs when $v^4-2\alpha v^2+3=0$, i.e., for 
\be
v=v_{s\pm}=\sqrt{\alpha\pm \sqrt{\alpha^2-3}}\,.
\ee
Similar to the Gauss--Bonnet case, one can find the corresponding singular points. However, for general $\alpha$ and non-zero $q$ the resultant expressions for $p_{s\pm}$ and  $t_{s\pm}$ are not very illuminating. A particularly interesting case occurs when $\alpha=\sqrt{3}$ (for which the
two thermodynamic singularities `coincide') and $q=0$. In this case we find 
\be\label{singularLovelock}
v_s=3^{1/4}\,,\quad t_s=\frac{d-2}{2\pi}3^{-1/4}\,,\quad p_s=\frac{(d-1)(d-2)}{36\pi}\sqrt{3}\,.
\ee  
It is easy to check that the corresponding black hole has zero mass $M=0$; such ``massless" black holes can occur for hyperbolic geometries with appropriate identifications \cite{Smith:1997wx,Mann:1996gj,Mann:1997jb}. This very special case shall be discussed in Sec.~\ref{Sec:alpha3}.

The positivity of entropy requires
\be
(d-4)(d-6)v^4-2\alpha(d-2)(d-6)v^2+3(d-2)(d-4)>0\,.
\ee
The corresponding admissible roots are 
\be
v_{1,2}=\sqrt{\frac{d-2}{d-4}\Bigl(\alpha\pm \sqrt{\alpha^2-\frac{3(d-4)^2}{(d-6)(d-2)}}\Bigr)}\,,
\ee
and coincide when
\be\label{alphaD}
\alpha=\alpha_d=\sqrt{\frac{3(d-4)^2}{(d-6)(d-2)}}\, .
\ee
For $\alpha < \alpha_d$ the entropy is always positive.
 The two conditions are displayed in Fig.~\ref{Fig:Vss}. As we can see, for $\alpha<\sqrt{3}$ the black hole entropy is always positive and there are no thermodynamic singularities. However, for $\alpha>\sqrt{3}$, we may have both positive and negative entropy black holes and  thermodynamic singularities may be present. 
\begin{figure}
\begin{center}
\rotatebox{-90}{
\includegraphics[width=0.34\textwidth,height=0.31\textheight]{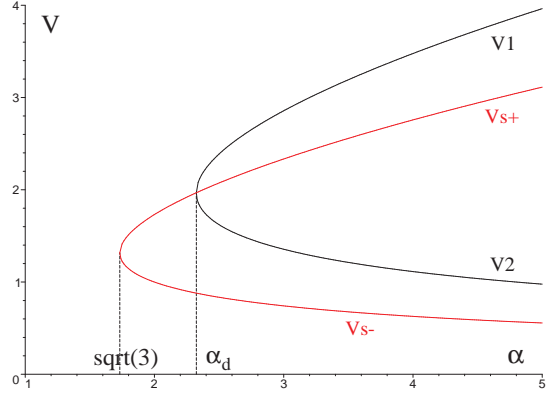}
}
\caption{{\bf Conditions for $\kappa=-1$ black holes.} The thermodynamic singularities occur on a thick red curve. The black hole entropy is positive to the left of the region outlined by thick black curve. We have chosen $d=7$. As $d$ increases, the red curve remains unchanged, whereas the black curve moves closer to it; the two curves coincide in the limit $d\to \infty$.
}  
\label{Fig:Vss}
\end{center}
\end{figure}

In the next two subsections we shall discuss the behavior of this equation in $d=7$ and $d=8$ dimensions, for various 
black hole topologies and various regions of the parameter space $(q,\alpha)$.

\subsection{Seven dimensions}
In seven dimensions we have the following equation of state:
\be
p=\frac{t}{v}-\frac{5\kappa}{\pi v^2}+\frac{2\alpha\kappa t}{v^3}-\frac{5\alpha}{2\pi v^4}
+\frac{3t}{v^5}+\frac{q^2}{v^{10}}\,.\label{p7}
\ee
Eqs.~\eqref{crit3} and \eqref{crit4} reduce to 
\be \label{tc7}
t_c =\frac{10}{\pi(v_c^4+6 \alpha  \kappa  v_c^2+ 15)}\Bigl[\alpha v_c+\kappa v_c^3-\frac{\pi q^2}{v_c^5}\Bigr]\,, 
\ee
and 
\be \label{d2p7}
v_c^{12}-3\alpha\kappa v_c^{10}+3(2\alpha^2-15)v_c^8
-15\alpha\kappa v_c^6-3\pi \kappa q^2 (3v_c^4+14\kappa\alpha v_c^2+25)=0\,.
\ee
The AdS asymptotics of various branches is displayed in Figs.~\ref{LovelockAsympt} and \ref{LovelockBHAsympt}, positive entropy condition as well as thermodynamic singularities are displayed in Fig.~\ref{Fig:Vss}, $\alpha_7$ given by \eqref{alphaD} now reads $\alpha_{7}=3\sqrt{3/5}$.

\subsubsection{Spherical case}
\begin{figure}
\begin{center}
\rotatebox{-90}{
\includegraphics[width=0.34\textwidth,height=0.31\textheight]{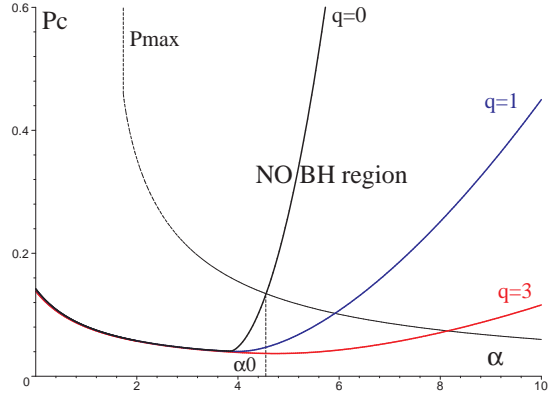}
}
\caption{{\bf Critical pressure: $d=7$ and $\kappa=+1$.} In $d=7$ and the spherical case we observe one critical point and the associated VdW behavior. The corresponding pressure $p_c$ is displayed for $q=0, 1, 3$. We observe that for $\alpha>\alpha_0\approx 4.55$ there exists a minimum charge $q_{min}$ such that for $q<q_{min}$ the critical pressure $p_c$ exceeds the corresponding 
maximal pressure $p_+$.   
}  
\label{Fig:Pc7d}
\end{center}
\end{figure} 

For black holes of spherical topology with charge or not, we confirmed numerically that in the range $\alpha\in (0,10)$, the equation of state admits exactly one critical point, characterized by the standard critical exponents \eqref{exponents}, and the system demonstrates ``classical Van der Waals'' behavior. However, as $\alpha$ increases, the corresponding critical pressure, see Fig.~\ref{Fig:Pc7d}, increases, and eventually exceeds the admissible pressure $p_+$, that is occurs for a compact space solution.  Alternatively, similar to the Gauss--Bonnet case, for each $\alpha>\alpha_0\approx 4.55$ there exists a minimum charge $q_{min}$ such that for $q<q_{min}$ the critical pressure $p_c$ exceeds the corresponding 
maximal pressure set by $p_+$.

\subsubsection{Hyperbolic case: multiple RPT}

The hyperbolic case, $\kappa=-1$, is more interesting.  We shall consider the uncharged $q=0$ case  in detail and then briefly mention what happens when nontrivial charge is added.
When $q=0$ and depending on the parameter $\alpha$ we may have up to two critical points (Fig.~\ref{Lovelock7dcritHU}) and  
get various physical situations as summarized in Table~\ref{tab:cases}. The corresponding $p-v$, $g-t$, and $p-t$ diagrams are displayed 
in Figs.~\ref{Lovelock7dcrit1}--\ref{Lovelock7dcrit5}. 
\begin{figure*}
\centering
\begin{tabular}{cc}
\rotatebox{-90}{
\includegraphics[width=0.34\textwidth,height=0.28\textheight]{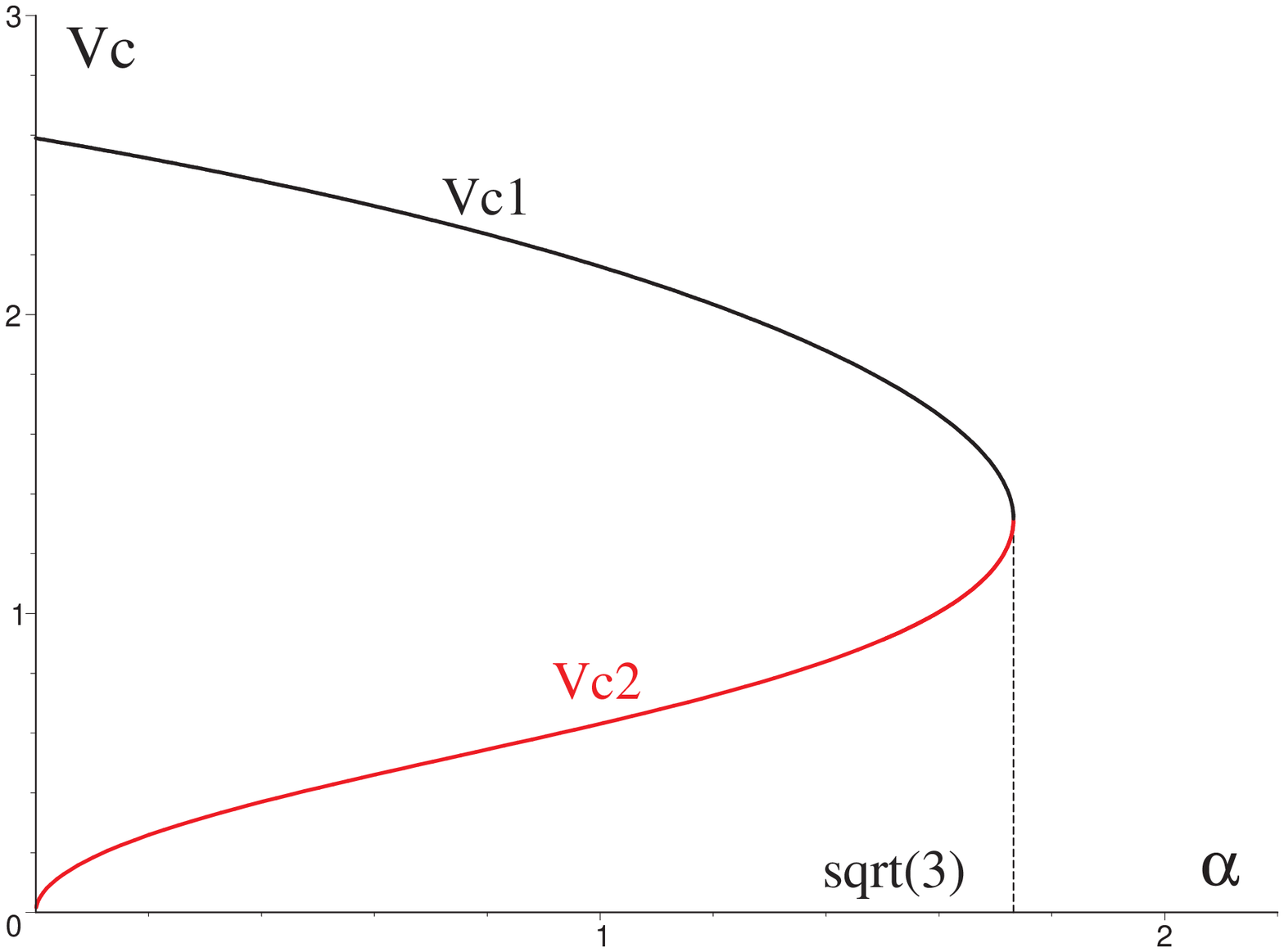}} &
\rotatebox{-90}{
\includegraphics[width=0.34\textwidth,height=0.28\textheight]{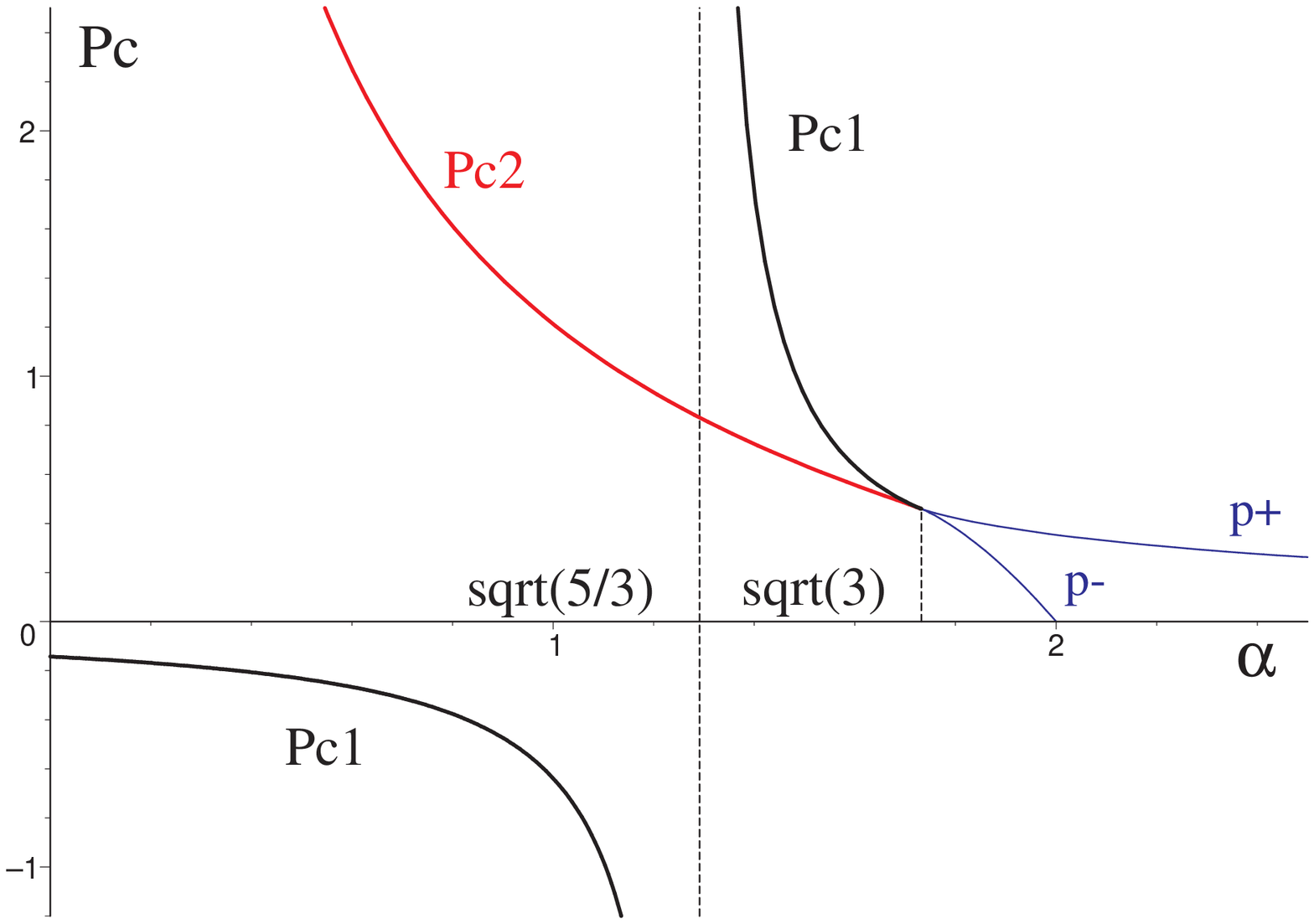}} \\
\end{tabular}
\caption{{\bf Critical points in $d=7$: $\kappa=-1, q=0$ case.}
Depending on the parameter $\alpha$ we may have up to two physical critical points. The special case is $\alpha=\sqrt{3}$ for which these two critical points `coincide' forming a special critical point discussed in the next subsection. 
}\label{Lovelock7dcritHU}
\end{figure*} 
\begin{table}
\begin{centering}
\begin{tabular}{|c|c|c|c|}
\hline
case & range of $\alpha$ & \# critical points & behavior\tabularnewline
\hline
I&$\alpha\in(0,\sqrt{5/3})$ & 1& VdW \tabularnewline
\hline
II&$\alpha\in(\sqrt{5/3},\sqrt{3})$ & 2& VdW \& reverse VdW\tabularnewline
\hline
III&$\alpha=\sqrt{3}$ & 1& special \tabularnewline
\hline
IV&$\alpha\in(\sqrt{3},3\sqrt{3/5})$ & 0& infinite coexistence line \tabularnewline
\hline
V&$\alpha>3\sqrt{3/5}$ & 0& multiple RPT, infinite coexistence line  \tabularnewline
\hline
\end{tabular}
\protect\caption{Types of physical behavior in $d=7, \kappa=-1, q=0$ case.\label{tab:cases}}
\end{centering}
\end{table}

\begin{figure*}
\centering
\begin{tabular}{cc}
\rotatebox{-90}{
\includegraphics[width=0.34\textwidth,height=0.28\textheight]{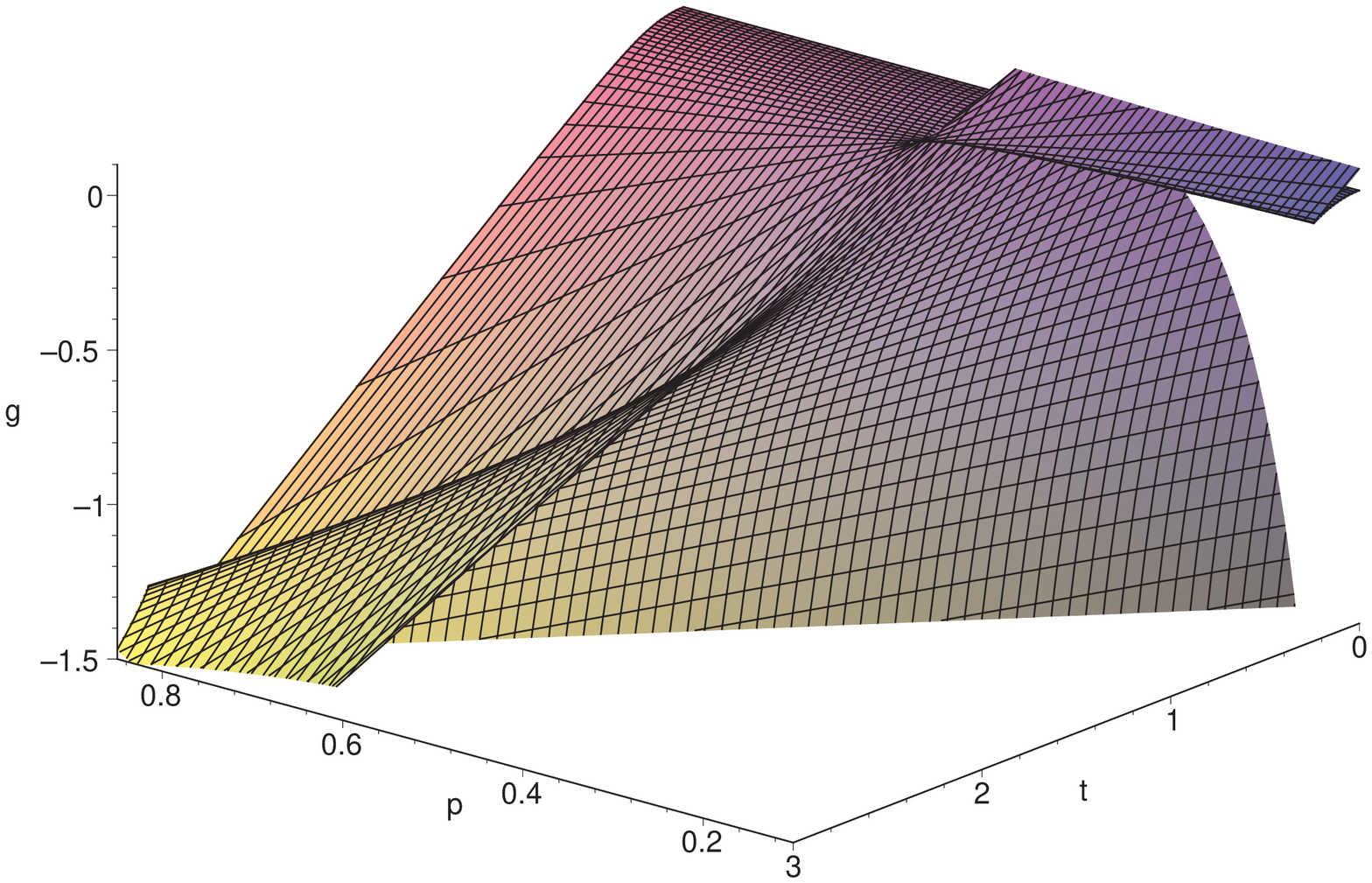}} &
\rotatebox{-90}{
\includegraphics[width=0.34\textwidth,height=0.28\textheight]{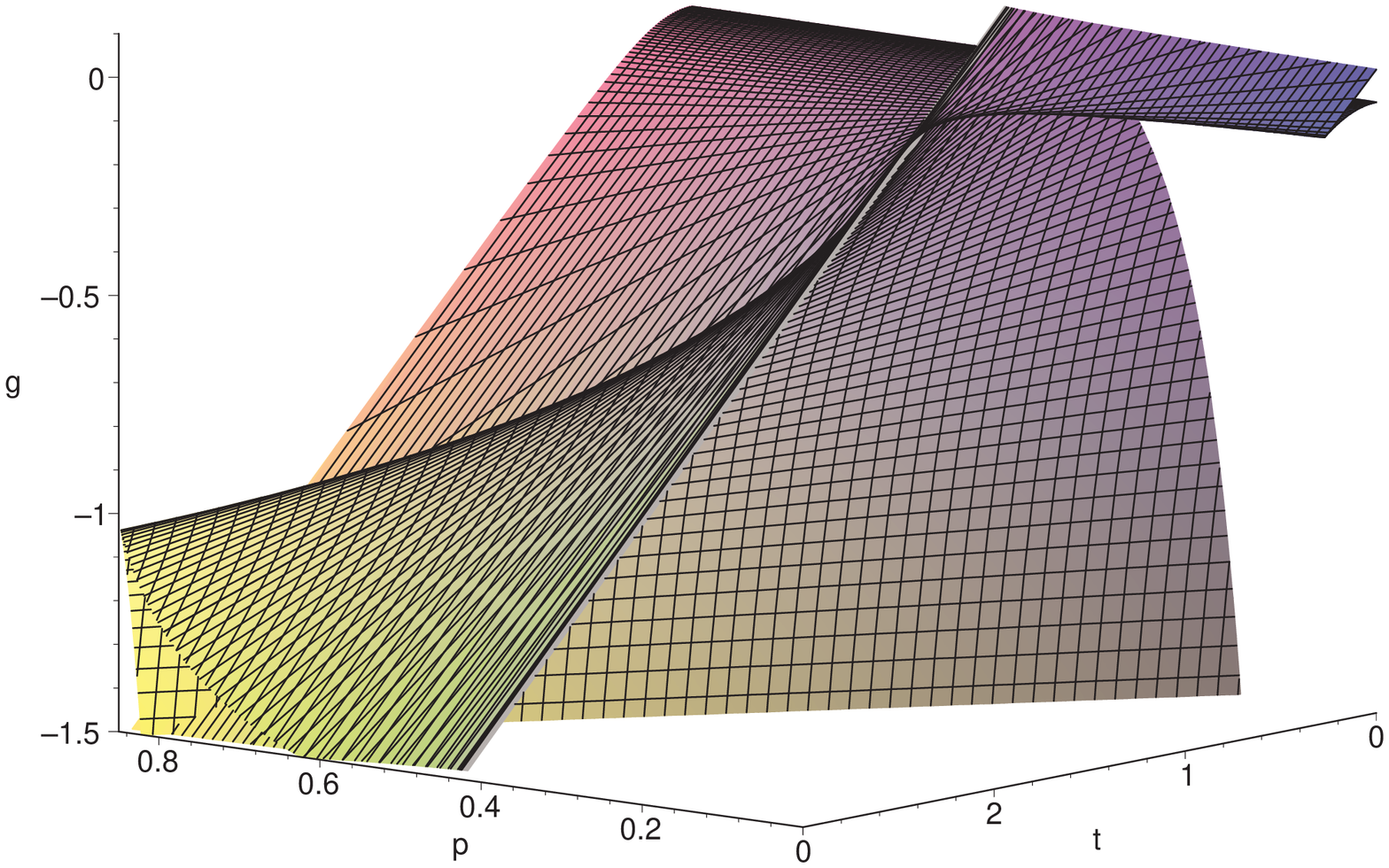}} \\
\rotatebox{-90}{
\includegraphics[width=0.34\textwidth,height=0.28\textheight]{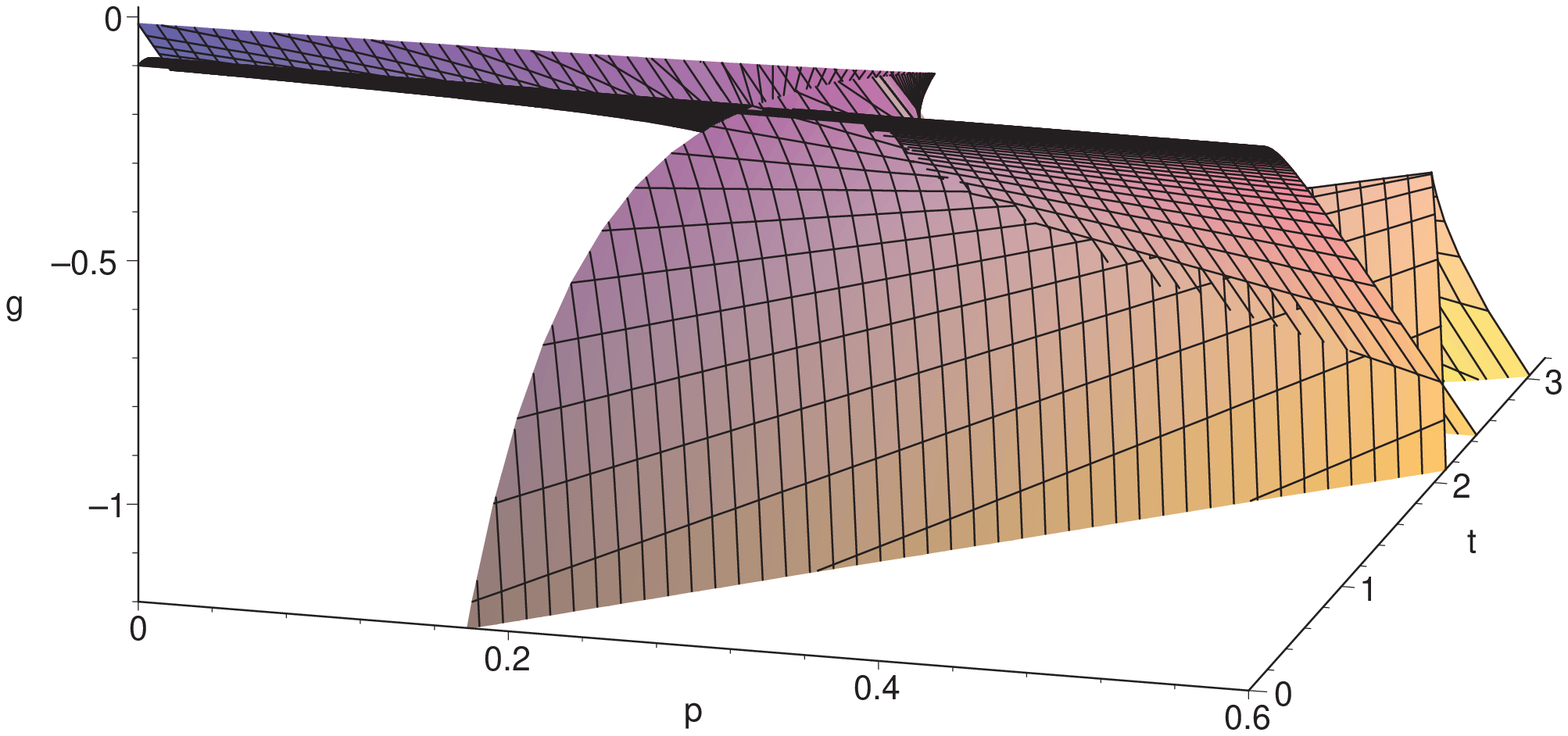}} & 
\rotatebox{-90}{
\includegraphics[width=0.34\textwidth,height=0.28\textheight]{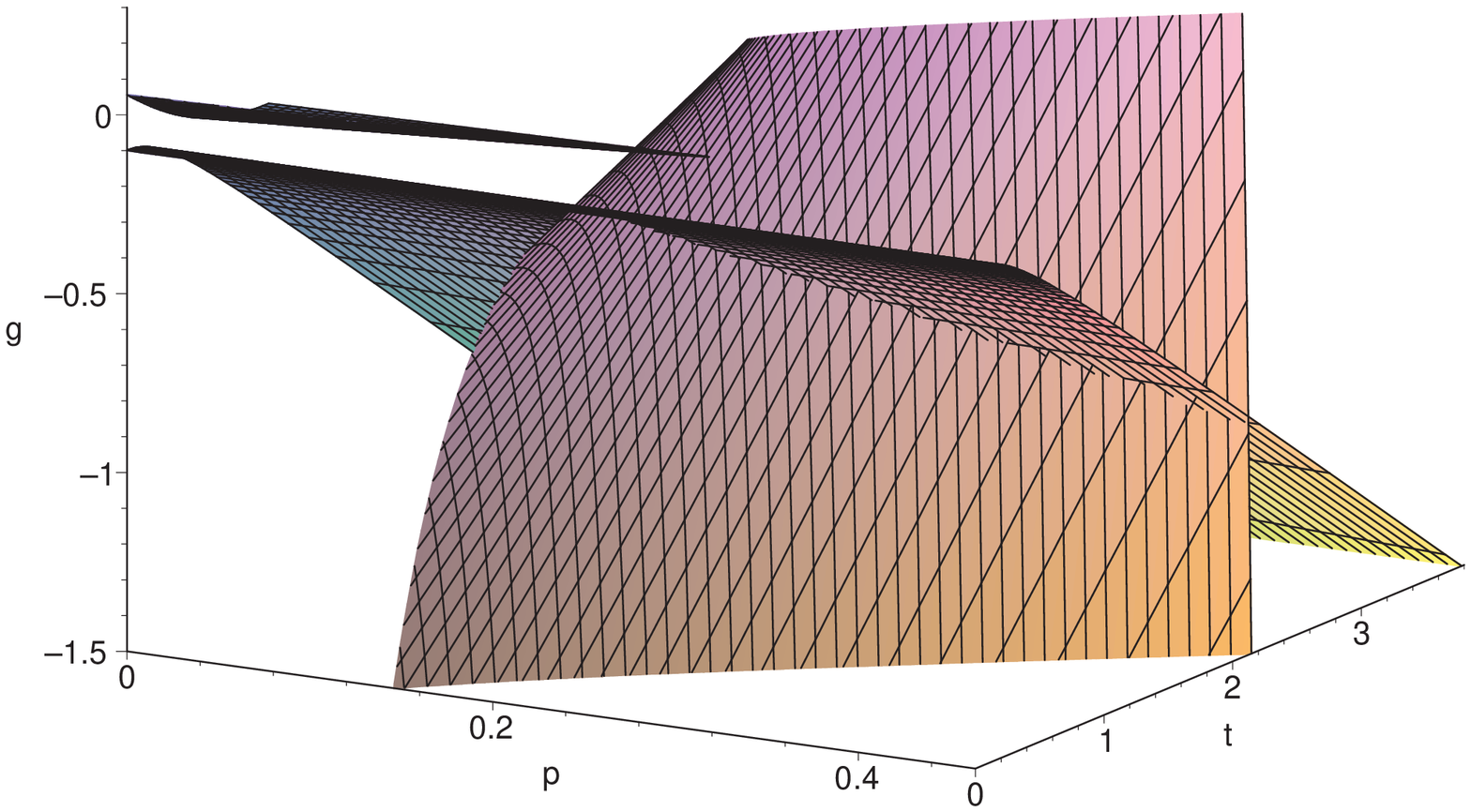}}\\
\end{tabular}
\caption{{\bf Gibbs free energy: uncharged $d=7, \kappa=-1$ case.}
The Gibbs free energy is displayed successively for $\alpha=1.65, \sqrt{3}, 1.85, 2.5$. In the $\alpha=1.65$ case we observe a presence of two swallowtails that never occur at the same pressure. The $\alpha=\sqrt{3}$ is a special case for which the previous swallowtails  emerge   from the same isolated critical point, characterized by non-standard critical exponents. For $\alpha=1.85\in(\sqrt{3},3\sqrt{3/5})$ the behavior of $g$ is quite complicated, however, the global minimum of $g$ corresponds to one possible first-order phase transition. Finally, for $\alpha=2.5>3\sqrt{3/5}$ the presence of negative entropy black holes effectively makes the admissible Gibbs `discontinuous'. Besides the standard first-order phase transition, we can also observe, in a small range of pressures, the `smooth' RPT and/or the zeroth-order phase transition, as displayed in Fig.~20. 
}  
\label{Lovelock7dcrit1}
\end{figure*}

\begin{figure*}
\centering
\begin{tabular}{cc}
\rotatebox{-90}{
\includegraphics[width=0.34\textwidth,height=0.28\textheight]{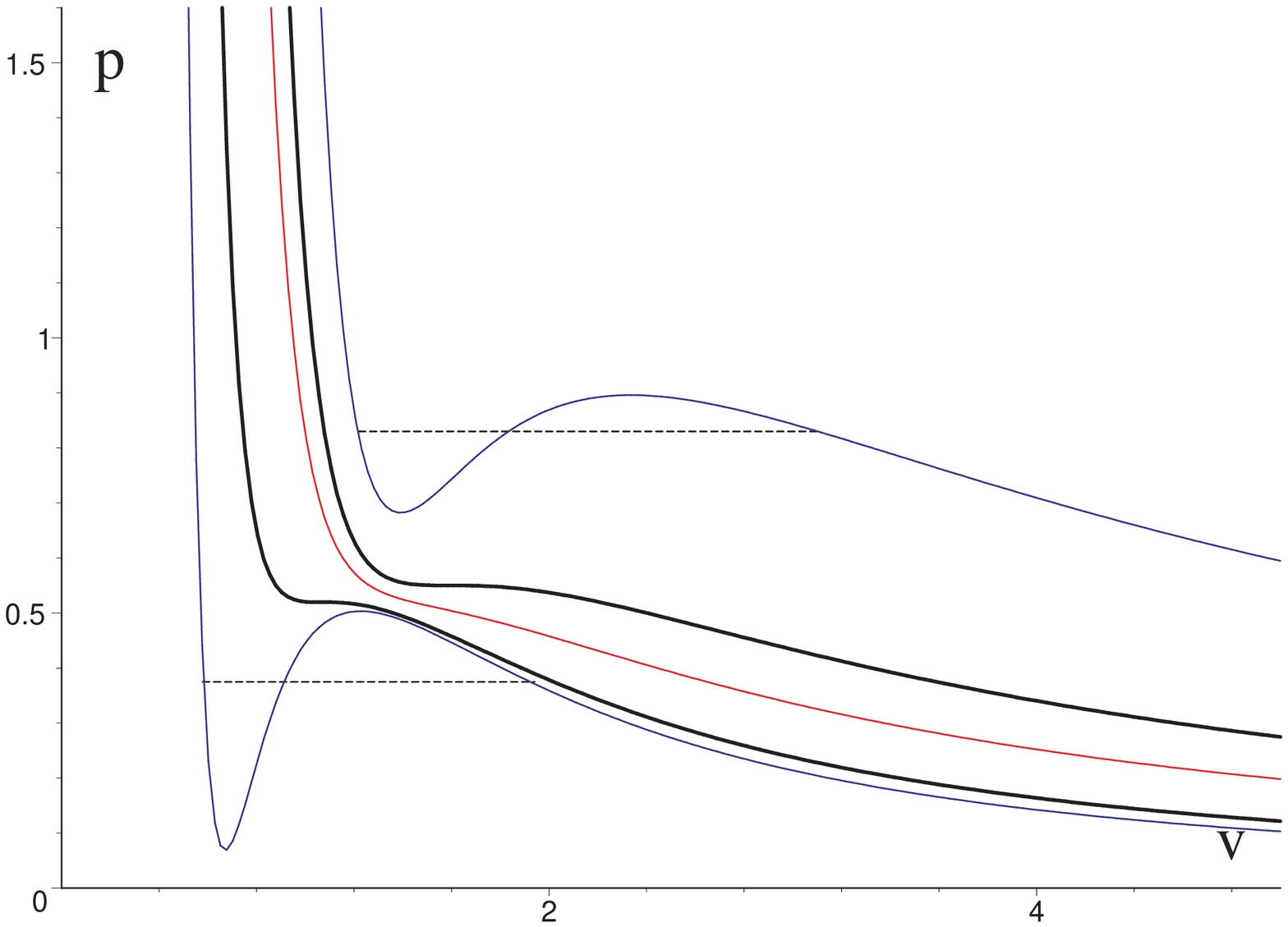}} &
\rotatebox{-90}{
\includegraphics[width=0.34\textwidth,height=0.28\textheight]{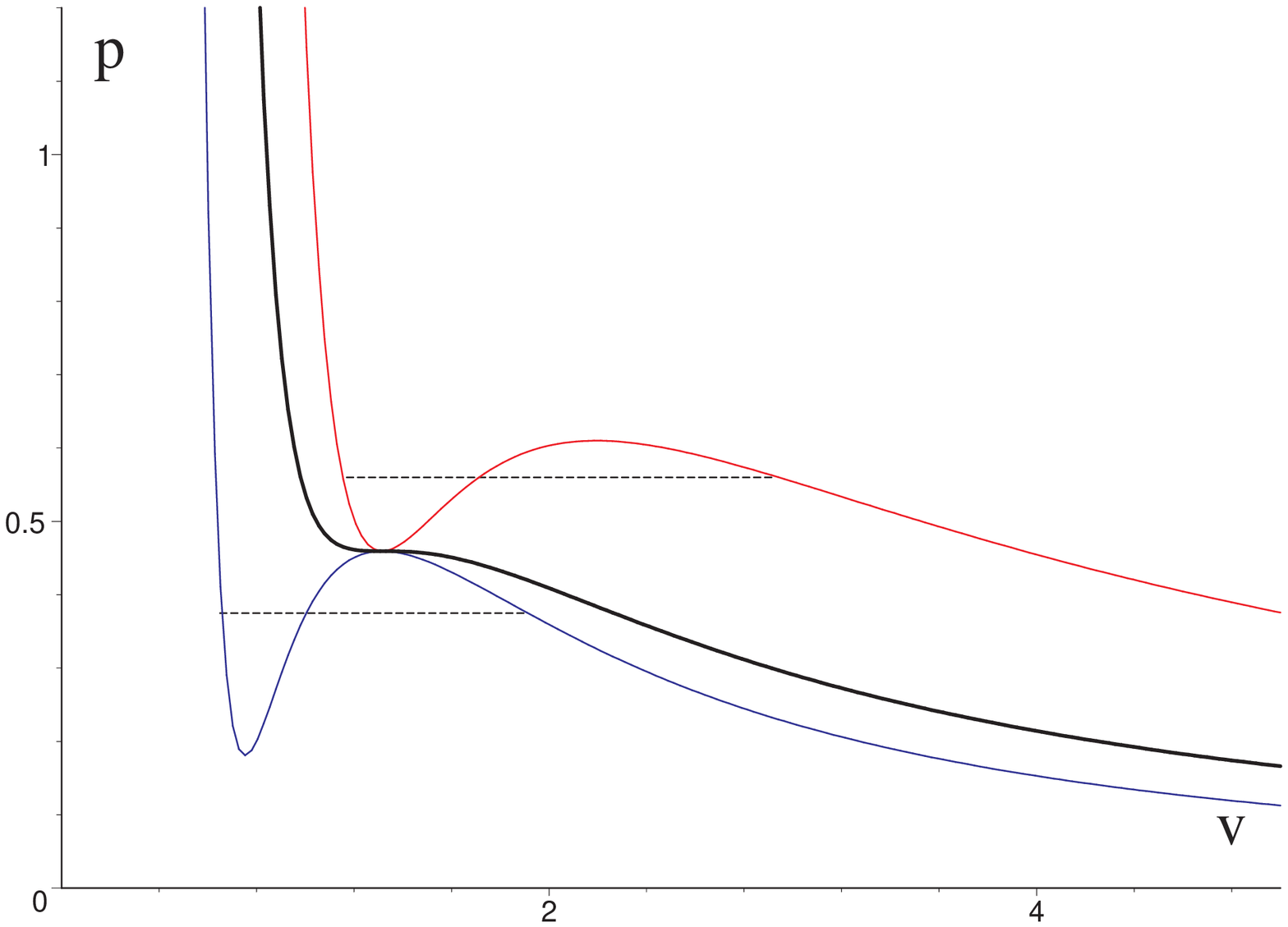}} \\
\rotatebox{-90}{
\includegraphics[width=0.34\textwidth,height=0.28\textheight]{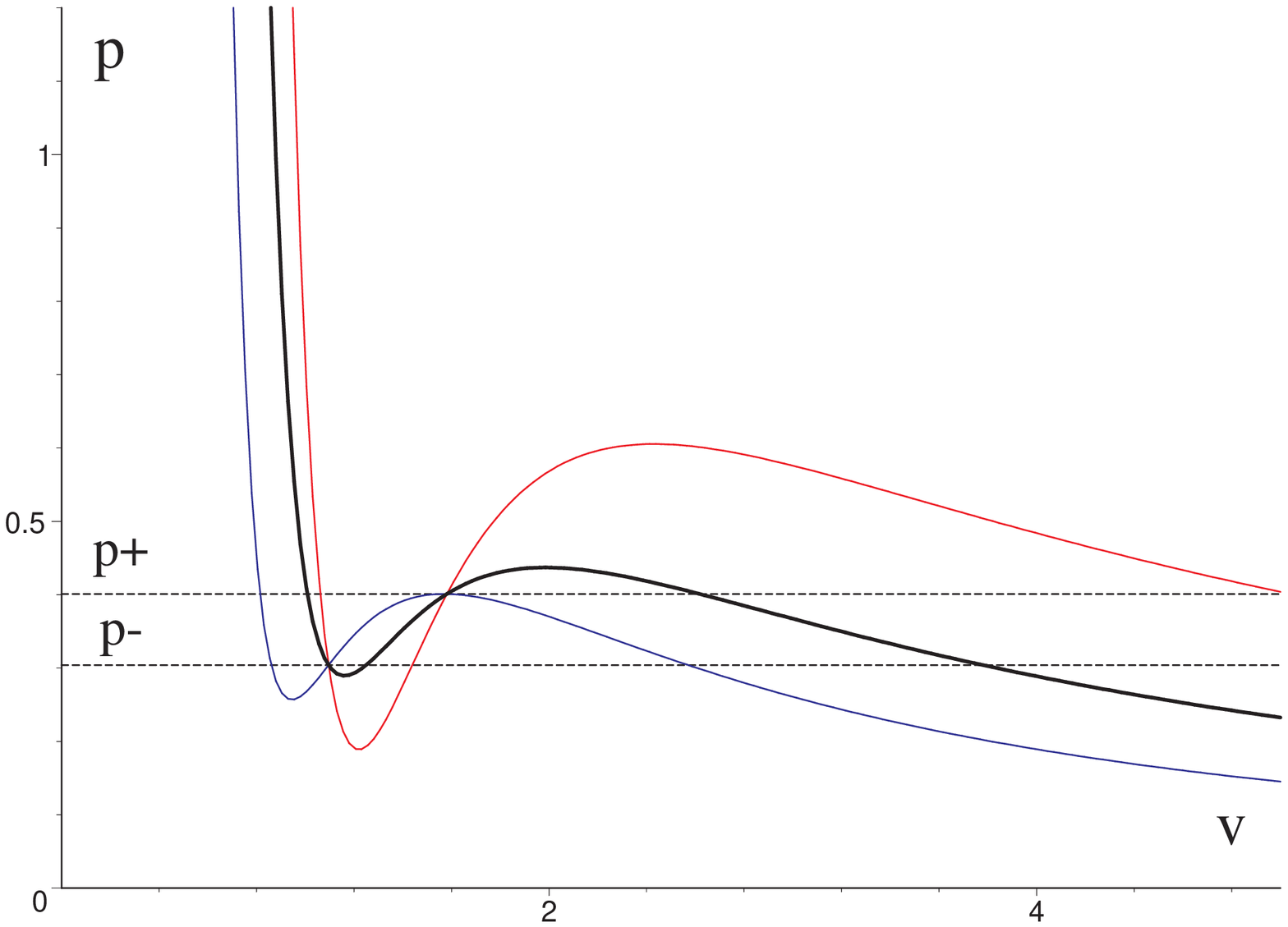}} & 
\rotatebox{-90}{
\includegraphics[width=0.34\textwidth,height=0.28\textheight]{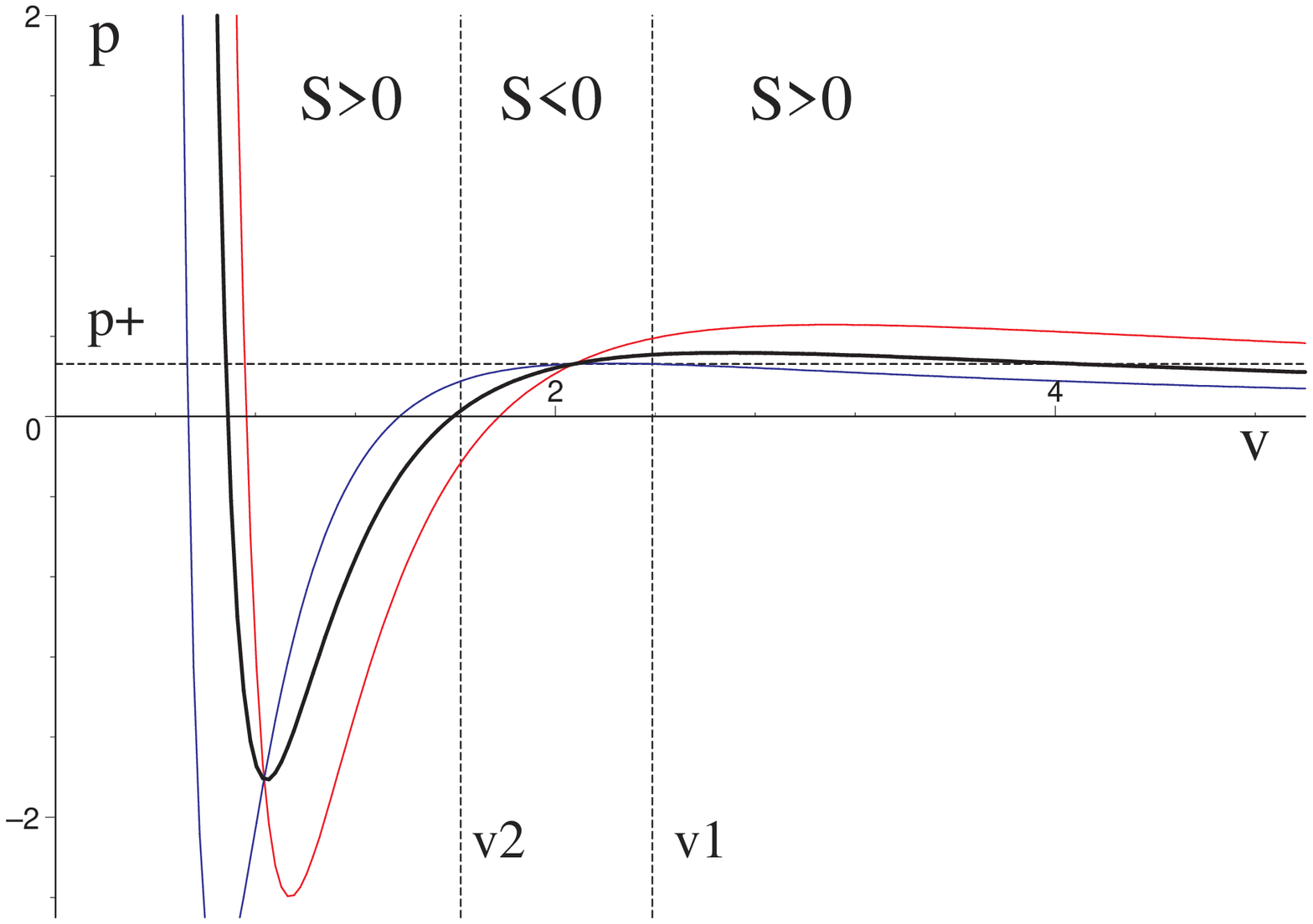}}\\
\end{tabular}
\caption{{\bf $p-v$ diagram: uncharged $d=7, \kappa=-1$ case.}
The $p-v$ diagram is displayed successively for $\alpha=1.65, \sqrt{3}, 1.85, 2.5$. In the $\alpha=1.65$ we observe two critical isotherms displayed by thick black curves, the isotherm with $(t_{c1}+t_{c2})/2$ is displayed by red curve and demonstrates the `ideal gas behavior', whereas the isotherms with $t>t_{c2}$ and $t<t_{c1}$ display oscillations that are replaced according to Maxwell's equal area law.  $\alpha=\sqrt{3}$ is a special case discussed in the next subsection. For $\alpha>\sqrt{3}$ we observe the presence of two thermodynamic singularities. The case $\alpha=2.5>3\sqrt{3/5}$ moreover displays the region of negative entropy black holes, in between $v_2$ and $v_1$. We also display $p_\pm$ in these cases. 
Note that the temperatures `reverse' in between the two thermodynamic singularities---hotter isotherms occur for lower pressures. }  
\label{Lovelock7dcrit2}
\end{figure*}

Namely, when $\alpha\in(0,\sqrt{5/3})$ we observe 1 physical critical point (with positive $p_c, v_c$ and $t_c$) and the associated VdW-like
first order small/large black hole phase transition terminating at a critical point characterized by the swallowtail critical exponents  \eqref{exponents}. Note however that, contrary to the $\kappa=+1$ case, the coexistence line terminates at a finite pressure $p$ as $t\to 0$.    
At $\alpha=\sqrt{5/3}$ an additional physical critical point emerges with `infinite' $t_c$ that becomes finite and positive as $\alpha$ increases; in the range $\alpha\in (\sqrt{5/3},\sqrt{3})$ we observe two critical points and the associated VdW and `reverse VdW' behavior, shown in Figs.~\ref{Lovelock7dcrit2}.

At $\alpha=\sqrt{3}$, the two critical points merge together and a qualitatively new behaviour emerges.  We shall postpone discussion of
this situation until later in this section.

Increasing $\alpha$ even further, in the region $\alpha\in(\sqrt{3}, 3\sqrt{3/5})$ we find thermodynamic singularities.\footnote{Associated with these singularities are the two reconnections of various branches---making the $g-t$ diagram quite complicated.} However, these do not exist in the branches globally minimizing the Gibbs free energy.  We therefore regard solutions in this range of $\alpha$ as having sensible thermodynamic behaviour, and we observe a first-order phase transition as displayed in Fig.~\ref{Lovelock7dcrit3} d.

Similar behavior persists even for $\alpha>3\sqrt{3/5}$. However in such a region some of the black holes may have negative entropy and hence are unphysical. Discarding such black holes, the Gibbs free energy is no longer continuous. Moreover, the hypersurface of large black holes displays interesting curved shape, see Fig.~\ref{Lovelock7dcrit4}, leading to a {\em multiple reentrant phase transition}. To understand this phenomenon, let us look more closely at the 2d $g-t$ diagram displayed on RHS of Fig.~\ref{Lovelock7dcrit4}. We observe that for 
$p\in(p_1\approx 0.23209,p_2\approx 0.23311)$ the branch of small black holes (displayed by a thick black curve that remains almost identical for various pressures) crosses twice the branches of large black holes---displayed by dashed colored curves---that moreover terminate at a finite temperature. This indicates that there will be two first-order phase transitions, possibly accompanied, for $p\in (p_0\approx 0.21809, p_2)$ and $t\in(0,t_z\approx 0.16864)$, by a {\em zeroth-order phase transition}, e.g. \cite{Altamirano:2013ane}. Consider the constant pressure $p'=1.002\times p_1\in(p_1, p_2)$ curves displayed in the diagram by thick black and dashed black lines. As the temperature decreases from say $t=0.4$, the system follows the lower dashed black curve being a large black hole, until at $t_3\approx 0.308$ the two branches cross and the system undergoes a first-order phase transition to a small black hole. As $t$ decreases even further the global minimum of $g$ corresponds to the small black hole on a thick black curve until at $t_2\approx 0.20$ another first-order phase transition, this time to a large black hole, occurs. Then the system follows the dashed black curve as a large black hole until this terminates at $t_1\approx 0.162$. If the temperature is decreased even further the system jumps to the thick black curve, undergoing the zeroth-order phase transition and becoming a small black hole again.  In summary, we observe reentrant large/small/large/small black hole phase transition. 
The corresponding $p-t$ phase diagram is displayed in Fig.~\ref{Lovelock7dcrit5}.

\begin{figure*}
\centering
\begin{tabular}{cc}
\rotatebox{-90}{
\includegraphics[width=0.34\textwidth,height=0.28\textheight]{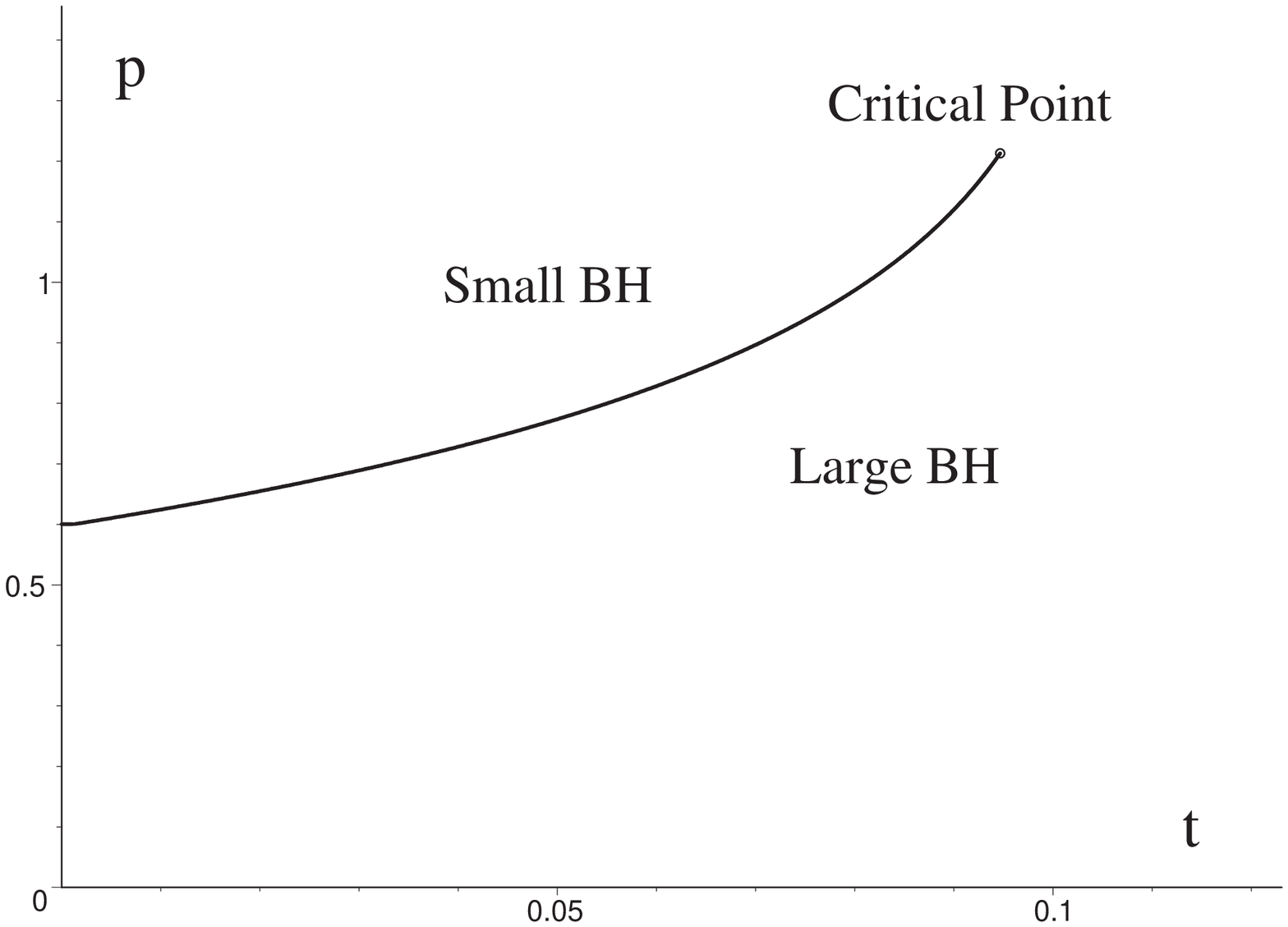}} &
\rotatebox{-90}{
\includegraphics[width=0.34\textwidth,height=0.28\textheight]{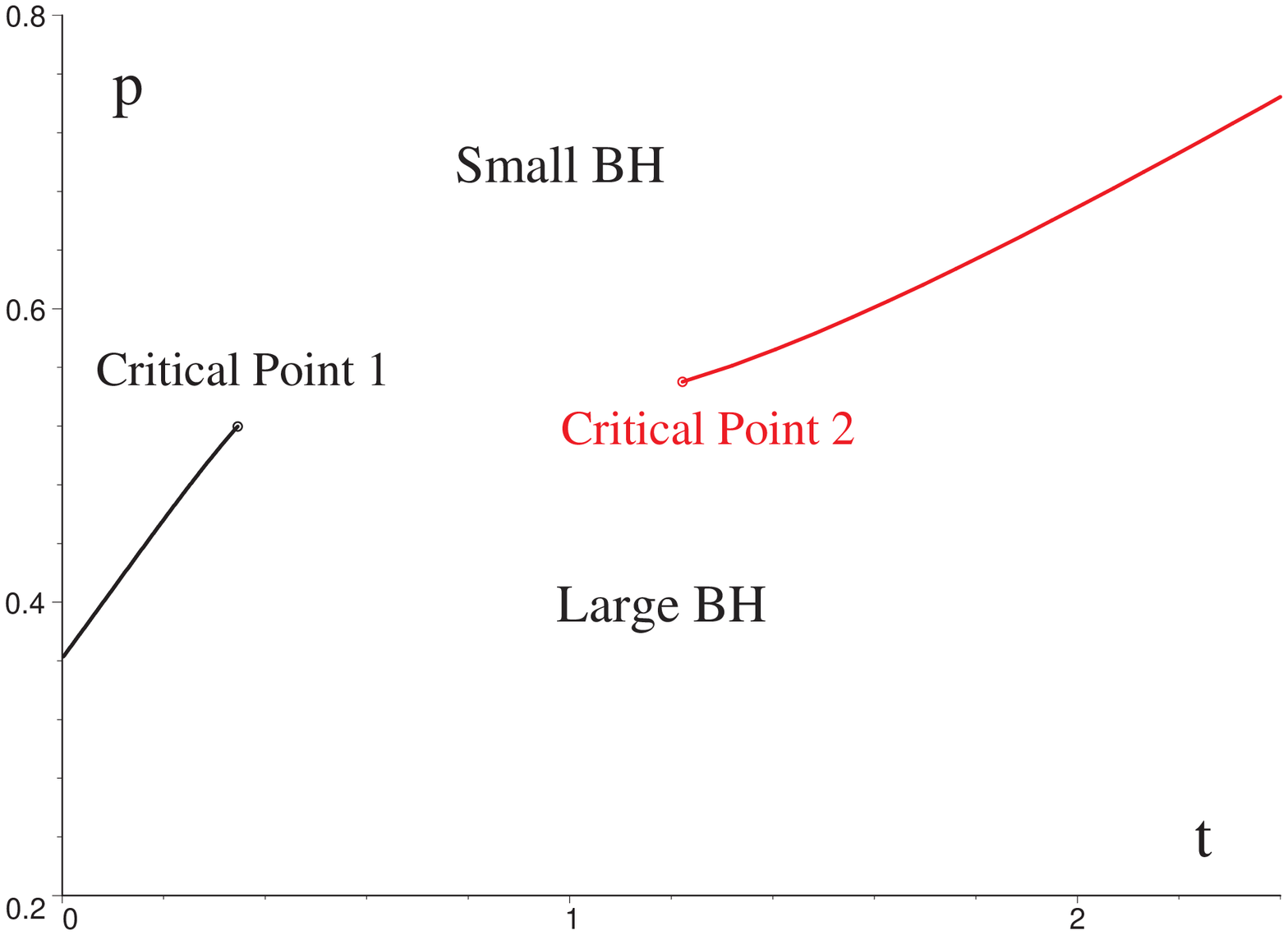}} \\
\rotatebox{-90}{
\includegraphics[width=0.34\textwidth,height=0.28\textheight]{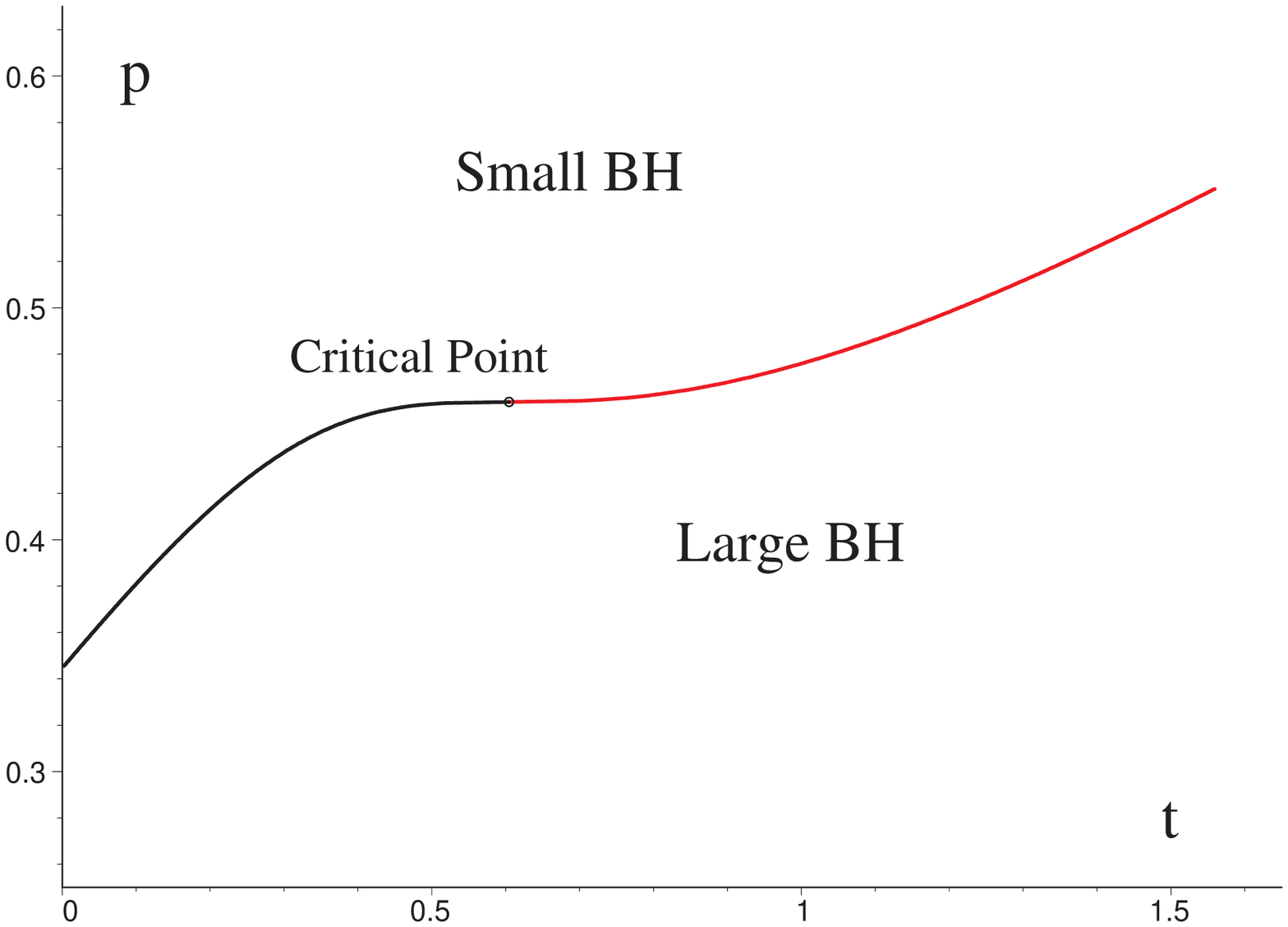}} & 
\rotatebox{-90}{
\includegraphics[width=0.34\textwidth,height=0.28\textheight]{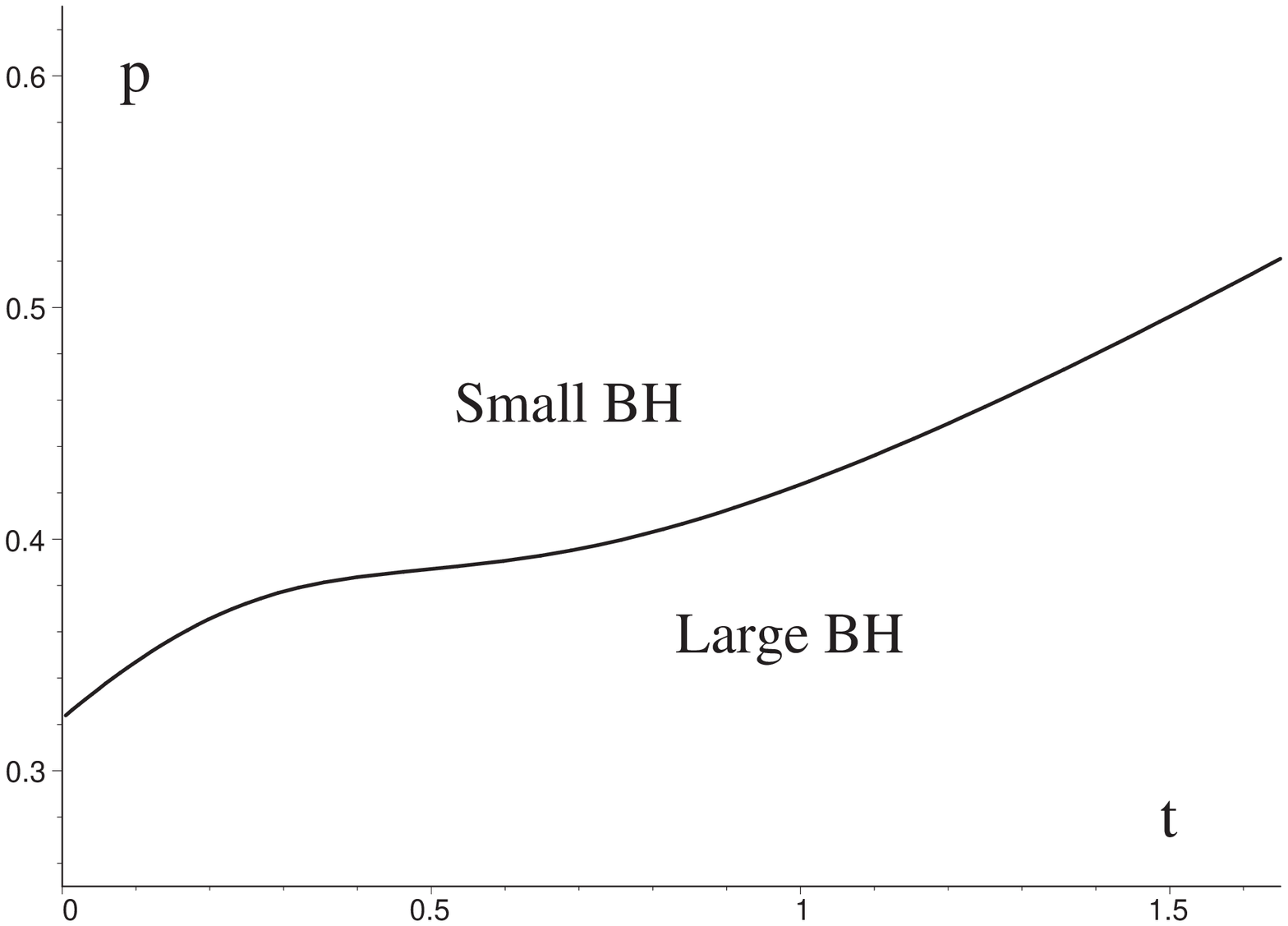}}\\
\end{tabular}
\caption{{\bf $p-t$ phase diagram for $\kappa=-1$.}
The distinct behaviour of the $p-t$ diagram is displayed for $\alpha=1, 1.65, \sqrt{3}, 1.85$. The $\alpha=\sqrt{3}$ case is a special case for which the two critical points `coincide' forming an isolated critical point discussed in Sec. ~4.5; for $\alpha\in(\sqrt{3}, 3\sqrt{3/5})$ we no longer observe a critical point, however, the first-order phase transition still persists. The case $\alpha>3\sqrt{3/5}$ demonstrates  reentrant behavior and is discussed in the next figure.  
}  
\label{Lovelock7dcrit3}
\end{figure*}

\begin{figure*}
\centering
\begin{tabular}{cc}
\rotatebox{-90}{
\includegraphics[width=0.34\textwidth,height=0.28\textheight]{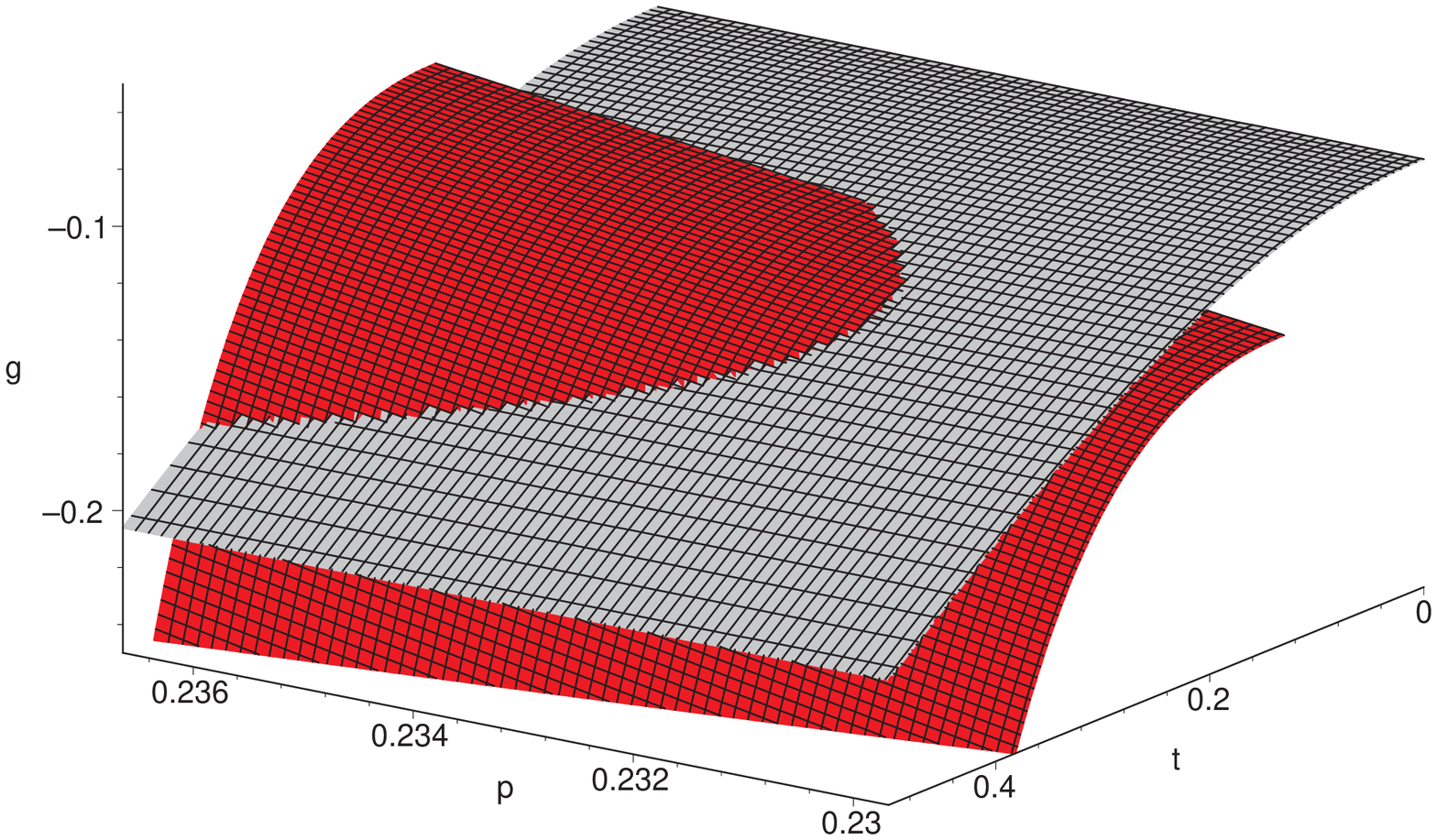}} &
\rotatebox{-90}{
\includegraphics[width=0.34\textwidth,height=0.28\textheight]{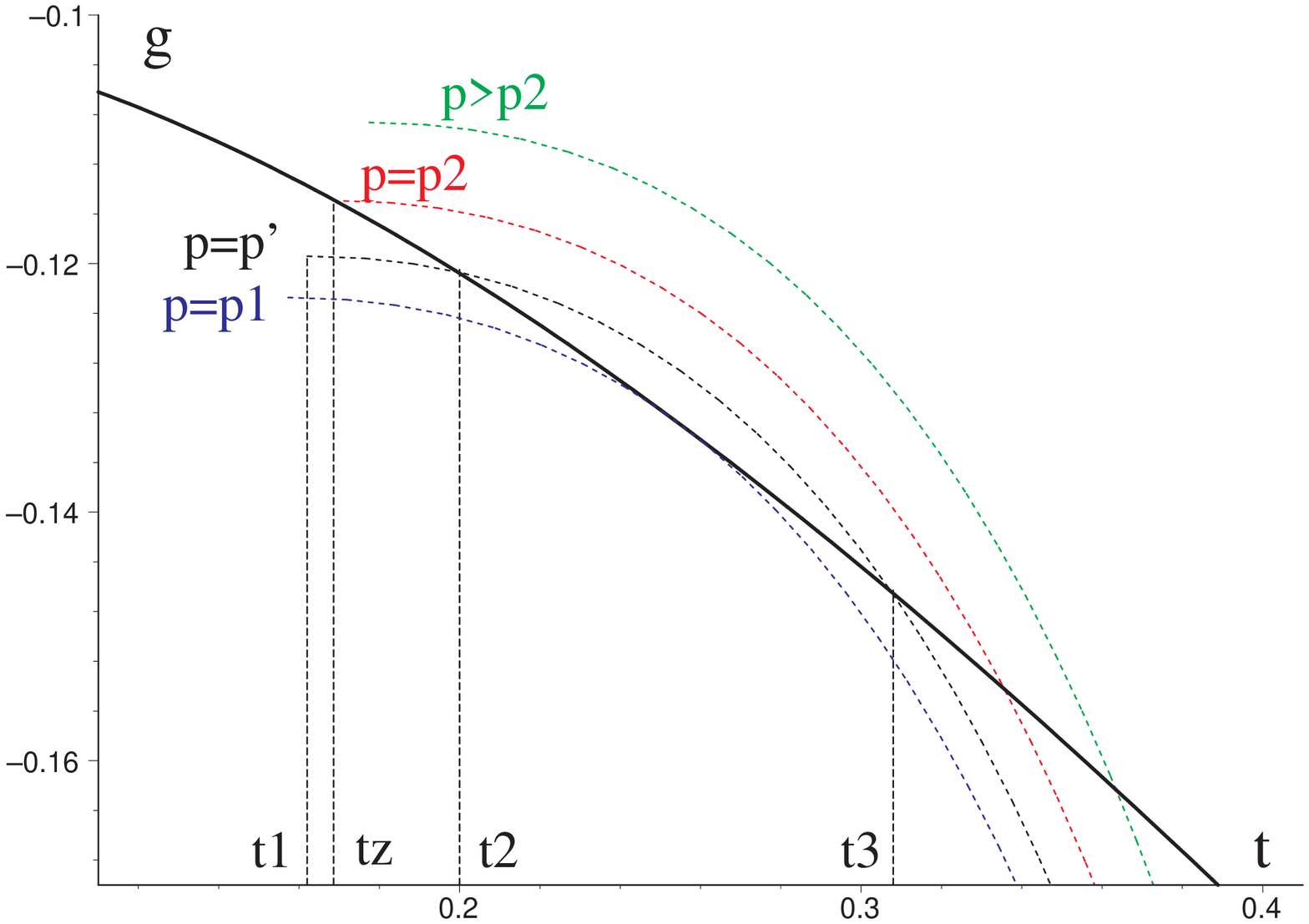}} \\
\end{tabular}
\caption{{\bf Multiple RPT: Gibbs free energy.} The Gibbs free energy is displayed for $d=7, q=0, \kappa=-1, \alpha=2.5>3\sqrt{3/5}$ case.
{\em Left} figure is a close up of  Fig.~17. {\em Right} figure represents $p=const.$ slices of the left one. We observe two branches: the branch of small black holes (displayed by a thick black curve---almost identical for various pressures) and the branch of large black holes displayed by dashed coloured curves. 
We do not display the surfaces of $g$ for negative entropy black holes.
Consequently, the Gibbs free energy seems discontinuous---the large black hole branch terminates at finite $t$. For $p\in (p_0\approx 0.218088, p_2\approx
0.2331108665$ we observe the zeroth-order phase transition. More interestingly, for any $p\in(p_1,p_2)$, the global minimum of $g$ alternates from branch to branch: small and large black hole branches double cross indicating the presence of multiple RPT behavior.  Namely, 
consider a constant pressure $p'=1.002\times p_1\in(p_1, p_2)$ displayed by thick black and dashed black curves. As the temperature decreases from say $t=0.4$, the system follows the lower dashed black curve being a large black hole, until at $t_3\approx 0.308$ the two branches cross and the system undergoes a first-order phase transition to a small black hole. As $t$ decreases even further the global minimum of $g$ corresponds to the small black hole on a thick black curve until at $t_2\approx 0.20$ another first-order phase transition, this time to a large black hole occurs. Then the system follows the dashed black curve as a large black hole until this terminates at $t_1\approx 0.162$. If the temperature is decreased even further the system jumps to the thick black curve, undergoing the zeroth-order phase transition and becoming a small black hole again. In summary, we observe reentrant large/small/large/small black hole phase transition. 
}  
\label{Lovelock7dcrit4}
\end{figure*}

\begin{figure*}
\centering
\begin{tabular}{cc}
\rotatebox{-90}{
\includegraphics[width=0.34\textwidth,height=0.28\textheight]{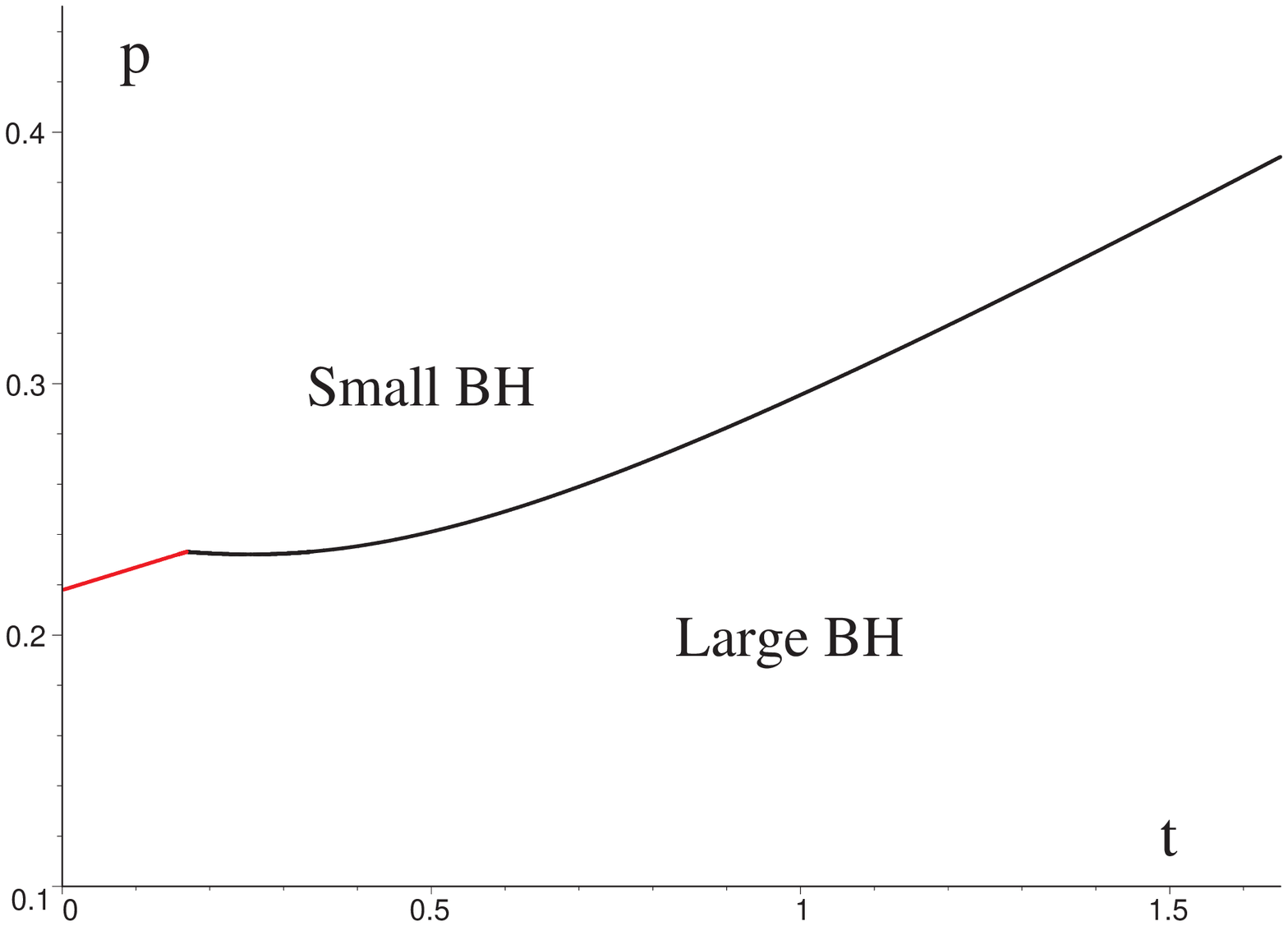}} &
\rotatebox{-90}{
\includegraphics[width=0.34\textwidth,height=0.28\textheight]{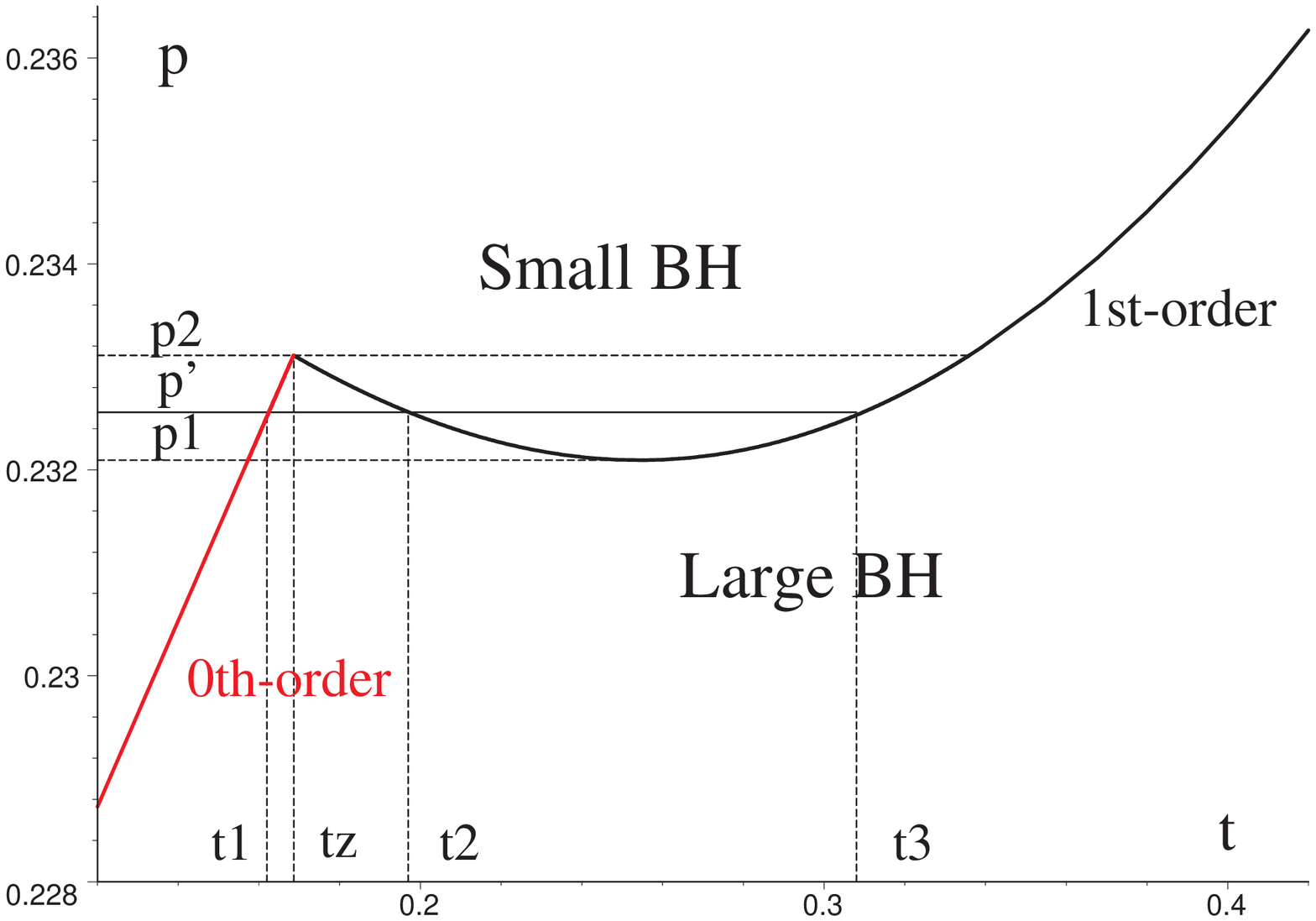}} \\
\end{tabular}
\caption{{\bf Multiple RPT: $p-t$ phase diagram.}
The $p-t$ phase diagram is displayed for $d=7, q=0, \kappa=-1, \alpha=2.5>3\sqrt{3/5}$.
The thick black curve displays the first-order phase transition between small and large black holes, the red curve stands for the corresponding  zeroth-order phase transition. 
{\em Right} figure represents a close up of the left figure. For a fixed pressure $p'\in (p_1, p_2)$ as temperature increases we may observe multiple phase transitions showing the reentrant behavior. Namely, we observe a phase transition from small black holes to large black holes, back to small black holes again, and finally to large black holes. The first transition is of the zeroth-order while the other two are of the first-order; the temperatures $t_1,t_2$ and $t_3$ coincide with those in Fig.~20. The zeroth-order phase transition terminates at $(t_z, p_2)$.
}  
\label{Lovelock7dcrit5}
\end{figure*}

\begin{figure*}
\centering
\begin{tabular}{cc}
\includegraphics[width=0.44\textwidth,height=0.28\textheight]{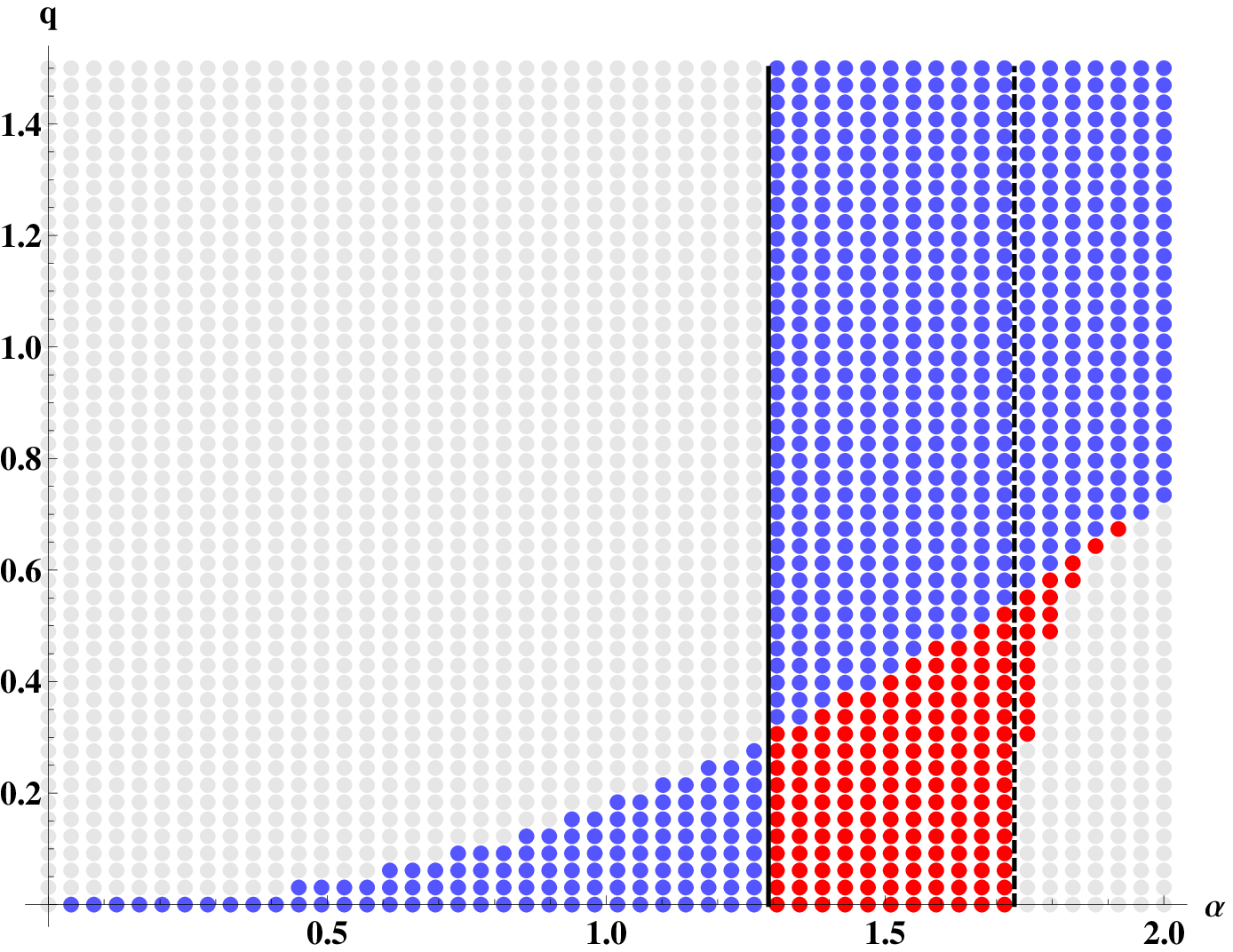} &
\rotatebox{0}{
\includegraphics[width=0.44\textwidth,height=0.28\textheight]{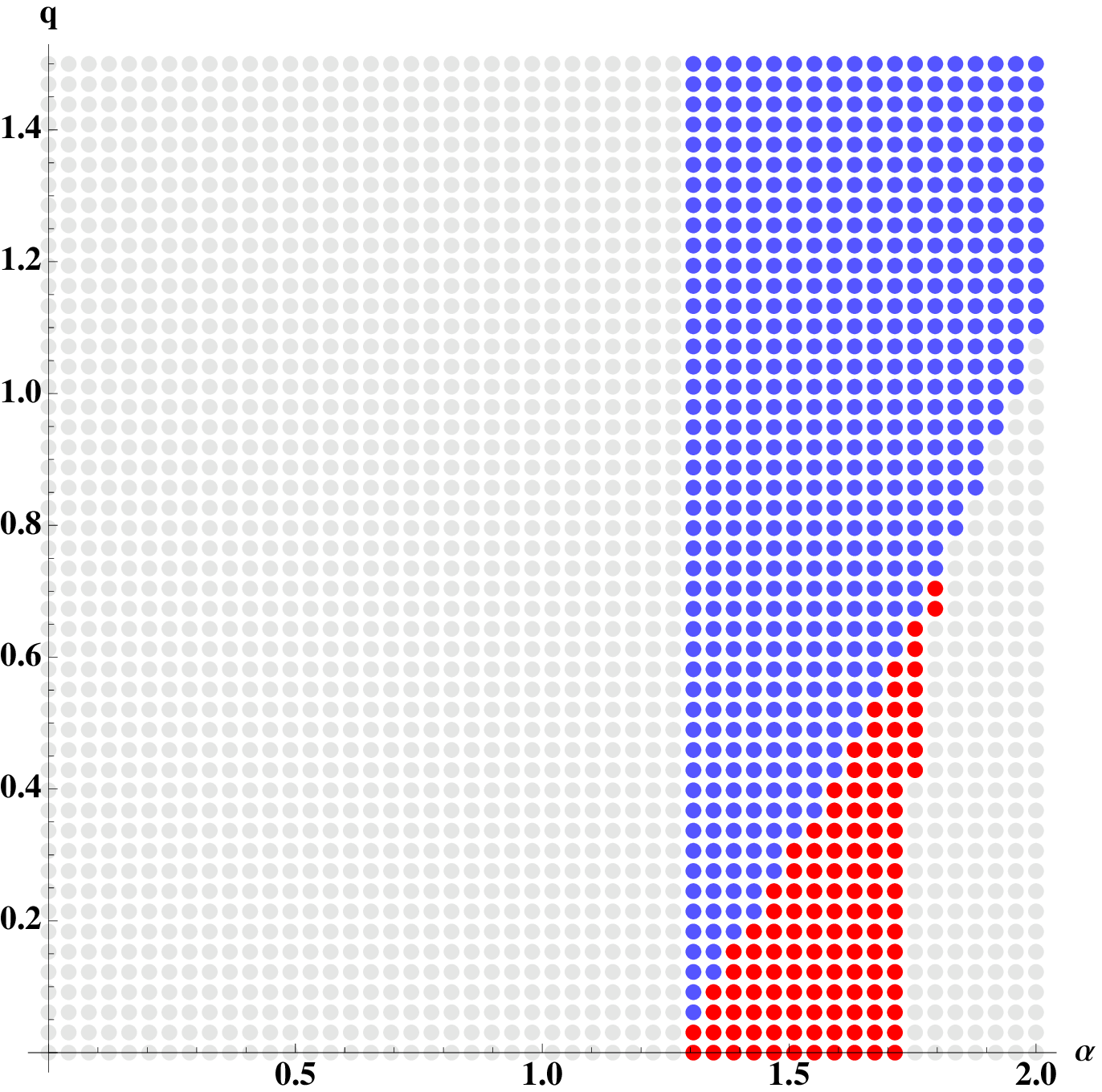}}\\
\end{tabular}
\caption{{\bf Critical points in $(q,\alpha)$-parameter space: $\kappa=-1$ case.}
The number of critical points with positive  $(p_c, v_c, t_c)$ is displayed in the $(q,\alpha)$-parameter space for $\kappa=-1$. Grey dots correspond to no critical points, blue to one critical point, and red to two; black solid and dashed lines highlight $\alpha=\sqrt{5/3}$ and $\alpha=\sqrt{3}$, respectively.
Contrary to $d=7$ ({\em left}) case, in $d=8$ ({\em right}) there are no critical points for $\alpha<\sqrt{5/3}$.
}  
\label{Fig:Lvq7a}
\end{figure*}

{\em Charged case.}
When $q$ is sufficiently small the behaviour is similar to the $q = 0$ case: namely we observe one critical point in the range $0<\alpha<\sqrt{5/3}$, two critical points for $\sqrt{5/3}<\alpha<\sqrt{3}$, and no critical points for $\alpha>\sqrt{3}$. 
More generally, the number of physical critical points in the $(q,\alpha)$-parameter space is displayed on LHS of Fig.~\ref{Fig:Lvq7a}. 
When $\alpha<\sqrt{5/3}$ we observe the standard VdW behavior in the blue region with one critical point and no critical behavior in the grey region. For $\alpha\in(\sqrt{5/3},\sqrt{3})$, as $q$ increases one of the two critical temperatures decreases and soon becomes negative. Consequently, the VdW-like swallowtail disappears and for large $q$ we observe only the reverse VdW behavior; this is displayed in Fig.~\ref{Fig:RVdW7d}. The situation for 
$\alpha\geq \sqrt{3}$ is rather complicated because of the presence of thermodynamic singularities. We postpone the detailed study of this case for future study.

\begin{figure*}
\centering
\begin{tabular}{cc}
\rotatebox{-90}{
\includegraphics[width=0.44\textwidth,height=0.28\textheight]{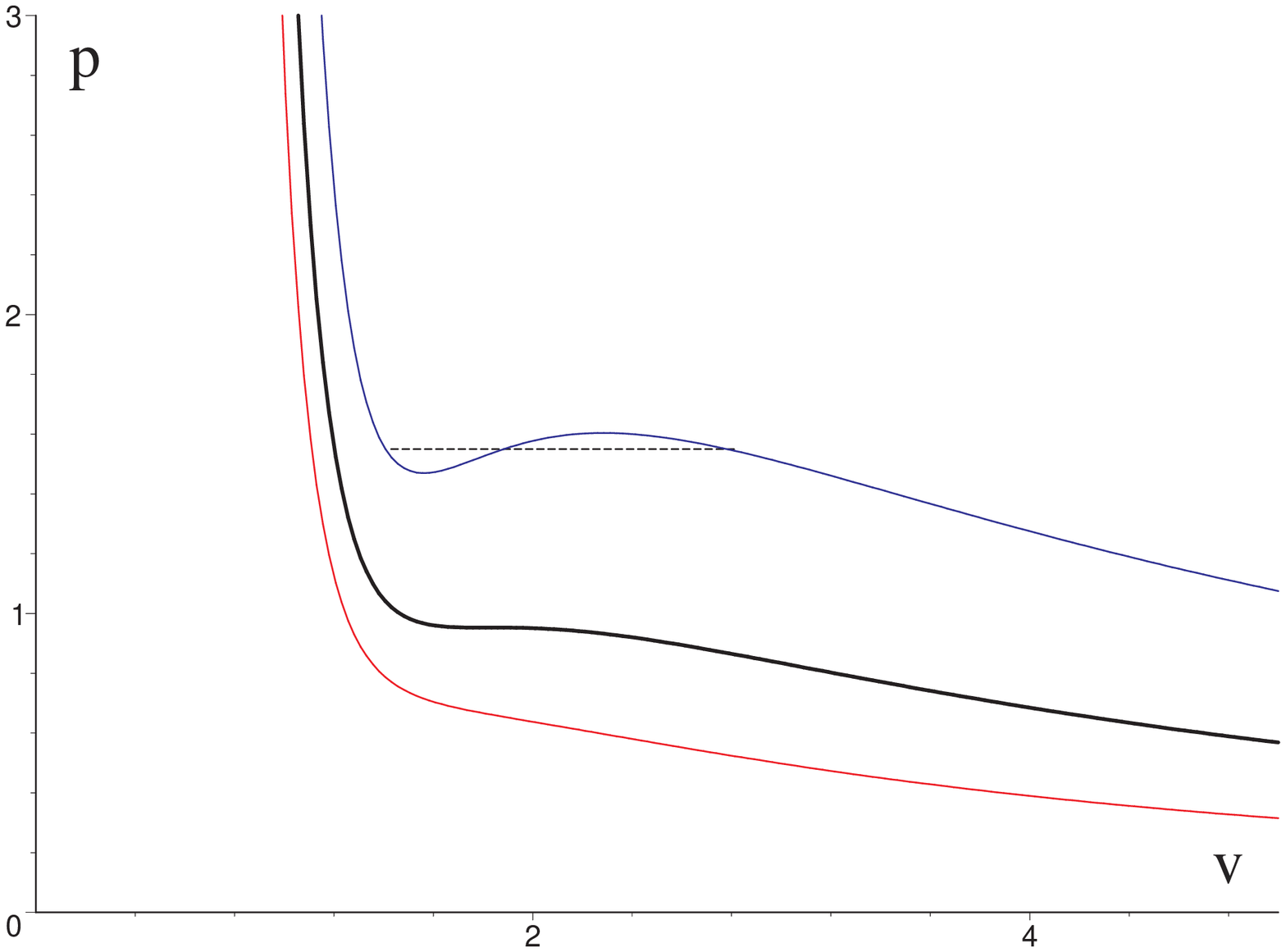}} &
\rotatebox{-90}{
\includegraphics[width=0.44\textwidth,height=0.28\textheight]{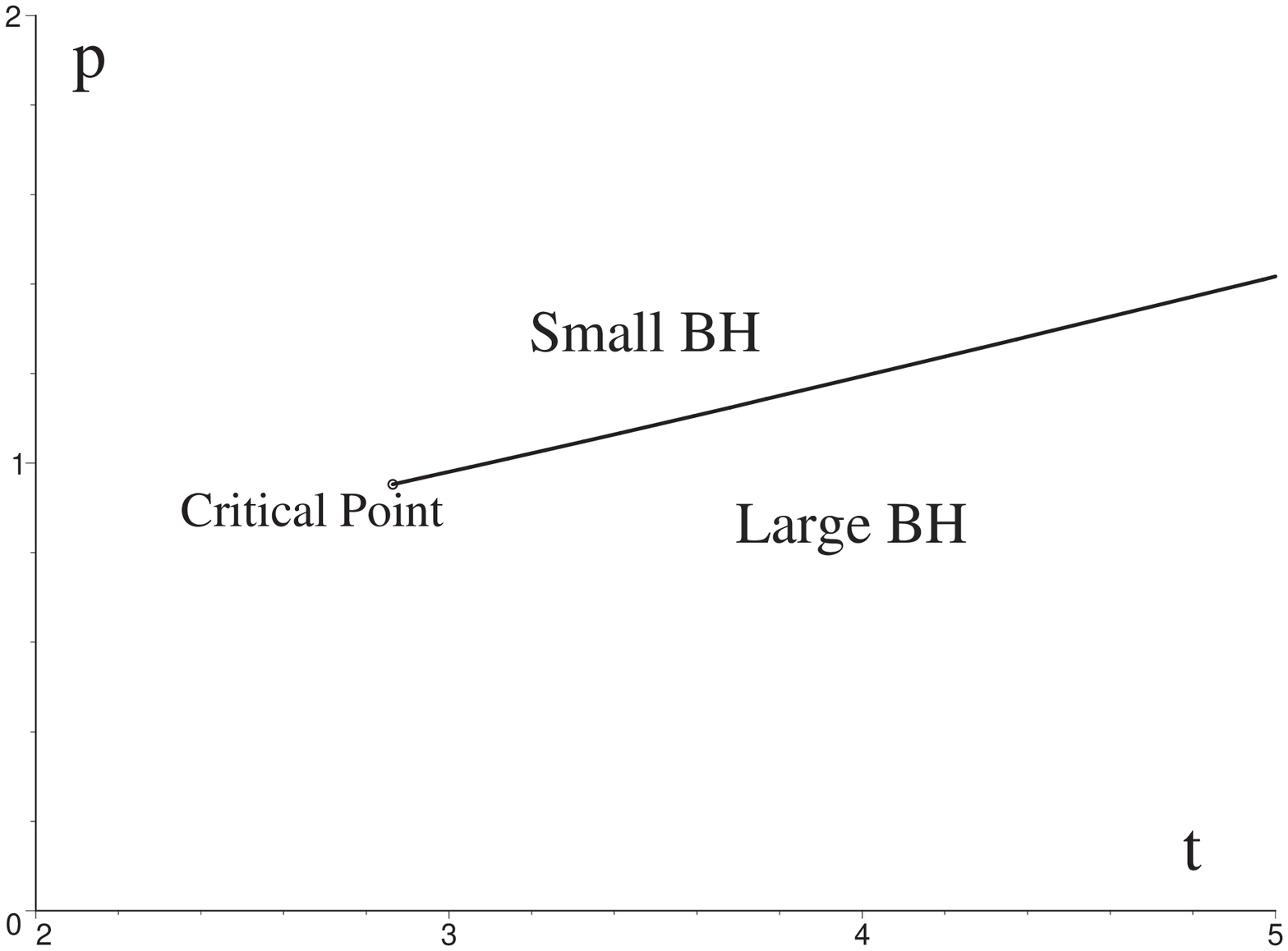}}\\
\end{tabular}
\caption{{\bf Reverse VdW behavior.}
The characteristic reverse VdW behavior is displayed for $d=7, \kappa=-1, q=1, \alpha=1.5$ case. {\em Left:} the $p-v$ diagram. {\em Right:} the $p-t$ phase diagram.  
}  
\label{Fig:RVdW7d}
\end{figure*}

\subsection{Eight dimensions}

In eight dimensions we have the following equation of state:
\be
p=\frac{t}{v}-\frac{15\kappa}{2\pi v^2}+\frac{2\alpha\kappa t}{v^3}-\frac{9\alpha}{2\pi v^4}
+\frac{3t}{v^5}-\frac{3\kappa}{2\pi v^6}+\frac{q^2}{v^{12}}\,.\label{p7}
\ee
Eqs.~\eqref{crit3} and \eqref{crit4} reduce to 
\be \label{tc7}
t_c =\frac{3}{\pi v_c(v_c^4+6 \alpha  \kappa  v_c^2+ 15)}\Bigl[3\kappa+6\alpha v_c^2+5\kappa v_c^4-\frac{4\pi q^2}{v_c^6}\Bigr]\,, 
\ee
and 
\be \label{d2p7}
5v_c^{14}-12\alpha\kappa v_c^{12}+6(6\alpha^2-35)v_c^{10}
-36\alpha\kappa v_c^8+45 v_c^6-4\pi \kappa q^2 (11v_c^4+54\kappa\alpha v_c^2+105)=0\,.
\ee
The AdS asymptotics and various black hole branches are qualitatively similar to those displayed in Figs.~\ref{LovelockAsympt} and \ref{LovelockBHAsympt}, positive entropy condition as well as thermodynamic singularities behave as in Fig.~\ref{Fig:Vss}, with $\alpha_8=2$.
The number of possible critical points with positive $(p_c, v_c, t_c)$ as we probe the $(q,\alpha)$-parameter space is displayed in Figs.~\ref{Fig:Lvq7a} and \ref{Fig:Lvq7b}.  

\begin{figure*}
\centering
\begin{tabular}{cc}
\includegraphics[width=0.44\textwidth,height=0.28\textheight]{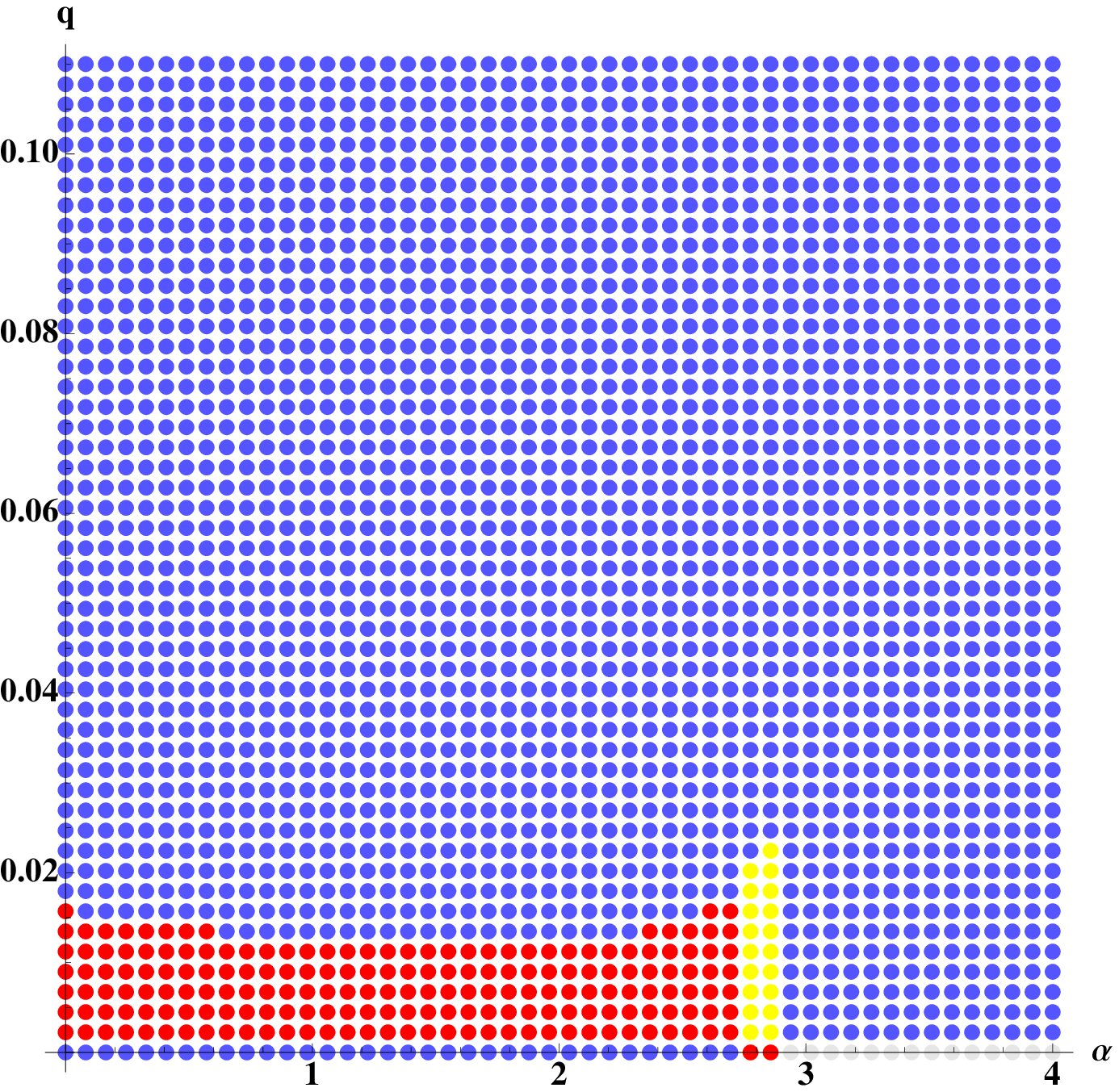} &
\rotatebox{0}{
\includegraphics[width=0.44\textwidth,height=0.28\textheight]{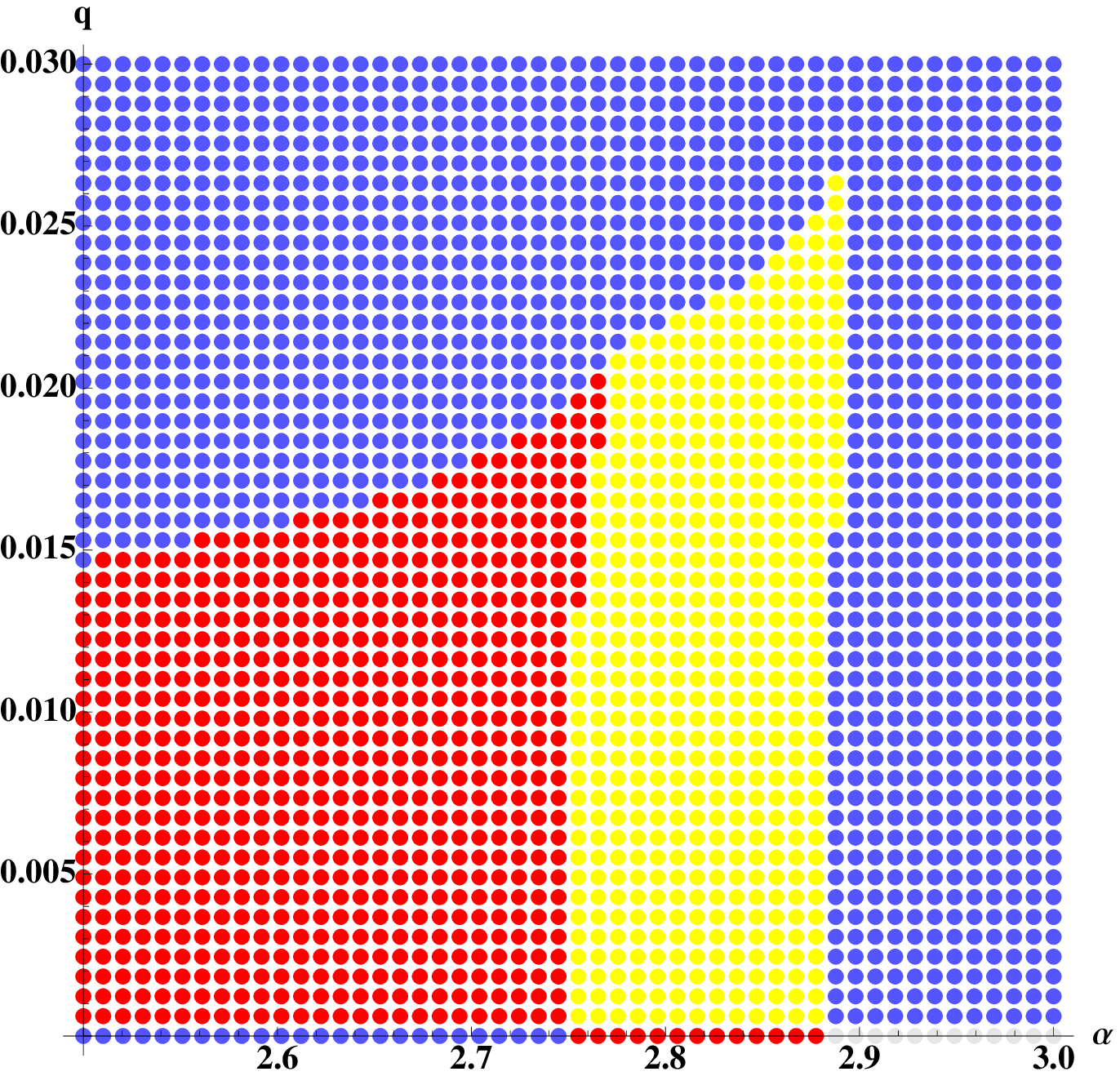}}\\
\end{tabular}
\caption{{\bf Critical points in $(q,\alpha)$-parameter space: $d=8, \kappa=+1$ case.} 
The number of critical points with positive  $(p_c, v_c, t_c)$ is displayed in the $(q,\alpha)$-parameter space; grey dots correspond to no critical points, blue to one critical point, red to two, and yellow to three. The corresponding diagram for $d=7$ is trivial (contains only the blue region with one critical point) and hence is not displayed. Although all critical points have positive  $(p_c, v_c, t_c)$, some $p_c$ may exceed the maximum pressure $p_+$ and hence occurs for a compact space.  Note also the qualitatively different behavior for $q=0$.
}  
\label{Fig:Lvq7b}
\end{figure*}

\subsubsection{Spherical case}

As with the Gauss--Bonnet $d=6$ case, the thermodynamic behavior is qualitatively different for uncharged and charged black holes. We shall not discuss this in full generality. Rather we concentrate on two important cases: i) reentrant phase transitions, present for uncharged black holes and ii) multiple first order phase transitions accompanied by a triple point, in the weakly charged case.  

\begin{figure*}
\centering
\begin{tabular}{cc}
\rotatebox{-90}{
\includegraphics[width=0.34\textwidth,height=0.28\textheight]{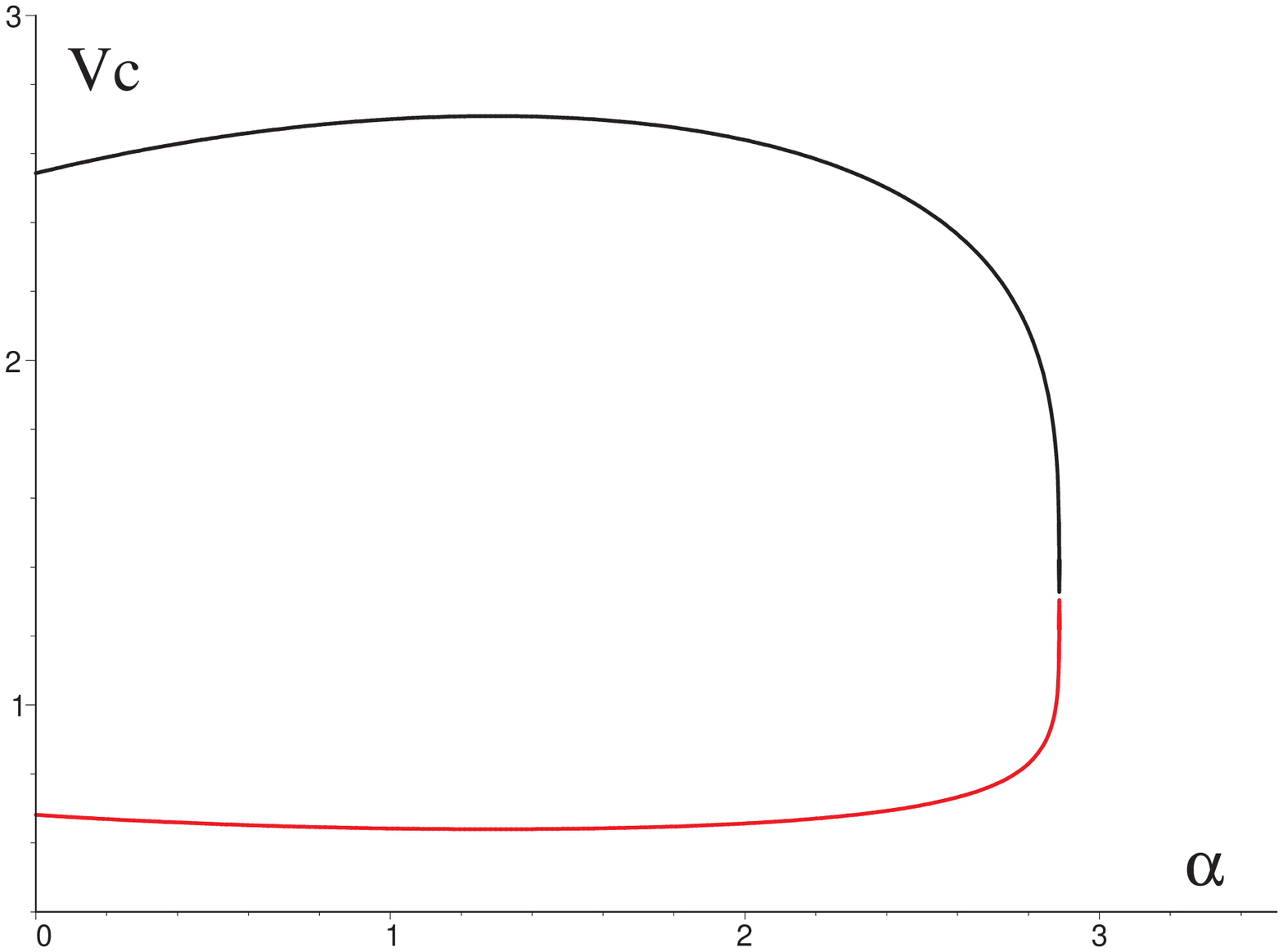}} &
\rotatebox{-90}{
\includegraphics[width=0.34\textwidth,height=0.28\textheight]{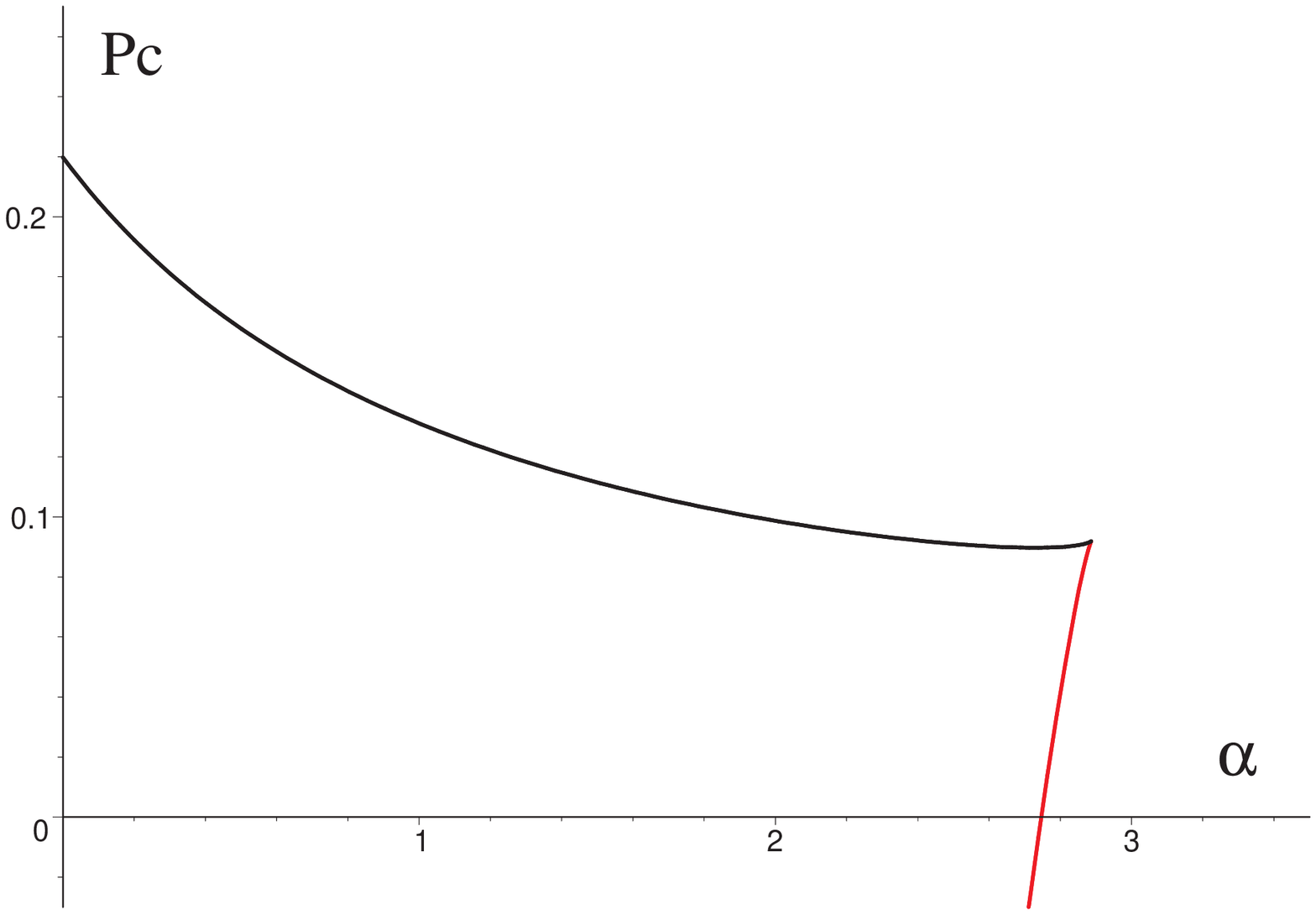}}\\
\end{tabular}
\caption{{\bf Critical points: $d=8, q=0, \kappa=1$.}
Critical volume $v_c$ and pressure $p_c$ are displayed as functions of $\alpha$. We observe that for $\alpha\in(\alpha_1\approx 2.747,\alpha_2\approx 2.886)$ we have two critical points with positive $(p_c, v_c,t_c$). However, only one of them occurs in a branch globally minimizing the Gibbs free energy. 
}  
\label{Fig:d8vc1}
\end{figure*}

\begin{figure*}
\centering
\begin{tabular}{cc}
\rotatebox{-90}{
\includegraphics[width=0.34\textwidth,height=0.28\textheight]{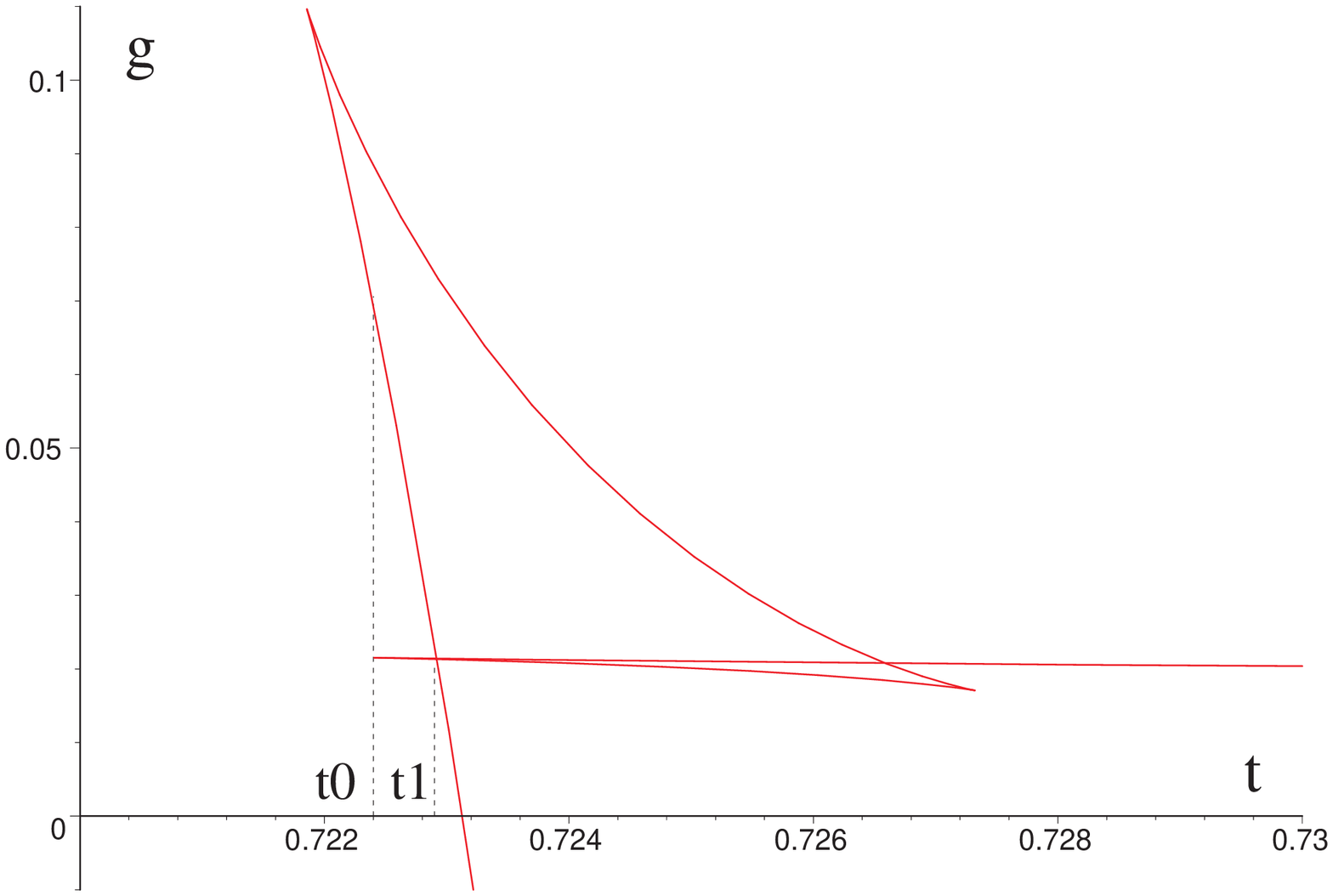}} &
\rotatebox{-90}{
\includegraphics[width=0.34\textwidth,height=0.28\textheight]{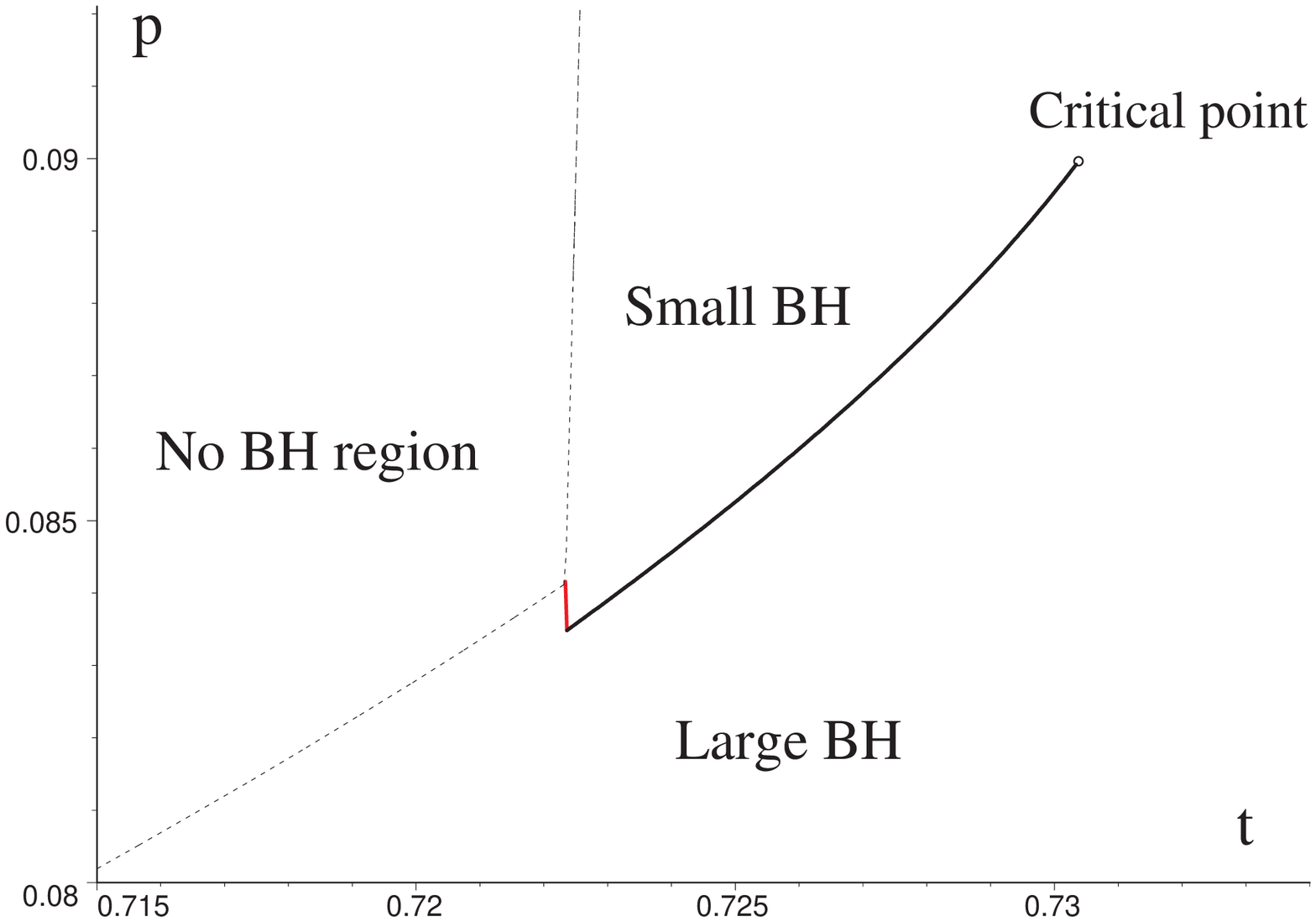}}\\
\end{tabular}
\caption{{\bf Reentrant phase transition: $d=8, q=0, \kappa=1$.}
{\em Left:} $g-t$ diagram. The characteristic behavior of the Gibbs free energy when the RPT is present is displayed for $p=0.08384$. The temperature  $t_1$ indicates the standard large/small BH first-order phase transition; $t_0$ the peculiar small/large BH zeroth-order phase transition. {\em Right:} $p-t$ diagram. 
The zeroth-order phase transition is displayed by thick red curve. The dashed curve outlines the `no black hole region'. We have set $\alpha=2.8$.
}  
\label{Fig:RPT8dBB}
\end{figure*} 

{\em Reentrant phase transition.}
When $q=0$ we may have up to two 
critical points, the corresponding $v_c$ and $p_c$ are displayed in Fig.~\ref{Fig:d8vc1}. Namely, for $\alpha<\alpha_1\approx 2.747$ we observe one critical point, for $\alpha_1<\alpha<\alpha_2\approx 2.886$ two critical points, and above $\alpha_2$ there are no critical points. Since we are in even dimension and in the absence of charge, small black holes may have arbitrarily high temperature \cite{Kastor:2011qp}. Consequently, for $\alpha<\alpha_2$ and certain range of pressures, we  observe a {\em reentrant phase transition}, similar to the one observed in \cite{Altamirano:2013ane}. For example, for $\alpha=2.8$
we display the characteristic behaviour of the Gibbs free energy on LHS of Fig.~\ref{Fig:RPT8dBB}. Looking at this figure, we observe that for high  temperature, large black holes (lower vertical curve) globally minimize the Gibbs free energy. As temperature decreases, at $t_1$ there is a first order phase transition to small black holes displayed by horizontal curve. Following this curve further, at $t=t_0$, this curve terminates and the system cannot be a small black hole anymore. Rather it undergoes a zeroth order phase transition and `jumps' to the upper vertical curve denoting the large black holes again. Hence as temperature monotonously changes from high to low the system undergoes phase transitions from
large to small and back to large black hole, a phenomenon known as a reentrant phase transition, seen  for  $d=6$ rotating black holes in the Einstein gravity  \cite{Altamirano:2013ane}.  
The corresponding $p-t$ phase diagram is displayed on r.h.s of Fig.~\ref{Fig:RPT8dBB}.

{\em Triple point.} When a small charge $q$ is added to the black hole, we find up to three critical points, shown in Fig.~\ref{Fig:RPT8da} for $\alpha=1$. Consequently the small/intermediate and intermediate/large black hole phase transitions as well as a triple point may be present. 
The Gibbs free energy exhibits two swallowtails that terminate at critical points on one side and merge together to form a triple point on the other side, see Fig.~\ref{Fig:RPT8db}. 
The triple point for example occurs  at
\be
q=0.012\,,\quad p_{3c}=0.03209\,,\quad t_{3c}=0.54729\,,
\ee
where black holes of three different sizes, $v_1=0.3948, v_2=0.4855, v_3= 9.7190$ `coexist'. Similar to the $d=6$ Gauss--Bonnet case, the $p-v$ diagram 
saturates `double Maxwell's equal area law', {in parallel with} Fig.~\ref{Fig:9}a. The corresponding phase diagram is displayed on RHS of Fig.~\ref{Fig:RPT8db}. 

\begin{figure*}
\centering
\begin{tabular}{cc}
\rotatebox{-90}{
\includegraphics[width=0.34\textwidth,height=0.28\textheight]{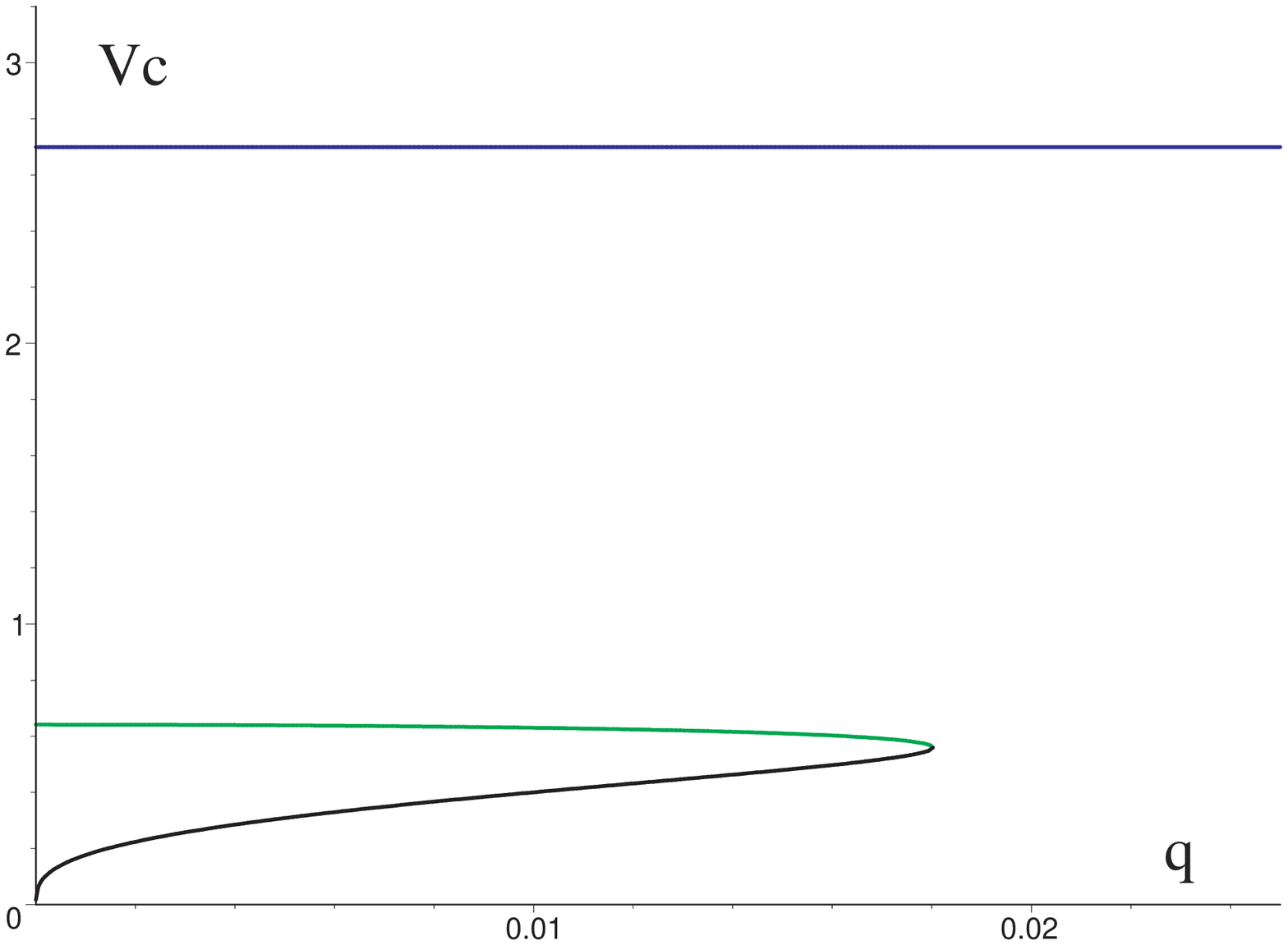}} &
\rotatebox{-90}{
\includegraphics[width=0.34\textwidth,height=0.28\textheight]{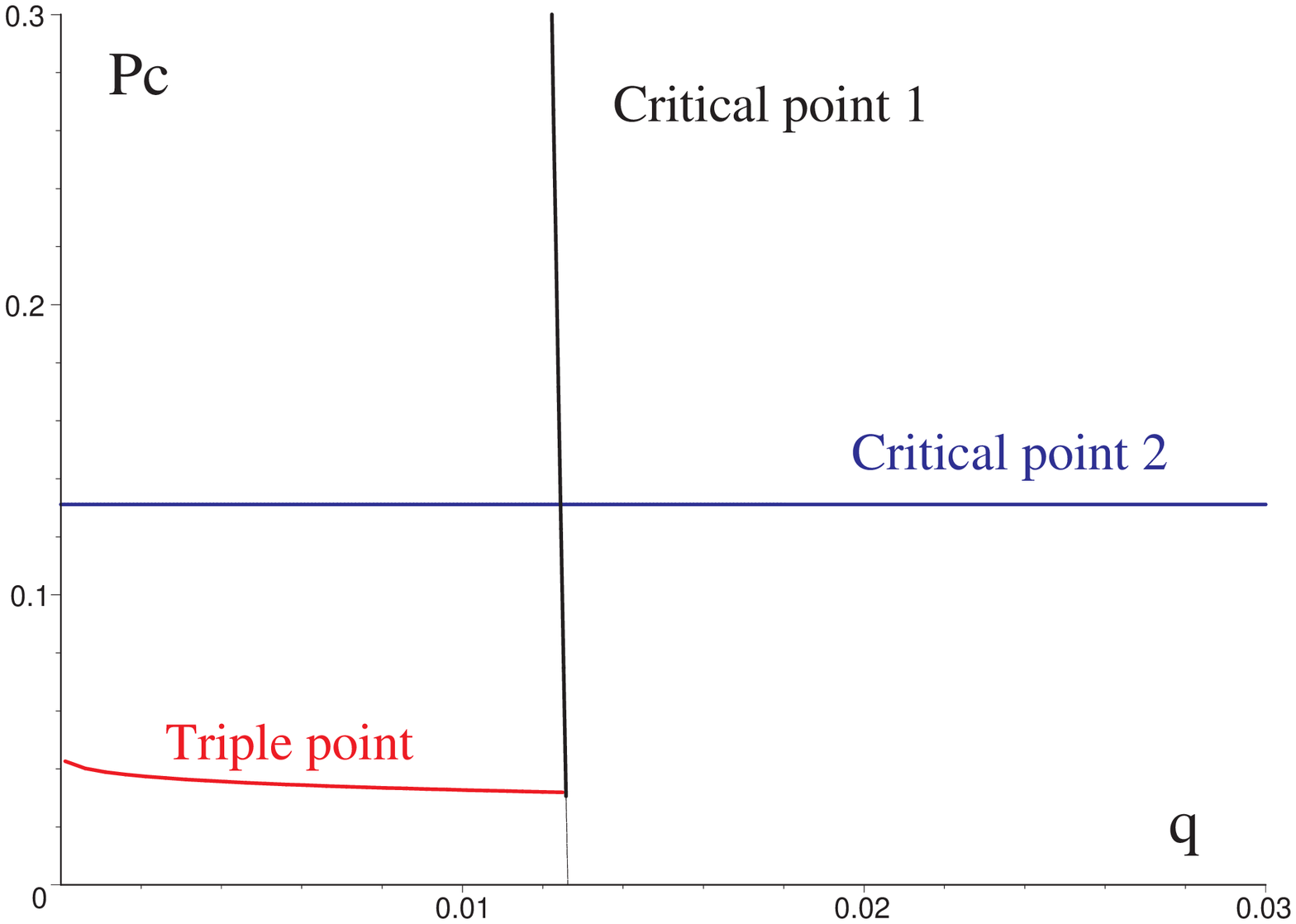}}\\
\end{tabular}
\caption{{\bf Critical points: $d=8,\kappa=1, \alpha=1$.}
{\em Left:} critical volume $v_c$ is displayed for $q\in(0, 0.03)$. {\em Right:} critical pressures. For certain range of $q$'s we observe the existence of a triple point.  
}  
\label{Fig:RPT8da}
\end{figure*} 

\begin{figure*}
\centering
\begin{tabular}{cc}
\rotatebox{-90}{
\includegraphics[width=0.34\textwidth,height=0.28\textheight]{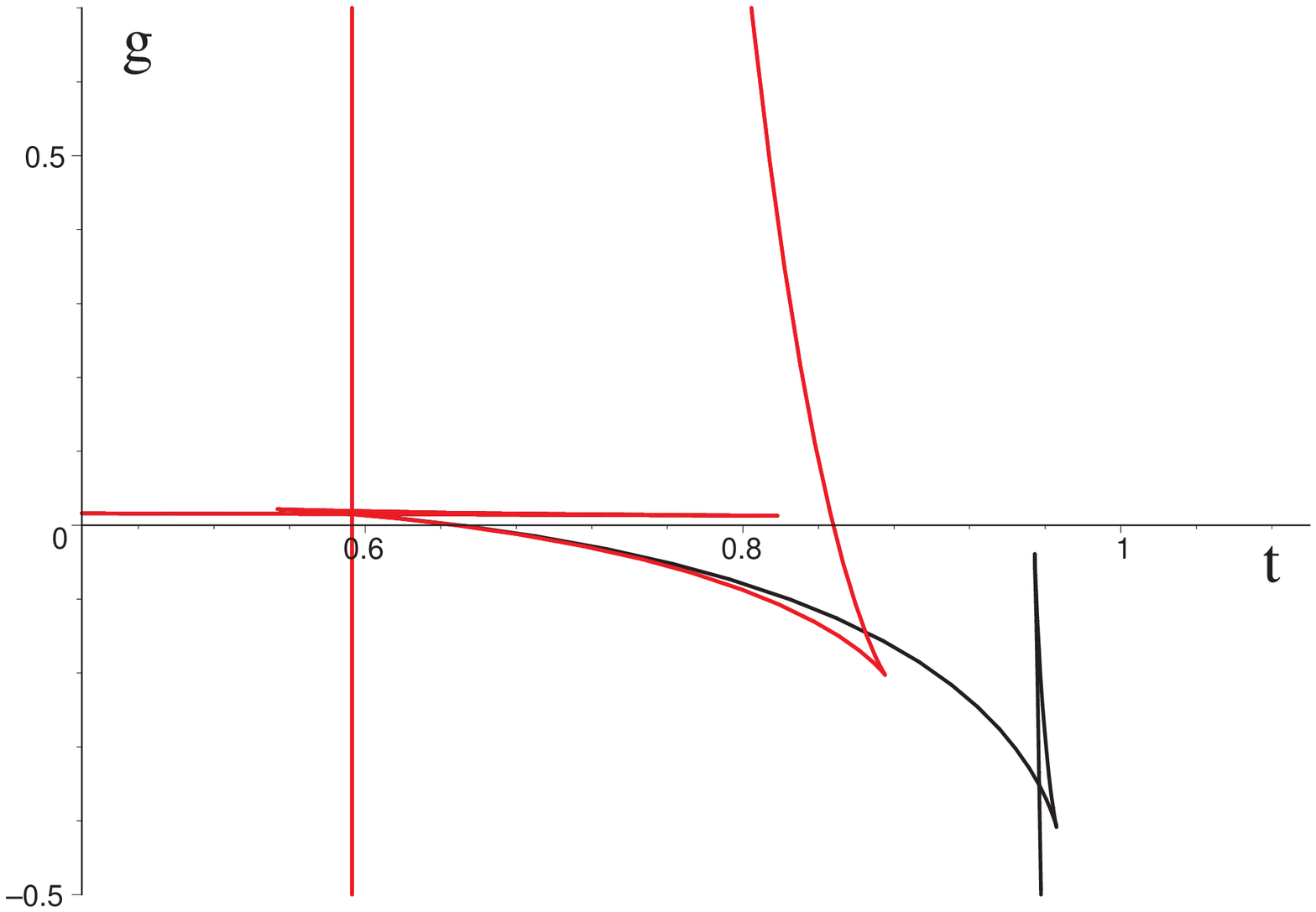}} &
\rotatebox{-90}{
\includegraphics[width=0.34\textwidth,height=0.28\textheight]{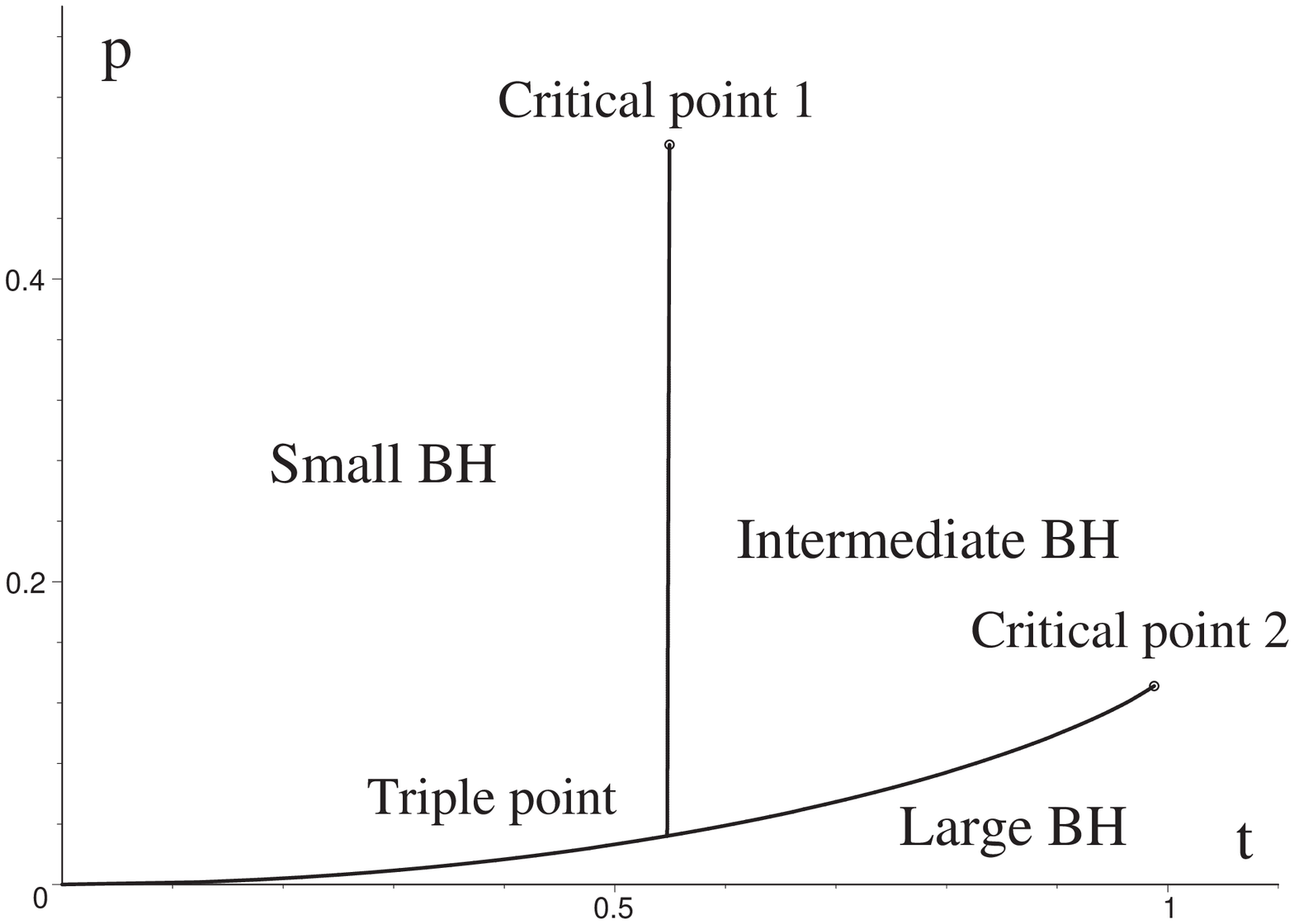}}\\
\end{tabular}
\caption{{\bf Triple point: $d=8,\kappa=1, \alpha=1$.}
{\em Left: $g-t$ diagram.} The characteristic double swallowtail indicating the presence of two first-order phase transitions is displayed  by thick black curve. When two such swallowtails `coincide' we observe a triple point. {\em Right: $p-t$ diagram.} The phase diagram possesses two first-order phase transitions that eventually terminate at critical points on one side and merge together to form a triple point on the other side. 
To signify these features we have set different charges in the two figures. Whereas left figure is displayed for $q=0.00161$
(for which the two swallowtails are apparent), for right figure we have set $q=0.012$ for which the two critical pressures are comparable. 
}  
\label{Fig:RPT8db}
\end{figure*}

\subsubsection{Hyperbolic case}

The thermodynamic behavior of $d=8$ hyperbolic Lovelock black holes is very similar to the $d=7$ case. We display the 
structure of possible critical points in the $(q,\alpha)$-parameter space on RHS of Fig.~\ref{Fig:Lvq7a}---apart from the absence of small blue region associated with the VdW behavior for $\alpha<\sqrt{3}$ the figures seem very much alike.
Specifically, for $q=0$ case the possible thermodynamic phenomena are summarized in table~\ref{Table3}.

\begin{table}
\begin{centering}
\begin{tabular}{|c|c|c|c|}
\hline
case & range of $\alpha$ & \# critical points & behavior\tabularnewline
\hline
I&$\alpha\in(0,\sqrt{5/3})$ & 0 & no critical behavior \tabularnewline
\hline
II&$\alpha\in(\sqrt{5/3},\sqrt{3})$ & 2& VdW \& reverse VdW\tabularnewline
\hline
III&$\alpha=\sqrt{3}$ & 1& special \tabularnewline
\hline
IV&$\alpha\in(\sqrt{3},2)$ & 0& infinite coexistence line \tabularnewline
\hline
V&$\alpha>2$ & 0& multiple RPT, infinite coexistence line  \tabularnewline
\hline
\end{tabular}
\protect\caption{Types of physical behavior in $d=8, \kappa=-1, q=0$ case.\label{Table3}}
\end{centering}
\end{table}

\subsection{$\alpha=\sqrt{3}$: isolated critical point}\label{Sec:alpha3}
A special case  occurs when the parameter $\alpha$ takes the particular value $\alpha=\sqrt{3}$;
 the system can be solved analytically and the solution  expressed in the simple form  
 \cite{Dehghani:2005vh, Dehghani:2009zzb}
\be
f = \kappa + \frac{r^2}{\sqrt{3\alpha_3}} \bigg[1-\Bigl(1-3\sqrt{3\alpha_3}\bigl(\alpha_0-
\frac{16\pi M}{(d-2)\Sigma_{d-2}^{(\kappa)}{r^{d-1}}}+\frac{8\pi Q^2}{(d-2)(d-3)r^{2d-4}}\bigr)\Bigr)^\frac{1}{3}\bigg]\,.
\ee
In what follows we concentrate on the $Q=0$ case. The equation of state and the Gibbs free energy are given by \eqref{EQSTATELOV} and \eqref{gLovelock}, taking the $\alpha=\sqrt{3}$ limit. While certain properties of this case in the 
context of $p-v$ criticality have been studied previously  \cite{Mo:2014qsa, Xu:2014tja, Belhaj:2014tga}, we here point out an interesting novel feature.  

For $\kappa=+1$, we find one physical critical point, with positive $(p_c, v_c, t_c)$, characterized by 
\bea\label{critPW}
v_{c} &=& 3^\frac{1}{4}\times \sqrt{\frac{d+3+2\tilde d}{d-3}}\,, \quad 
t_{c}  = \frac{3^\frac{3}{4}(d-2)(d-3)^2(d-2+\tilde d)}{6\pi \bigl(\tilde d+3d-6\bigr)\sqrt{(d+3+2\tilde d)(d-3)}}\,,  \nonumber \\
p_{c}  &=& \frac{\sqrt{3}(d-2)(d-3)^2\bigl[3(d+1)(69-5d)(d-2)-\tilde d(d^2-160d+255)\bigr]}{36\pi (\tilde d+3d-6)(d+3+2\tilde d)^3}\,, \qquad
\eea
where $\tilde d=\sqrt{(d-2)(12-d)}$. Obviously, there is no solution for $d>12$ and so the range of $d$ admitting critical points is $7\leq d \leq12$. In fact, we find that there is no criticality associated with the critical point in $d=12$ (we have a cusp similar to one in the $d=6$ Gauss--Bonnet case). In $d=10, 11$ an additional critical point emerges which, however, occurs in a branch that does not globally minimize the Gibbs free energy.   Recapitulating, for $\kappa=+1$  critical behavior occurs in $d=7,8,9,10,11$ dimensions:  in $d=7$ the critical point is associated 
with the VdW behavior, whereas in $d=8,9,10,11$ we observe a reentrant phase transition similar to the one studied in the previous subsection for $d=8$.

To study the nature of the critical point \eqref{critPW} we study its {\em critical exponents}.  The standard procedure is to Taylor expand the equation of state around this critical point. By introducing the new variables 
\be\label{omega}
\omega=\frac{v}{v_c}-1\,,\quad \tau=\frac{t}{t_c}-1\,,
\ee
we find that the equation of state expands as
\be
\frac{p}{p_c}=1+A\tau+B\tau \omega+C\omega^3+\dots\,,
\ee
with $A,B,C$ non-trivial $d$-dependent constants---implying the standard critical exponents \eqref{exponents}.

The situation is considerably different for $\kappa=-1$.  Here in any dimension $d\geq 7$ we find a single critical point \cite{Mo:2014qsa, Xu:2014tja}
at
\be\label{SpecialC}
v_c=3^{1/4}\,,\quad t_c=
\frac{d-2}{2\pi} 3^{-1/4}\,,\quad p_c= 
\frac{(d-1)(d-2)}{36\pi}\sqrt{3}=p_+=p_-\,.    
\ee
Note that this critical point occurs exactly at the thermodynamic singular point \eqref{singularLovelock}, and implies that the black hole is massless ($M=0$). 
This 
leads to a very peculiar behaviour as described below. Namely, in the $p-v$ diagram this critical point corresponds to a place where various isotherms merge together as displayed in Fig.~\ref{Lovelock7dcrit2}b.\footnote{ {Contrary to the previous cases concerning thermodynamic singularities, the isotherms do not cross here but rather merge and depart again. This is a direct consequence of the fact that in fact two thermodynamic singularities coincide at  $v=v_s$.}
}
The Gibbs free energy displays two swallowtails, both emanating from the same origin given by \eqref{SpecialC}, shown in Fig.~\ref{Lovelock7dcrit1}b. In the $p-t$ diagram,  Fig.~\ref{Lovelock7dcrit3}c,
we consequently observe an {\em isolated critical point}. Such a critical point is special as can be seen from the following expansion of the equation of state, using variables \eqref{omega}: 
\be\label{expansionCritical}
\frac{p}{p_c} = 1 + \frac{24}{d-1}\tau \omega^2  -8 \frac{(d-4)}{(d-1)} \omega^3 + \cdots
\ee
Together with the fact that the specific heat at constant volume, $C_v\propto T\frac{\partial S}{\partial T}\Bigr|_v\propto |t|^{-\tilde{\alpha}}$, identically vanishes for our Lovelock black holes as $S=S(V,\alpha_2,\alpha_3,\dots)$, we conclude that the critical exponents associated with the isolated critical point are 
\be\label{exponentsIsolated}
\tilde{\alpha}=0\,,\quad \tilde{\beta}=1\,,\quad \tilde{\gamma}=2\,,\quad \tilde{\delta}=3\,.
\ee
Such exponents are different from the swallowtail exponents \eqref{exponents}. This means that not all {\em scaling relations} remain valid for our critical point. In fact we find that the following scaling relation: 
\be
\tilde{\gamma}= \tilde{\beta}(\tilde{\delta}-1)
\ee
remains valid, whereas the equality between $2-\tilde{\alpha}=2\tilde{\beta}+\tilde{\gamma}$ is violated; in other words three instead of two of the critical exponents $\tilde{\alpha},\tilde{\beta}, \tilde{\gamma}$ and $\tilde{\delta}$ are independent.


All this looks rather interesting. However, the presence of the thermodynamic singularity that occurs exactly at $v=v_c$ seems puzzling. Is there any pathology hiding in the black hole spacetime? In particular are the black holes at the critical point and its nearby vicinity non-singular at the horizon? To answer this question, we decided to study the tidal forces that an observer falling through the black hole horizon would experience. Such forces are determined by the orthonormal components of the Riemann tensor and, in our case, depend on $f'(r_+)$ and $f''(r_+)$.\footnote{%
Alternatively, one may want to study various curvature invariants, for example the Kretschmann scalar, given by \eqref{Kretsch}. In either case the conclusions remain qualitatively the same.}
 One can show that both these quantities when expressed as functions of $p$ and $v$ are smooth and finite at $p=p_c$, however they diverge at $v=v_c$ for pressures slightly off $p_c$. For example we have
\bea
f''(v_c+dv, p_c+dp)&=&\frac{3^\frac{7}{4}\pi^2}{25}\frac{(dp)^2}{(dv)^5}+
\frac{9\sqrt{3}\pi^2}{10}\frac{(dp)^2}{(dv)^4}
+\frac{3^\frac{1}{4}57\pi^2}{20}\frac{(dp)^2}{(dv)^3}\nonumber\\
&&+\frac{\pi\sqrt{3}(8+13\sqrt{3}\pi dp)}{8}\frac{dp}{(dv)^2}+
\frac{3^\frac{3}{4}\pi(40\sqrt{3}+63\pi dp)}{40}\frac{dp}{dv}\nonumber\\
&&+\frac{\sqrt{3}(800+1100\sqrt{3}\pi dp+957\pi^2(dp)^2)}{1200}+O(dv,dp)\,.\qquad
\eea  
However, such behavior is not fatal and in fact occurs for the Gibbs free energy when expressed as function of $p$ and $v$ as well.
It is simply an artifact of the fact that at $v=v_c$ and $p\neq p_c$ the Gibbs free energy suffers from an infinite jump while at the same time the temperature blows up---implying we are infinitely far from the critical point,
\bea
g(v_c+dv,p_c+dp)&=&-\frac{3^\frac{d+5}{4}}{6(d-6)(d-4)(d-2)}\frac{dp}{(dv)^2}-\frac{2\times 3^\frac{d+4}{4}}{3(d-6)(d-4)(d-2)}\frac{dp}{dv}\nonumber\\
&&-\frac{3^\frac{d-3}{4}}{\pi (d-4)(d-6)}+O(dp,dv)\,,\nonumber\\
t(v_c+dv,p_c+dp)&=&\frac{3^\frac{3}{4}}{4}\frac{dp}{(dv)^2}+\frac{\sqrt{3}dp}{dv}+\frac{3^\frac{1}{4} 23}{16}dp+t_c+O(dp,dv)\,.
\eea    
 To cure this it is simply enough to express the Gibbs free energy as a function of $p$ and $t$, its natural variables, leaving a nice `smooth' expansion of $g$ around the critical point \eqref{SpecialC}.
Similarly, one can show that both $f'$ and $f''$ have nice expansion around the critical point\footnote{A very simple way
to check these statements, at least for $t=t_c$, is to use the expansion \eqref{expansionCritical} and the fact that at $t=t_c$, $dp\propto (dv)^{\tilde{\delta}}$ with $\tilde{\delta}=3$. Plugging this into the expansions of $f^{(n)}$ for $n=1,2,3,4$ we find that all such expansions are  finite and well-behaved.}.

To summarize, the thermodynamics around the special isolated critical point \eqref{SpecialC} seems well defined and
we have not found anything pathological about the corresponding black hole spacetimes. 
One can also show that the branches of black holes that globally minimize the Gibbs free energy (and possess non-negative
temperature) have always non-negative specific heat $C_P$ and hence are
locally thermodynamically stable, while at the critical point
we find $C_P = 0$. Further discussion of this interesting isolated point can be found in \cite{Dolan:2014vba}.

\section{Summary and Conclusions}\label{secSummary}

Our comparison of the thermodynamics of black holes in Gauss-Bonnet and 3rd-order Lovelock gravity has indicated that the latter contains interestingly and qualitatively new thermodynamic behaviour.

As in previous works, we have seen various thermodynamic phenomena, such as Van der Waals behaviour, reentrant phase transitions (RPT), and tricritical points. All these phenomena naturally and generically occur in the context of 3rd-order Lovelock gravity. For example we confirmed the existence of a tricritical point in $d=8,9,10$ dimensions in the case of charged black holes and the existence of RPT in $d=8,9,10,11$ dimensions for the electrically neutral ones. Moreover, we have seen 
`multiple RPT' behaviour, in which the Gibbs free-energy is continuous at the phase transition point.  This feature has not previously been noted.

In the case of hyperbolic $\kappa=-1$ black holes we generically find thermodynamic singularities, in which all isotherms cross at a particular value of $v$ in the $p-v$ diagram. The corresponding Gibbs free energy suffers from `infinite jump' and undergoes `reconnection'.  In particular, we may observe a form of swallowtail in which one end of the swallowtail `goes to infinity'.  Since the global minimum of the Gibbs free energy is always well-defined, we can still make sense of thermodynamics. We also observe regions where black holes have negative entropy. We have excluded these from thermodynamic considerations. However there has been a recent proposal in which negative entropy is interpreted in terms of heat
flow out of a volume \cite{Johnson:2014xza,MacDonald:2014zaa, Johnson:2014pwa}.  It would be interesting to see if a similar interpretation holds for Lovelock black holes.

We also further elucidated thermodynamic behaviour when $\alpha=\sqrt{3}$ and $\kappa=-1$ for 3rd-order uncharged Lovelock black holes 
\cite{Mo:2014qsa, Xu:2014tja}. In this interesting special case we find that the 
equation of state has non-standard expansion about a special critical point. Rather than $p/p_c=1+A\tau+B\tau\omega+C\omega^3+\dots$  (characteristic for mean field theory critical exponents and  swallowtail catastrophe behaviour) we obtain
\be
\frac{p}{p_c}=1+\frac{24}{d-1}\tau \omega^2-\frac{8(d-4)}{d-1}\omega^3+\dots\,,
\ee  
suggesting a violation of certain scaling relations  and non-standard critical exponents. We shall discuss this  feature of Lovelock gravity   further in \cite{Dolan:2014vba}. 

Future work could also include the following.  Beyond $p=p_{max}$ the asymptotic structure of the spacetime changes, being
compact for $p>p_{max}$.   Perhaps there is a phase transition to such solutions?  One might also consider a possibility of identifying
an effective cosmological constant, rather than the bare cosmological constant,  with pressure \cite{Cai:2013qga} and describing the thermodynamics from that perspective, analogous to the approach taken for boson stars \cite{Henderson:2014dwa}.

Finally, an interesting question is whether one could observe a {\em quatrocritial point},  in which four first-order phase transitions coalesce. Such a point would correspond to three swallowtails merging together, or alternatively, to three Van der Waals oscillations of a single isotherm such that the equal areas occur for the same pressure.  A necessary (not sufficient) condition for the existence of a quatrocritical point is the existence of 3 maxima and 3 minima for a single isotherm; in other words $\partial p/\partial v=0$ would have to have 6 solutions. 
This can never happen for $U(1)$ charged Gauss--Bonnet black holes. However, it might in principle occur for 3rd-order Lovelock black holes, though 
our results and preliminary study in higher dimensions shows this unlikely. 
Whether or not a quatrocritical point can be found for (possibly higher-order) Lovelock black holes remains to be seen.

\section*{Acknowledgments}

AMF acknowledges support received from the Helmholtz International Center for FAIR (HIC for FAIR), H-QM and the Department
of Physics and Astronomy, University of Waterloo,Waterloo, ON, Canada for the kind hospitality during the period of work on this project.
DK is supported by Perimeter Institute.  
Research at Perimeter Institute is supported by the Government of
Canada through Industry Canada and by the Province of Ontario through the
Ministry of Research and Innovation. This work was supported in part
by the Natural Sciences and Engineering Research Council of Canada.


\providecommand{\href}[2]{#2}\begingroup\raggedright\endgroup

\end{document}